\documentclass[times,preprint]{aastex631}

\newcommand{\texthide}[1]{}
\usepackage{xspace}
\usepackage{natbib}
\usepackage{subfigure}
\usepackage{savesym}
\savesymbol{tablenum}
\usepackage{siunitx}
\usepackage{amsmath}
\usepackage{datetime}
\usepackage{threeparttable}
\usepackage{tabularx}
\usepackage{longtable}
\usepackage{multirow}
\usepackage{graphicx}
\usepackage{colortbl,array,xcolor} 
\usepackage{capt-of}

\defcitealias{Wang2021}{Paper I}
\defcitealias{Wang2020}{W20}

\usepackage[encapsulated]{CJK}
\usepackage{ucs}
\usepackage[utf8x]{inputenc}
\newcommand{\cntext}[1]{\begin{CJK*}{UTF8}{gbsn}#1\end{CJK*}}
\newcommand{\jptext}[1]{\begin{CJK*}{UTF8}{min}#1\end{CJK*}}

\restoresymbol{SIX}{tablenum}
\newdateformat{monthyeardate}{%
	\monthname[\THEMONTH] \THEYEAR}

\received{}
\revised{}
\accepted{}
\submitjournal{ApJ}
\shorttitle{}
\shortauthors{Wang et al.}

\begin{document}

\title{Architecture of planetary systems predicted
  from protoplanetary disks observed with ALMA II:
  evolution outcomes and dynamical stability}

\correspondingauthor{Shijie Wang}
\email{shijie.wang13@alumni.imperial.ac.uk}

\author[0000-0002-5635-2449]{Shijie Wang (\cntext{汪士杰})}
\affil{Department of Physics, The University of Tokyo, Tokyo 113-0033, Japan}

\author[0000-0001-7235-2417]{Kazuhiro D. Kanagawa (\jptext{金川和弘})}
\affiliation{Research Center for the Early Universe, School of Science, The University of Tokyo, Tokyo 113-0033, Japan}
\affiliation{College of Science, Ibaraki University, Mito 310-0056, Japan}

\author[0000-0002-4858-7598]{Yasushi Suto (\jptext{須藤靖})}
\affiliation{Department of Physics, The University of Tokyo,
	Tokyo 113-0033, Japan}
\affiliation{Research Center for the Early Universe, School of Science, The University of Tokyo, Tokyo 113-0033, Japan}

\begin{abstract}
Recent ALMA observations on disk substructures suggest the presence of
embedded protoplanets in a large number disks. The primordial
configurations of these planetary systems can be deduced from the
morphology of the disk substructure and serve as initial conditions
for numerical investigation of their future evolution. Starting from
the initial configurations of 12 multi-planetary systems deduced from
ALMA disks, we carried out two-stage N-body simulation to investigate
the evolution of the planetary systems at the disk stage as well as
the long term orbital stability after the disk dispersal. At the disk
stage, our simulation includes both the orbital migration and
pebble/gas accretion effects. We found a variety of planetary systems
are produced and can be categorised into distant giant planets,
Jupiter-like planets, Neptune-like planets and distant small
planets. We found the disk stage evolution as well as the final
configurations are sensitive to both the initial mass assignments and
viscosity. After the disk stage, we implement only mutual gravity
between star and planets and introduce stochastic perturbative
forces. All systems are integrated for up to \SI{10}{Gyr} to test
their orbital stability. Most planetary systems are found to be stable
for at least \SI{10}{Gyr} with perturbative force in a reasonable
range. Our result implies that a strong perturbation source such as
stellar flybys is required to drive the planetary system unstable. We
discuss the implications of our results on both the disk and planet
observation, which may be confirmed by the next generation telescopes
such as {JWST} and {ngVLA}.
\end{abstract}


\section{Introduction\label{sec:intro}}

For the last few decades, we have witnessed numerous unexpected and
surprising discoveries in exoplanetary researches. The observed
exoplanetary population shows a wide distribution in size, mass,
semi-major axis, eccentricity and inclinations
\citep[e.g.][]{Henry2000,Charbonneau2000,
  Queloz2000,Borucki2010,Ricker2014,Winn2014}, as exemplified by the
discovery of close-in Hot Jupiters \citep[e.g.,][]{Mayor1995},
far-away gas giants \citep[e.g.,][]{Chauvin2004}, planets with large
spin-orbit misalignment \citep[e.g.,][]{Winn2005,Winn2009},
multi-planetary systems consisting of earth-like planets
\citep[e.g.,][]{Gillon2017}, and widely separated super-Jupiters
\citep[e.g.,][]{Marois2008} among others. The observational
breakthroughs have been followed by significant progress in
theoretical understanding of formation and evolution of planetary
systems, including but not limited to planetary migration, dynamical
instability, planetary population synthesis model, tidal
circularisation, secular perturbation, and pebble accretion
\citep[e.g.,][]{Lissauer1993,Lin1993,Rasio1996,
  Tanaka1999,Wu2003,Ida2004,Alibert2005,Nagasawa2008,
  Mordasini2009,Ormel2010,Naoz2011,Lambrechts2012}. Nevertheless,
there is no established formation model that consistently explains the
universality and diversities in the observed properties of
exoplanetary systems.

While it is widely believed that planets form from protoplanetary
disks (PPDs), the initial conditions responsible for planetary systems
are highly uncertain, and reliable theoretical predictions are
difficult.  Due to the intrinsic difficulty in predicting the
architecture of protoplanetary systems from PPDs, it has been
conventional to examine the outcome of different evolution models of
planets by employing somewhat artificial initial
conditions. Therefore, even if such approaches successfully reproduce
certain properties of specific populations of observed systems, it is
not clear to what extent it can be interpreted as a strong argument in
favour of the proposed theoretical model unless the initial condition
itself is justified.

This dilemma is being overcome by the high-angular resolution imaging
achieved by the Atacama Large Millimeter Array (ALMA), which has
revealed a variety of substructures in the nearby PPDs \citep[e.g.,][]
{Partnership2015,Andrews2018,Huang2018a,Long2018,VanderMarel2019}. In
particular, their concentric rings and gaps are exactly what have been
predicted by theories: sufficiently massive planets exert torques on
the surrounding disk and create dust density gaps and associated rings
\citep[e.g.,][]{Lin1993,Paardekooper2004,Crida2006}. Thus, the first
discovery of such substructures for the HL Tau system
\citep{Partnership2015} has attracted a lot of attention. The growing
number of the PPDs with such substructures may be used as
observationally motivated initial conditions for the protoplanetary
systems, which bypass unavoidable theoretical uncertainties mentioned
above.

According to the widely accepted interpretation, the mass and location
of the unseen embedded planets may be basically estimated from the
radius and width of the gaps. One serious drawback is the fact that
although ALMA can identify the morphology of the dust component of the
PPDs with high-spatial resolution, it is difficult to detect the faint
gas emission with the resolution as high as that of the dust
emission. Since the opening mechanisms for gas gaps and dust gaps are
physically different, the estimated mass for the embedded planets
depends on the assumption whether the gas/ring structures are present
in the dust component alone
\citep[e.g.,][]{Dipierro2016,Rosotti2016,Dipierro2017}, or similarly
exist for the gas component as well
\citep[e.g.,][]{Kanagawa2016,Dong_Fung2017,Zhang2018}.

The impact of this fundamental limitation on the planetary mass
prediction has been largely overlooked before, but was examined
recently by \cite{Wang2021} (hereafter
\citetalias{Wang2021}). According to the pebble accretion model
\citep[e.g.][]{Morbidelli2012,Lambrechts2014,Bitsch2018}, a
protoplanet cannot open even a shallow gas gap before its mass reaches
a threshold value, called the pebble isolation mass, $M_{\rm
  iso}$. Thus, $M_{\rm iso}$ serves as a boundary to distinguish the
gaps exist in the dust disk alone or in both the gas and dust
disks. In other words, the mass of embedded planets $M_{\rm p, dust}$
estimated from the dust gap alone assumption should be less than
$M_{\rm iso}$, while that estimated from the gas-dust gap assumption
$M_{\rm p, gas}$ \citep{Kanagawa2016} should exceed $M_{\rm
  iso}$. \citetalias{Wang2021} applied this consistency relation for
12 ALMA disks with clear multiple gaps, and estimated the planetary
masses in a way that is consistent to the planet formation scenario.

The main purpose of the current paper is to follow the evolution of
the 12 embedded multi-planetary systems starting from the
observational motivated initial conditions, predict their outcomes,
and compare with the architecture of the observed exoplanetary
systems. This approach has been attempted earlier for the HL Tau
system by \cite{Wang2020} (hereafter \citetalias{Wang2020}), in which
its prominent three gaps have been assumed to be carved by three
protoplanets with masses exceeding $M_{\rm iso}$.  We found that the
resulting planetary systems are basically stable over \SI{10}{Gyrs},
and they resemble several widely-separated giant planets observed by
direct imaging including HR 8799, $\beta$ Pictoris and PDS 70. In the
present paper, we extend \citet{Wang2020} to the 12 ALMA disks with
multiple-gap substructure. This allows us to interpret the previous
result in a more general context, not particular to the specific case
of the HL Tau.

The rest of the paper is organized as follows. Section
\ref{sec:planetary_systems} provides an overview of our investigation
and summarises the results of \citetalias{Wang2021} that the masses
and locations of the possible protoplanets embedded in the 12 ALMA
disks, which serve as the initial conditions for the simulations in
the present paper. Section \ref{sec:accretion-migration} describes our
disk-planet interaction model including the pebble and gas accretion
and orbital migration The results of simulations at the end of the
disk-planet interaction stage are presented in section
\ref{sec:rslt_disk_stage}, and are employed as the initial conditions
for the purely N-body simulations after the disk dispersal performed
in section \ref{sec:rslt_stochastic_force}.  In particular, we include
the stochastic perturbative force due to residual planetesimals so as
to examine the fate of few-body protoplanetary systems in a more
realistic manner.  The stability of such systems is presented in
section \ref{sec:result-stability}, and section \ref{sec:discussion}
discusses the validity, consistency and implications of the
results. Finally section \ref{sec:summary} is devoted to the summary
and conclusion of the paper.

\section{Simulation and initial conditions}
\label{sec:planetary_systems}
\subsection{Overview}
\label{sec:plt_sys_overview}
\begin{figure}[h]
	\centering
	\includegraphics[width=0.9\linewidth]{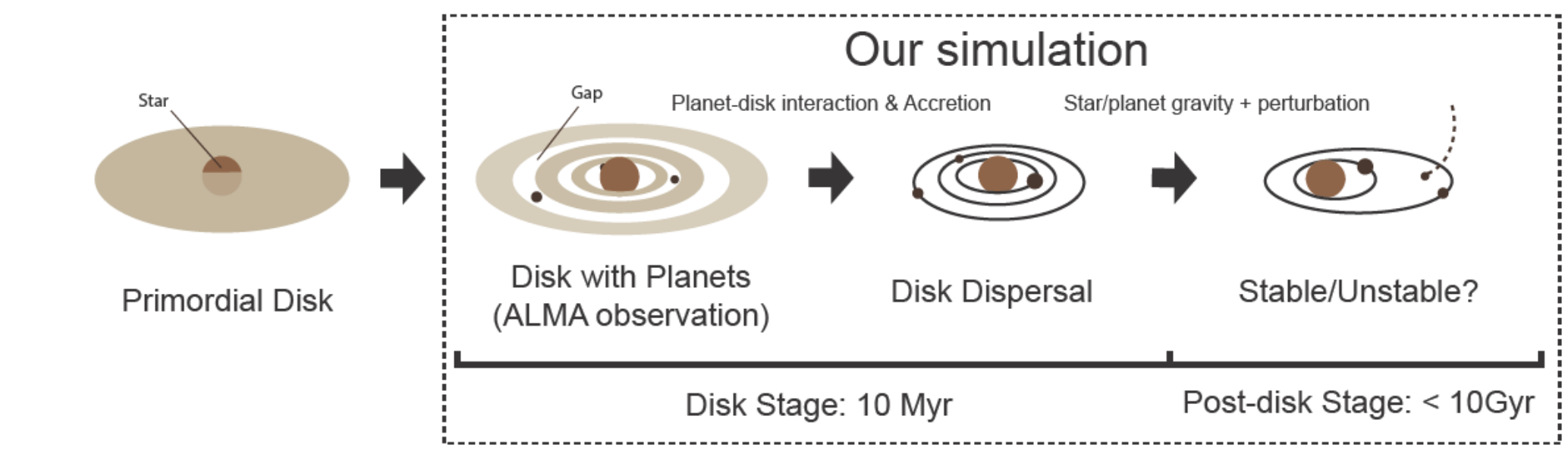}
	\caption{Schematic illustration of our simulation
			starting from the observed ALMA disks.}
	\label{fig:schematic_flow}
\end{figure}
Figure~\ref{fig:schematic_flow} illustrates the overall picture of our
simulation presented in this paper.  During the evolution of the
primordial disk around a protostar, protoplanets form and open clear
gap and ring structures in the dust disk as identified by ALMA. The
locations and masses of those embedded planets are estimated in
\citetalias{Wang2021} from the observed disk properties, as described
in the next subsection. The initial conditions for our simulation are
based on these predictions.
	
Our simulation consists of two different stages. First at the disk
stage, we compute the evolution of the system by taking account of the
mutual gravity among the central star and other planets, planet-disk
interaction, and the resulting migration and mass accretion of each
planet. The evolution model and the results are described in sections
\ref{sec:accretion-migration} and \ref{sec:rslt_disk_stage},
respectively. After the gas component of the disk is dispersed, we
switch to the second stage simulation, in which we consider the
gravitational interaction among the star and planets, and neglect
planet-disk interaction.

The above two-stage simulation method is basically the same as that
adopted by \citetalias{Wang2020}, but with improvements in several
aspects. Firstly, we predict the mass of the planet embedded in the
observed gap using a more sophisticated method that considers the
different gap-opening mechanisms in the gas and dust disks, as
presented in \citetalias{Wang2021}. Secondly, we take account of the
pebble accretion for planets below the pebble-isolation mass during
the disk stage simulation. Finally, in the post-disk stage simulation,
we introduce the stochastic perturbations due to numerous
planetesimals around each planet, in addition to the mutual gravity
among the star and other planets.

\subsection{Initial condition for protoplanets embedded
  in the observed ALMA disk systems}
\label{sec:ini_planetary_mass}

In \citetalias{Wang2021}, we found that the mass of protoplanets
embedded in the ALMA disks sensitively depends on the assumption
whether or not the unobserved gas component has the similar gap
structure as the dust component. The difference leads to fairly
different predictions for the final outcome of the planetary
systems. Therefore, we briefly summarize the main conclusion of
\citetalias{Wang2021} here, and present how we assign initial
conditions for the 12 ALMA disks that we consider in the present
analysis.

The estimated planetary masses in \citetalias{Wang2021} are based on
the observed gap substructures in PPDs reported by \cite{Long2018},
the Disk Substructures at High Angular Resolution Project (DSHARP)
\citep{Andrews2018}, and \cite{VanderMarel2019}. In this paper, we
select all disks (twelve in total) that exhibit multiple-gaps in
\citetalias{Wang2021}; out of the 12 disks, five disks have two gaps,
six disks have three gaps, and the remaining one (HD 163296) has four
gaps. The parameters for those disks are listed in Table 1 of
\citetalias{Wang2021}.

We list the mass and outer radius of the disks in Table
\ref{tab:fidu_mass_table}, which are relevant for determining the
initial surface density in section \ref{sec:inicons}. The outer radius
of the disk is based on the disk radius at millimetre wavelength given
by \cite{VanderMarel2019} and \cite{Huang2018}. The total disk mass is
also adopted from the the respective reference if provided, otherwise
it is calculated from the \SI{1300}{\micro\metre} flux using the
relation provided in \cite{Cieza2008}:
\begin{equation}
	\label{eqn:disk_mass_from_flux}
	M_{\rm disk} = \SI{1.7e-4}{M_\odot}
	\left(\frac{F_{1300\mu{\rm m}}}{\rm mJy}\right)
	\left(\frac{d}{\SI{140}{pc}}\right)^2.
\end{equation}
Since the above equation assumes the disk is optically thin, the disk mass may be underestimated when the disk is optically thick as suggested by \cite[e.g.,][]{Liu2019,Zhu2019}. Moreover, we assume the gas-to-dust mass ratio to be \num{100}, which also involves a large uncertainly. We will discuss its implication on the disk mass in section~\ref{sec:caveats}.

For the HL Tau disk, we adopt the disk mass estimated by
\cite{Kwon2015}, same as we adopted in \citetalias{Wang2020}.

Here we briefly show the method that \citetalias{Wang2021} proposed to
predict planetary mass by classifying the gaps into three groups
(there is no gap corresponding to Group IV in our selected ALMA
disks). There are three characteristic mass scales relevant in the
present analysis. If the planet is embedded in a dust gap alone in
which the gas component does not open a significant gap, the
corresponding planet is expected to have the mass
\citep[e.g.,][]{Lodato2019}:
\begin{equation}
	\label{eqn:pltmass_Hill}
	M_{\rm p, dust}(R_{\rm gap})= 6.01M_\oplus
	\left(\frac{M_*}{M_\odot}\right)
	\left(\frac{\Delta_{\rm gap}/R_{\rm gap}}{0.1}\right)^{3}, 
\end{equation}
where $ R_{\rm gap} $ and $ \Delta_{\rm gap} $ are the radius and
width of the observed gap, and $ M_* $ is the mass of the central
star. In contrast, if the gas component exhibits the same gap
structure as observed for the dust component, the mass of such a
planet is predicted and empirically given by \cite{Kanagawa2016} as
\begin{align}
	\label{eqn:pltmass_Kanagawa}
	M_{\rm p, gas}(R_{\rm gap}) = 175M_\oplus \left(\frac{M_*}{M_\odot}\right)
	\left(\frac{\Delta_{\rm gap}/R_{\rm gap}}{0.5}\right)^2
	\left(\frac{h_{\rm gap}/R_{\rm gap}}{0.05}\right)^{3/2}
	\left(\frac{\alpha}{\num{e-3}}\right)^{1/2},
\end{align}
where $ \alpha $ is the assumed disk viscosity.We follow
\cite{Dong2018a} and estimate the gas scale height $h_{\rm gap} $ at $
R_{\rm gap} $ the stellar temperature model, assuming a locally
isothermal disk and ideal gas equation of state \citepalias[see
  section 2,][]{Wang2021}. Adopting this model results in a vertically
flared disk with the flaring index equal to $ 0.25 $
\begin{equation}
	\label{eqn: asp_ratio_power_law}
	\frac{h}{R} = \left(\frac{h}{R}\right)_{\SI{100}{au}}
	\left(\frac{R}{\SI{100}{au}}\right)^{0.25}.
\end{equation}
The important scale that distinguishes between the above two cases is
the pebble isolation mass $M_{\rm iso}$. When a planet becomes more
massive than $M_{\rm iso}$, it starts to open a shallow gas gap that
generates a pressure maximum at its vicinity. Thus the pebble accretion
is terminated, and instead, run-away gas accretion onto the planet is
triggered by the cooling. The planet grows quickly in mass to open a
deep gap in both the gas and dust disks. Therefore, $M_{\rm p,
  dust}<M_{\rm iso}$ and $M_{\rm p, gas}>M_{\rm iso}$ should
essentially correspond to the conditions for the dust-only gaps and
both dust and gas gaps, respectively.

We adopt a fitting formula by \cite{Bitsch2018} for the pebble
isolation mass at $R$:
\begin{align}
	\label{eqn:peb_iso_mass}
	M_{\rm iso}(R) &= 25 M_{\oplus}\left(\frac{M_*}{M_\odot}\right)
	\left(\frac{h/R}{0.05}\right)^3f(\alpha,\eta), \\
\end{align}
where
\begin{align}
	\label{eq:f-alpha-eta}
	f(\alpha,\eta) &=
	\left[0.34\left(\frac{-3.0}{\log\alpha}\right)^4+0.66\right]
	\left(1-\dfrac{\eta + 2.5}{6}\right),
\end{align}
with $\eta = \partial\ln P/\partial\ln R$ being the logarithmic
pressure gradient.

\begin{figure}
	\centering
	\includegraphics[width=0.8\linewidth]{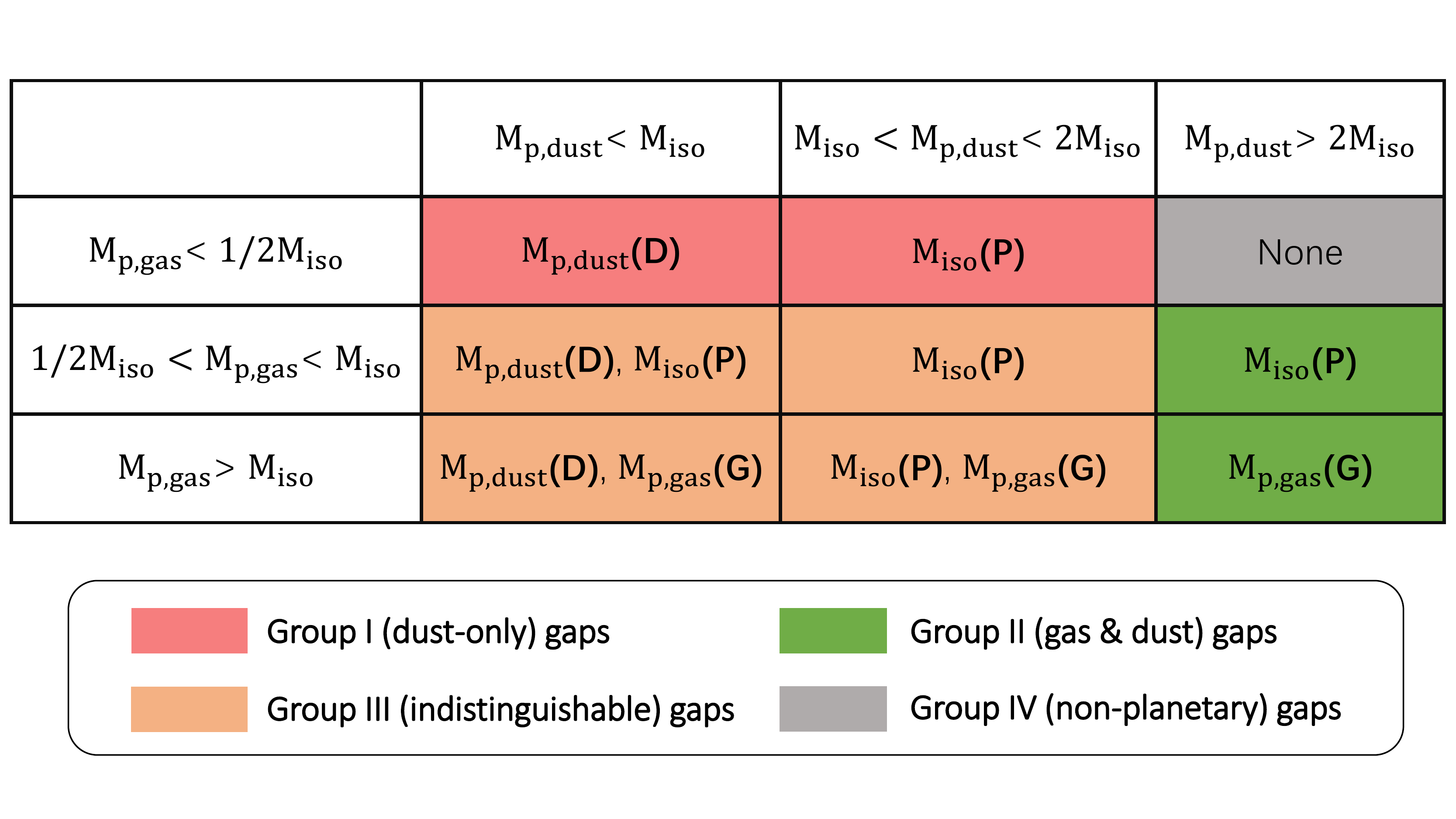}
\caption{Criteria for the planetary mass assignment and gap
          classification. Groups I gaps only open in the dust disk;
          Group II gaps open in both the gas and dust disk; Group III
          gaps are indistinguishable gaps that can be interpreted as
          either dust-only gaps or gas gaps; Group IV gaps have
          non-planetary origin. No gap belongs to Group IV in our
          sample.}
	\label{fig:massruletab}
\end{figure}

In reality, however, the values of $M_{\rm p,dust}$, $M_{\rm p,gas}$
and $M_{\rm iso}$ suffer from fairly large theoretical and
observational uncertainties. Therefore, in \citetalias{Wang2021} we
allow a fudge factor of two for the estimated masses, and classify the
observed ALMA gaps into the three groups: Group I (dust-only), Group
II (gas \& dust) and Group III (indistinguishable) gaps. The gap
classification criteria are summarised in Figure
\ref{fig:massruletab}.

In this paper, since our migration and accretion recipes are
determined by whether the planetary mass is above or below $ M_{\rm
  iso} $ without the fudge factor of two, we further refine the mass
assignment of \citetalias{Wang2021} to make the evolution of the
respective gap interpretation consistent with $ M_{\rm iso} $. The
respective criteria for the mass assignment are also summarised in
Figure \ref{fig:massruletab}. For example, for Group I gaps, we adopt
$M_{\rm p, dust}$ as the initial planetary mass only if $M_{\rm
  p,dust} < M_{\rm iso}$; otherwise, $M_{\rm iso}$ is adopted
instead. Similar rule is also applied to Group~II gaps, and for
Group~III gaps it is possible to have two possible initial mass
assignments. For convenience, in this paper, we use symbols D (dust),
G (gas), and P (pebble-isolation mass) to indicate that the mass of
the planet is set to be $M_{\rm p,dust}$, $M_{\rm p, gas}$, and
$M_{\rm iso}$ at the initial epoch of our simulations. Note that here
we do not consider Group~IV (non-planetary origin) because no gap
belongs to this group in \citetalias{Wang2021}.

\begin{table*}[!ht]
	\centering
	\caption{Disk parameters and initial planetary mass at fiducial $ \alpha = \num{e-3}$.}
	\resizebox{\linewidth}{!}{
		\begin{threeparttable}
			\begin{tabular}{lllcclclclcl}
				\multirow{3}{*}{Disk(set)}&\multirow{2}{*}{$ M_{\rm disk} $}& \multirow{2}{*}{$ R_{\rm out} $}& \multicolumn{2}{c}{Planet 1} & \multicolumn{2}{c}{Planet 2} & \multicolumn{2}{c}{Planet 3} & \multicolumn{2}{c}{Planet 4} \\
				&&& \multicolumn{2}{c}{Location(au)} & \multicolumn{2}{c}{Location(au)} & \multicolumn{2}{c}{Location(au)} & \multicolumn{2}{c}{Location(au)} \\
				& ($ \si{M_\odot} $) & (\si{au})&Type & Mass$ (\si{M_J}) $& Type& Mass$ (\si{M_J}) $& Type& Mass$ (\si{M_J}) $& Type & Mass$ (\si{M_J}) $          \\
				\hline
				\hline
				\textit{AA TAU}    & 0.0162 & 150 & \multicolumn{2}{c}{72}  & \multicolumn{2}{c}{118} &           &             &          &              \\
				\hline
				AA TAU(DD)                          &        &     & D         & 0.090       & D        & 0.031        &           &             &          &              \\
				AA TAU(PD)                          &        &     & P         & 0.175       & D        & 0.031        &           &             &          &              \\
				\hline
				\hline
				\textit{AS 209}    & 0.0366 & 139 & \multicolumn{2}{c}{8.7} & \multicolumn{2}{c}{61}  & \multicolumn{2}{c}{93}  &          &              \\
				\hline
				AS 209(GPD)                         &        &     & G         & 0.434       & P        & 0.193        & D         & 0.114       &          &              \\
				AS 209(GPP)                         &        &     & G         & 0.434       & P        & 0.193        & P         & 0.265       &          &              \\
				AS 209(GGD)                         &        &     & G         & 0.434       & G        & 0.200        & D         & 0.114       &          &              \\
				AS 209(GGP)                         &        &     & G         & 0.434       & G        & 0.200        & P         & 0.265       &          &              \\
				\hline
				\hline
				\textit{CI TAU}    & 0.0308 & 174 & \multicolumn{2}{c}{14}  & \multicolumn{2}{c}{48}  & \multicolumn{2}{c}{119} &          &              \\
				\hline
				CI TAU(GPD)                         &        &     & G         & 0.656       & P        & 0.127        & D         & 0.108       &          &              \\
				CI TAU(GPP)                         &        &     & G         & 0.656       & P        & 0.127        & P         & 0.250       &          &              \\
				CI TAU(GGD)                         &        &     & G         & 0.656       & G        & 0.132        & D         & 0.108       &          &              \\
				CI TAU(GGP)                         &        &     & G         & 0.656       & G        & 0.132        & P         & 0.250       &          &              \\
				\hline
				\hline
				\textit{DL TAU}    & 0.0374 & 147 & \multicolumn{2}{c}{39}  & \multicolumn{2}{c}{67}  & \multicolumn{2}{c}{89}  &          &              \\
				\hline
				DL TAU(DPG)                         &        &     & D         & 0.091       & P        & 0.142        & G         & 0.271       &          &              \\
				DL TAU(PPG)                         &        &     & P         & 0.095       & P        & 0.142        & G         & 0.271       &          &              \\
				\hline
				\hline
				\textit{DoAr 25}   & 0.0406 & 165 & \multicolumn{2}{c}{98}  & \multicolumn{2}{c}{125} &           &             &          &              \\
				\hline
				DoAr 25(DD)                         &        &     & D         & 0.071       & D        & 0.009        &           &             &          &              \\
				\hline
				\hline
				\textit{Elias 24}  & 0.0506 & 150 & \multicolumn{2}{c}{55}  & \multicolumn{2}{c}{94}  &           &             &          &              \\
				\hline
				Elias 24(GD)                        &        &     & G         & 0.529       & D        & 0.033        &           &             &          &              \\
				\hline
				\hline
				\textit{GO TAU}    & 0.0098 & 144 & \multicolumn{2}{c}{59}  & \multicolumn{2}{c}{87}  &           &             &          &              \\
				\hline
				GO TAU(GD)                          &        &     & G         & 0.262       & D        & 0.043        &           &             &          &              \\
				\hline
				\hline
				\textit{GY 91}     & 0.0191 & 140 & \multicolumn{2}{c}{41}  & \multicolumn{2}{c}{69}  & \multicolumn{2}{c}{107} &          &              \\
				\hline
				GY 91(DDD)                          &        &     & D         & 0.143       & D        & 0.011        & D         & 0.008       &          &              \\
				GY 91(PDD)                          &        &     & P         & 0.174       & D        & 0.011        & D         & 0.008       &          &              \\
				\hline
				\hline
				\textit{HD 143006} & 0.0139 & 82  & \multicolumn{2}{c}{22}  & \multicolumn{2}{c}{51}  &           &             &          &              \\
				\hline
				HD 143006(GG)                       &        &     & G         & 2.977       & G        & 0.264        &           &             &          &              \\
				\hline
				\hline
				\textit{HD 163296} & 0.0633 & 169 & \multicolumn{2}{c}{10}  & \multicolumn{2}{c}{48}  & \multicolumn{2}{c}{86}  & \multicolumn{2}{c}{145} \\
				\hline
				HD 163296(GGDD)                     &        &     & G         & 0.320       & G        & 0.995        & D         & 0.254       & D        & 0.030        \\
				HD 163296(GGPD)                     &        &     & G         & 0.320       & G        & 0.995        & P         & 0.406       & D        & 0.030        \\
				\hline
				\hline
				\textit{HL TAU}    & 0.105  & 80  & \multicolumn{2}{c}{12}  & \multicolumn{2}{c}{32}  & \multicolumn{2}{c}{82}  &          &              \\
				\hline
				HL TAU(GPP)                         &        &     & G         & 1.618       & P        & 0.221        & P         & 0.445       &          &              \\
				\hline
				\hline
				\textit{V1094 SCO} & 0.0401 & 290 & \multicolumn{2}{c}{60}  & \multicolumn{2}{c}{103} & \multicolumn{2}{c}{174} &          &              \\
				\hline
				V1094 SCO(DGD)                      &        &     & D         & 0.070       & G        & 0.365        & D         & 0.023       &          &              \\
				V1094 SCO(PGD)                      &        &     & P         & 0.138       & G        & 0.365        & D         & 0.023       &          &             \\
				\hline              
			\end{tabular}
			\begin{tablenotes}
				\item Notations for adopted planetary mass: G: $ M_{\rm p, gas} $, P: $ M_{\rm iso}
				$, D : $ M_{\rm p, dust} $.
				\item \textbf{References.} \cite{Huang2018};\cite{Long2018}; \cite{VanderMarel2019}.
			\end{tablenotes}
		\end{threeparttable}
	}
	\label{tab:fidu_mass_table}
\end{table*}

Table \ref{tab:fidu_mass_table} shows all sets of the initial planetary mass adopted when $ \alpha $ is the fiducial value $ \num{e-3} $. If one or more gaps in a disk system have two mass interpretations, we explore all the possible combinations of mass assignments. For example, both gap 2 and gap 3 in CI~Tau are Group~III gaps and have two mass assignments (`PG' and `DP'), therefore there are in total four combinations (`GPD',`GPP',`GGD',`GGP') for CI~Tau. The adopted $ \alpha $ can change both the classification of the gaps and the predicted planetary masses, especially when the predicted planetary mass is close to the respective $ M_{\rm iso} $.The planetary masses are estimated assuming each gap is opened by a single planet, and planets are located at the mid-points of the corresponding gaps. Note that the planetary masses adopted for HL~Tau in Table~\ref{tab:fidu_mass_table} are about two times as large as those adopted in \citetalias{Wang2020}, mainly because of the different
$\alpha$ values that we adopted ($\alpha=10^{-3}$ in Table~\ref{tab:fidu_mass_table}, but $\alpha < 6\times 10^{-4}$ in \citetalias{Wang2020}). If adopting the same value of $\alpha$, we obtain the similar masses as \citetalias{Wang2020}.

As seen in Table~\ref{tab:fidu_mass_table}, the planets located at the outer region are categorised as `P' or `D' in the most cases, which indicates smaller planets that carve no or very shallow gas gaps, as also discussed in \citetalias{Wang2021}. This may be consistent with the recent Molecules with ALMA at Planet-forming Scales (MAPS) observation that has observed gas structures of five disks (IM~Lup, GM~Aur, AS~209, HD~163296, and MWC~480) \citep{Oberg2021}. Because of the relatively large beam size as compared with that of the dust observations, this observation focuses on the structures in relatively outer region. In this scale, the fraction of overlapping between gas and dust substructures is $\sim 50\%$  \citep{Law2021}. \cite{Jiang2022} has shown that this fraction is similar to that predicted by the null hypothesis, and it may indicate no significant correspondence between them in the observed scale.
	
Moreover, we can directly compare our model with the observations of MAPS in the disks of HD~163296 and AS~209 that are involved in both samples. In the case of HD~163296, three gaps are detected in CO observation at 44~au, 93~au, and 148~au \citep{Zhang2021}, and our predictions are `G', `P' and `D', respectively (see Table~\ref{tab:fidu_mass_table}). The locations of the centre of the gaps are similar to these of dust gaps, but the observed shapes of outer two gaps are much different in gas and dust, which may be consistent with our prediction that shows no significant correspondence between gas and dust gaps. For the 44~au gap, the CO gap is observed at the similar location and it has the similar shape with the dust gap, which agrees our prediction. In the case of AS~209, the gas can be depleted in the inner 10~au \citep{Bosman2021} and it is consistent with the our prediction that the innermost planet at 8.7~au is labelled as `G', indicating a giant planet with a gas gap. For outer two gaps, \cite{Alarcon2021} has suggested that the smaller planets are preferred, which is also in good agreement with our prediction.
	
We should mention the comparison with the prediction of \citetalias{Wang2021} in other disks of MAPS sample (IM~Lup and MWC~480). In the case of IM~Lup, \citetalias{Wang2021} predicted `D'-type planet and no gas structure at 117~au, and no significant CO gap is detected in the observation \citep{Zhang2021} and it agrees with the prediction. In the case of MWC~480, a CO gap is detected at the similar location of the dust gap at 78~au, which agrees with \citetalias{Wang2021} that predict a `G'-type planet with a gas gap. As discussed above, the presence and/or absence of CO gas gaps for those disks are largely consistent with the prediction of \citetalias{Wang2021} within the uncertainty, even if not perfect.  Given the resolution of the CO observation, the current level of agreement is encouraging.
	
Moreover, we can directly compare our model with the observations of MAPS in the disks of HD~163296 and AS~209 that are involved in both samples. In the case of HD~163296, three gaps are detected in CO observation at 44~au, 93~au, and 148~au \citep{Zhang2021}, and our predictions are `G', `P' and `D', respectively (see Table~\ref{tab:fidu_mass_table}). The locations of the center of the gaps are similar to those of dust gaps, but the observed shapes of outer two gaps are much different in gas and dust, which may be consistent with our prediction that shows no significant correspondence between gas and dust gaps. For the 44~au gap, the CO gap is observed at the similar location and it has the similar shape with the dust gap, which agrees with our prediction. In the case of AS~209, the gas can be depleted in the inner 10~au \citep{Bosman2021} and it is consistent with our prediction that the innermost planet at 8.7~au is labelled as `G', indicating a giant planet with a gas gap. For the outer two gaps, \cite{Alarcon2021} has suggested that the smaller planets are preferred, which is also in good agreement with our prediction.
	
We should mention the comparison with the prediction of \citetalias{Wang2021} in other disks of MAPS sample (IM~Lup and MWC~480). In the case of IM~Lup, \citetalias{Wang2021} predicted `D'-type planet with no gas structure at 117~au, and no significant CO gap is detected in the observation \citep{Zhang2021} and it agrees with the prediction. In the case of MWC~480, a CO gap is detected at the similar location of the dust gap at 78~au, which agrees with \citetalias{Wang2021} that predict a `G'-type planet with a gas gap. As discussed above, the presence and/or absence of CO gas gaps for those disks are largely consistent with the prediction of \citetalias{Wang2021} within the uncertainty, even if not perfect.  Given the resolution of the CO observation, the current level of agreement is encouraging.

\section{Evolution of protoplanets in a gas disk:
  orbital migration and mass accretion} \label{sec:accretion-migration}

At the disk stage indicated in Figure \ref{fig:schematic_flow}, the
embedded protoplanets interact with the disk, in addition to the
mutual gravitational interaction with the central star and other
planets. While the detailed evolution needs to be studied with
intensive hydrodynamical simulations, we decided to adopt an empirical
parameterized model to describe the planet-disk interaction
\citepalias{Wang2020}. Specifically, we employ a cylindrical
coordinate, and compute the position vector of the $i$-th planet from
\begin{equation}
	\label{eqn:disk_eom}
	\boldsymbol{\ddot{R}}_i = \boldsymbol{f}_{\rm{grav},i}
	+ \boldsymbol{f}_{\rm{a},i} +\boldsymbol{f}_{\rm{e},i},
\end{equation}
where $ \boldsymbol{f}_{\rm{grav},i} $ is the gravity exerted by the
central star and other planets, $\boldsymbol{f}_{\rm{a},i} $ is the
migration force, and $\boldsymbol{f}_{\rm{e},i} $ is the eccentricity
damping force, acting on the $i$-th particle per unit mass.  The
latter two forces represent the planet-disk interaction, and are
discussed further in section~\ref{sec:method_mig}.

We simultaneously compute the evolution of the planetary mass due to
the accretion of pebbles and gas from the disk.  We describe how we
model these accretion processes in section~\ref{sec:method_acc},
followed by the disk surface density model in
section~\ref{subsec:disk-profile}.

\subsection{Planetary migration model \label{sec:method_mig}}

We adopt the same migration model as implemented in
\citetalias{Wang2020}. Based on the hydrodynamical simulation results
of \cite{Kanagawa2018}, this model considers the gas depletion effect
due to the gap opening and empirically parameterizes the planetary
migration in terms of a dimensionless factor $ K_{i} $ that
characterizes to the depth of the $i$-th gap
\begin{align}
	\frac{\Sigma_{\rm min,i}}{\Sigma_{g}(R_i)} = \frac{1}{1+0.04K_i},
	\label{eqn:siggap}
\end{align}
where $\Sigma_{min}$ and $ \Sigma_g $ are the minimum gas surface
density inside the gap and unperturbed surface density.

The factor $ K_i $ is written as \citep[e.g.][]{Kanagawa2016,Kanagawa2018}
\begin{align}
	\label{eq:def-K}
	K_i = \left(\frac{M_i}{M_*}\right)^2
	\left(\frac{h_i}{R_i}\right)^{-5}\alpha^{-1} ,
\end{align}
where $M_i$ is the time-dependent mass of the $i$-th planet, and $
h_i/R_i $ is the aspect ratio at its location.

Then the migration timescale of the $i$-th planet is empirically
approximated as
\begin{equation}
	\label{5.2taui_tsc}
	\tau_{a,i} = \dfrac{1+0.04K_i}{\gamma_{L,i}
		+ \gamma_{C,i}\exp(-K_i/K_{t})}\tau_{0,i}(R_i),
\end{equation}
where $\gamma_{C,i}=\Gamma_{C,i}/\Gamma_{0,i}$ and
$\gamma_{L,i}=\Gamma_{L,i}/\Gamma_{0,i}$, with $ \Gamma_{C,i}$,
$\Gamma_{L,i}$, and $\Gamma_{0,i}$ being the co-rotation, Lindblad,
and characteristic torques for the $i$-th planet
\citep{Paardekooper2011}. Further details can be found in section~3.2
of \citetalias{Wang2020}.

Since $\alpha=10^{-3}$ is adopted in our fiducial case, our model does not take into account of the additional torques that may arise in low-viscous or inviscid disks \citep[e.g.,][]{Li2009,Fung2018,McNally2019}. We also neglect the non-isothermal corotation torque \citep{Paardekooper2011} as the planets that we consider in this paper are mostly locating at the outer region of the disk, where the cooling timescale is shorter than the dynamical time scale. Assuming a disk with a constant opacity and fiducial disk profiles adopted in our simulation, a rough order of magnitude estimation considering the radiative cooling alone shows that the orbital time scale $ \Omega_K^{-1} $ is nearly one order of magnitude longer than the cooling time scale $ \tau_{\rm cool} $ at \SI{25}{au}, and the ratio $ \Omega_K^{-1}/\tau_{\rm cool} $ increases with $ R $ because the cooling time scale $ \tau_{\rm cool} \propto R^{-0.5} $. Formal analysis and calculation in \cite{Bitsch2014} also shows that the non-isothermal effect is more effective in the inner disk with $R < \SI{10}{au} $. It is therefore reasonable to adopt the locally isothermal assumption in our context.

Moreover, where the thermal torque \citep[e.g.,][]{Benitez-Llambay2015,Masset2017} can affect the migration of planet as small as a few Earth masses, it does not significantly change our results because the estimated planetary masses are generally much larger than Earth mass, as shown in Table~\ref{tab:fidu_mass_table}. Nevertheless, it is still worth to note that a few small planets in GY 91, DoAr 25, AA Tau and V1094 SCO systems are more susceptible to the impact of the thermal torque due to the high pebble accretion rate, and as a result their inward migration may be slowed down. Such non-isothermal effects are particularly important in the inner disk and deserve careful consideration in the future study concerning the evolution of close-in low mass planets.

Our model also assumes that the migration time scale is always longer than the gap opening time scale. It is worth to note that the migration can be faster than the gap opening when the surface density is high enough, as investigated by \cite{Kanagawa2021a}. However, in the context of our simulation, the typical disk surface density is not high enough to produce such a high migration speed. In the Type I migration regime, the migration is faster than the gap opening only if the surface density is larger than $ \SI{126}{gcm^{-2}} $ for a $ \SI{e-4}{M_{\odot}} $ planet at \SI{50}{au} \citep{Kanagawa2021a}, with $ h/R =0.08 $ and $ \alpha = \num{e-3} $. However, the typical surface density experienced by such a planet in our simulation is around $ 9 $ to $ \SI{0.9}{gcm^{-2}} $  (\num{e-6} to $ \SI{e-7}{M_\odot au^{-2}} $), meaning that the migration time scale $ \tau_{\rm Type I} $ is one or two orders of magnitude longer than the gap opening time scale $ \tau_{\rm open} $.  In the Type II migration regime ($ K \gg 25 $), the ratio $ \tau_{\rm Type II}/\tau_{\rm open} $ continues to increase as the disk surface density decays (see equation \ref{eqn:exp_decay_diskprofile}). Therefore, it is reasonable to assume that the migration is slower than the gap opening process in our simulations.
\subsection{Mass accretion onto planets \label{sec:method_acc}}

Pebble accretion and gas accretion are the two major channels for the
mass growth of protoplanets. The standard core accretion theory states
that a planet forms a rocky core via pebble accretion first
\citep[e.g.,][]{Ormel2010}. After the planetary mass exceeds $ M_{\rm
  iso} $, the run-away gas accretion process sets in.  Since our
present simulation considers the initial mass of planets ranging from
$ \SI{e-3}{M_J} $ to $ \sim \text{a few } \si{M_J} $, we take account
of both pebble and gas accretion processes. On the other hand, the
pebble accretion is not considered in \citetalias{Wang2020}, because
\citetalias{Wang2020} assume that the embedded planets in the HL Tau
system have opened gas gaps with initial masses larger than $M_{\rm
  iso}$.

\subsubsection{Pebble accretion \label{sec:peb_acc}}

When a planet is small, its gravity is too weak to capture the gas,
and therefore its mass grows mainly through the accretion of pebbles.
We follow \cite{Lambrechts2012}, and assume that the pebbles
can accrete on the $i$-th planet only within the effective radius
$r_{\rm acc,peb}$ from its location $R_i$:
\begin{align}
	\label{eqn:peb_acc_rad}
	r_{\rm acc,peb}(R_i) = \left(\frac{St}{0.1}\right)^{1/3}R_{H}(R_i), 
\end{align}
where $R_{H}(R_i)$ is the Hill radius of the planet at $R_i$ and $St$
is the Stokes number of the pebbles. Also we assume a non-stratified
disk with the turbulent diffusivity comparable to the turbulent
(kinematic) viscosity. In this case, the pebble scale height $h_{\rm peb}$ is approximately written in terms of the gas scale height $h$ \citep{Youdin2007} as
\begin{align}
	\label{eqn:peb_scaleheight}
	h_{\rm peb}(R_i) = \sqrt{\frac{\alpha}{St}} h(R_i).
\end{align}

If $r_{\rm acc,peb}> h_{\rm peb}$ ($r_{\rm acc,peb}< h_{\rm peb}$),
the accretion effectively proceeds in two (three) dimensions.  Thus we
adopt the following mass growth rate
\citep[e.g.][]{Lambrechts2012,Ormel2017,Johansen2019}:
\begin{align}
	\label{eqn:peb_mass_rate}
	\dot{M}_{\rm{peb},i} = 
	\left\{\begin{aligned}
		&2\Omega_{k,i} [r_{\rm acc,peb}(R_i)]^2\Sigma_{\rm peb}(R_i)
		& {\rm for}~ r_{\rm acc,peb}(R_i)> h_{\rm peb}(R_i)\\
		&6\pi\Omega_{k,i}[r_{\rm acc,peb}(R_i)]^3\rho_{\rm peb}(R_i)
		& {\rm for}~ r_{\rm acc,peb}(R_i)< h_{\rm peb}(R_i)
	\end{aligned}\right.
\end{align}
where $ \Omega_{k,i} $, $ \Sigma_{\rm peb}(R_i) $ and $ \rho_{\rm
  peb}(R_i) $ are the Keplerian angular velocity, pebble surface
density and mid-plane pebble density at $ R_i $, respectively. The
mid-plane pebble density $ \rho_{\rm peb}(R_i) $ is computed from the
pebble surface density $ \Sigma_{\rm peb}(R_i)$ using the relation $
\rho_{\rm peb}(R_i) = \Sigma_{\rm peb}(R_i)/(\sqrt{2\pi}h_{\rm
  peb}(R_i))$.

We note that the above two expressions are discontinuous at $r_{\rm
  acc,peb}(R_i)= h_{\rm peb}(R_i)$ because they correspond to the
respective asymptotic limits of 2D and 3D pebble accretion regimes,
strictly speaking. An interpolation between the two may be adopted to
describe the transition of accretion regimes, but it does not change
the results in practice. Therefore, we adopt these discontinuous
expressions of equation (\ref{eqn:peb_mass_rate}), same as the
original expression in the previous literature
\citep[e.g.,][]{Ormel2017}.

The pebble accretion process continues until the planetary mass
exceeds the pebble isolation mass $M_{\rm iso}$, at which the planet
is massive enough to generate a pressure bump that prevents the
pebbles from approaching the planet by effectively trapping
them. Therefore, after the outer planet reaches its pebble isolation
mass, the pebble flux onto the inner planet is also inhibited due to
the presence of the pressure bump. This pebble shielding effect is
incorporated by setting $ \dot{M}_{\rm peb} = 0$ in the region
interior to the planet that reaches pebble-isolation mass, and we turn
off the pebble accretion of the inner planets.

\subsubsection{Gas accretion}

When the planetary mass exceeds $M_{\rm iso}$, the planet grows only
via gas accretion. \cite{Tanigawa2016} investigated the gas accretion
from the disk and found out the accretion occurs effectively at
certain places around two Hill radii from the planet. Their model
expresses the accretion rate as the product of the accretion area per
unit time $ D $ and the surface density around the planet
(approximately the minimum surface density inside the gap $
\Sigma_{\rm min} $):
\begin{align}
	\label{eqn:accmodel}
	\dot{M}_{\rm{gas}, i} = D_i\Sigma_{{\rm min},i}, 
\end{align}
where $\Sigma_{{\rm min},i} $ is given by equation~(\ref{eqn:siggap}) in
terms of the gas surface density profile $ \Sigma_g(R_i) $, and 
\begin{align}
	D_i = 0.29\left(\frac{h_i}{R_i}\right)^{-2}
	\left(\frac{M_i}{M_*}\right)^{4/3}R_i^2\Omega_{K,i},
\end{align}
with $M_i$ being the mass of the $i$-th planet.

\subsection{Gas and pebble surface density profiles in the disk}
\label{subsec:disk-profile}

\subsubsection{Gas surface density profile}

We adopt the same model of gas surface density profile hosting
multiple planets as \citetalias{Wang2020}. For a quasi-steady gas disk
with the star and each planet taking in mass and angular momentum via
gas accretion, the gas surface density $ \Sigma_g(R) $ between the $ n
$th and $ (n+1) $th planets is given by
\begin{align}
	\Sigma_g(R) = \frac{\dot{M}_{\rm{gas},*}}{3\pi\nu(R)}
	\left(1-\sqrt{\frac{R_*}{R}}\right)
	+ \sum\limits_{i=1}^{n}\frac{\dot{M}_{\rm{gas},i}}{3\pi\nu(R)}
	\left(1-\sqrt{\frac{R_i}{R}}\right)
	\label{eqn:multip_sigmaunp}
\end{align}
where $ R_n \leq R < R_{n+1} $, $ \dot{M}_{\rm{gas},*} $ and
$\dot{M}_{\rm{gas},i}$ are the gas accretion rates of the star and the
$ i $-th planet, $ \nu(R)~\equiv~\alpha c_s(R) h(R)$ is the kinematic
viscosity with $c_s$ being the sound speed. The stellar accretion rate
$ \dot{M}_{\rm{gas},*} $ is the remaining gas flux after the global
gas influx $ \dot{M}_{\rm glob} $ is consumed due to the gas accretion
of all relevant planets:
\begin{equation}
	\label{eqn:glob_acc_rate}
	\dot{M}_{\rm{gas},*} \equiv 
	\dot{M}_{\rm glob} - \sum\limits_{i=1}^{N}\dot{M}_{\rm{gas},i}.
\end{equation}
If the mass of the $ i $-th planet is below $ M_{\rm iso} $ and
undergoes pebble accretion, we set $ \dot{M}_{\rm{gas},i} = 0$. The
mass and location of all the planets determine how the global gas
influx $ \dot{M}_{\rm glob} $ is distributed among planets and stars:
strong gas accretion of the outer planet consumes a large fraction of
the $ \dot{M}_{\rm glob} $ and thus lower the surface density at the
inner region, effectively quenching both migration and mass growth of
the inner planet. Such an effect is important and will be further
discussed in section \ref{sec:mass_assign_evo}.

To account for the decay of the disk due to both photoevaporation and
accretion, we set the the global accretion rate exponentially decay
with the e-folding gas dispersal timescale of $\tau_{\rm disk}$:
\begin{equation}
	\label{eqn:exp_decay_diskprofile}
	\dot{M}_{\rm glob}(t) = \dot{M}^{\rm ini}_{\rm glob}e^{-t/\tau_{\rm disk}},
\end{equation}
with $ \tau_{\rm disk}$ fixed throughout each realization. As will be
shown in section \ref{sec:inicons}, $ \dot{M}^{\rm ini}_{\rm glob} $
is determined by the observed disk size and mass given in table
\ref{tab:fidu_mass_table}. Thus, when $ R $ is sufficiently larger
than the radius of the outermost planet, the surface density profile $
\Sigma_g(R) \simeq \dot{M}_{\rm glob}/(3\pi\nu)$ is reduced to the
unperturbed disk profile given by e.g., \cite{Pringle1981}.

In our multiplanetary system, $\Sigma_{\rm g}$ of the inner
planet can be affected by the accretion of the outer planet. Since
the $\Sigma_{\rm g} (R)$ is determined by the mass flux at $R$. the
accretion of the outer planet at $R_{\rm p}$ reduces the total mass
influx and then effectively lowers the $\Sigma_{\rm g}$ for $R <
R_{\rm p}$. In this way, the accretion of the outer planet can
affect the $\Sigma_{\rm g}$ as well as $\Sigma_{\rm min}$ of the
inner planet.

In this setup, the accretion rate of any individual planet cannot
exceed $\dot{M}_{\rm glob}$ (global mass inflow). If an outer planet
experiences very rapid gas accretion by taking nearly 100 percent of
$\dot{M}_{\rm glob}$, the surface density inner to this planet will be
lowered to nearly zero. The schematic view as well as detailed
calculations can be referred to Figure 1, Figure 3, section 3.4 and appendix of \citetalias{Wang2020}.

Our model always assumes the disk is in a steady state and the change of $\Sigma$ is instantaneous for simplicity. In reality, however, the change of the gas surface density due to the gas accretion of the outer planets is expected to be delayed due to the finite time of the mass transfer. Our model may either underestimate or overestimate the timescales of the migration and accretion in the inner region of the disk due to this simplification.

\subsubsection{Pebble surface density profile}

The pebble surface density $ \Sigma_{\rm peb} $ at the location of the
$ i $-th planet is given from the flux of the pebbles as
\begin{equation}
	\label{eqn:peb_profile}
	\Sigma_{\rm peb,i} = \frac{\dot{M}_{\rm peb,glob}}{2\pi R
		\left|v_{\text{drift},i}\right|},
\end{equation}
where $ \dot{M}_{\rm peb,glob} $ is the global pebble accretion rate
and $ v_{\text{drift},i} $ is the pebble drifting speed. Unlike the
gas profile, we neglect the feedback of pebble accretion of
an individual planet on the global pebble profile. We further assume that $ \dot{M}_{\rm peb,glob} $ is proportional to $\dot{M}_{\rm glob} $:
\begin{equation}
	\label{eqn:peb_gas_glob_relation}
	\dot{M}_{\rm peb,glob}  = \chi \dot{M}_{\rm glob}(t),
\end{equation}
where the proportional constant $ \chi$ is set to the global pebble to
gas ratio (we adopt $ \chi=0.01 $).

The pebble-drifting speed $ v_{\text{drift},i} $ can be expressed in
terms of the Stokes number \citep{Weidenschilling1977}:
\begin{equation}
	\label{eqn:vdrift}
	v_{\rm drift,i} = -\frac{2\Delta v_{\text{sub},i}}{St+St^{-1}}
	+\frac{u_{R,i}}{1+St^2},
\end{equation}
where $ \Delta v_{\rm sub} $ is the relative velocity between pebbles
and sub-Keplerian gas, $ u_{R} $ is the radial drifting speed of the
gas, which are given respectively by \citep[e.g.][]{Birnstiel2012,
  Ormel2017,Johansen2019}
\begin{gather}
	\Delta v_{\text{sub},i} = -\frac{1}{2}\frac{h_i^2}{R_i}
	\frac{\partial \ln P_i}{\partial \ln R_i}\Omega_{K,i}\\
	u_{R,i} = -\frac{3}{2}\alpha\frac{h_i^2}{R_i}\Omega_{K,i}.
\end{gather}

Equation~(\ref{eqn:vdrift}) suggests that pebble drifting speed
strongly depends on the Stokes number, i.e., the aerodynamical
property of the pebbles. \cite{Ormel2017} pointed out that $ St \sim
\num{e-3}-\num{e-1} $ is the optimal range for efficient pebble
accretion, since pebbles with very small $ St ( <\num{e-5} $) or $
St\sim 1 $ are difficult to be accreted onto the planet:
aerodynamically small pebbles well couple with the gas and thus hard
to settle to the planet, while pebbles with $ St\sim 1 $ is hard to be
captured due to its fast drifting speed. For simplicity, we adopt a
constant $ St $ in each simulation run and set the fiducial $ St $
value to be \num{e-2}.

\subsection{Numerical simulations and disk parameters \label{sec:inicons}}

\begin{table}
	\centering
	\caption{Summary of the disk parameters}
	\resizebox{0.6\linewidth}{!}{%
		\begin{threeparttable}[t]
			\begin{tabular}{llc}
				\hline
				\hline
				{\textbf{Notation}} & {\textbf{Meaning}} & $ \textbf{Value}^* $ \\
				\hline
				$ \alpha $ & {Viscosity} &$ \num{5e-4},\underline{\boldsymbol{\num{e-3}}},\num{2e-3}$ \\   
				$ \tau_{\rm disk} $& {Disk decay lifetime}&$ \underline{\SI{2}{Myr}} $, \SI{3}{Myr}\\
				$ f $ & Flaring index& $ 0.25 $\\
				$ St $& Stokes Number& $ 0.01 $\\
				$\chi$&	Pebble-to-gas ratio	& $ 0.01 $\\
				\hline
			\end{tabular}%
			\begin{tablenotes}
				\item *the underscored is the fiducial value
			\end{tablenotes}
	\end{threeparttable}}
	\label{tab:common_disk_para}
\end{table}

Our simulations are performed using the public $ n $-body code
\texttt{rebound} \citep{Rein2012} and its extension
\texttt{reboundx} \citep{Tamayo2020}. As already introduced in section \ref{sec:plt_sys_overview}, our simulation consists of two stages: the
disk stage and post-disk stage. At the disk stage, the
\texttt{ias15} integrator \citep{Rein2014} is used, and the
planetary migration is computed taking account of both planet-disk
interaction and mass accretion (see sections
\ref{sec:method_mig} and \ref{sec:method_acc}).

We list all the disk parameters that are fixed constant throughout
each simulation in Table \ref{tab:common_disk_para}. To be consistent
with the temperature model that we adopted for the mass prediction,
the flaring index is fixed to \num{0.25} (equation~\ref{eqn:
asp_ratio_power_law}). The Stokes number $ St $ and pebble-to-gas
ratio $ \chi $ are only relevant to the pebble accretion.  We found
that the architecture does not significantly change even if we
vary both $ St $ and $ \chi $ by a factor of two. This is because
the migration timescale is much shorter than that of accretion due
to the inefficient pebble accretion when the planetary mass is
small. Thus we fix these two parameters and focus on the impact of
others.

For each ALMA disk system, we have surveyed different initial
mass assignments, $ \alpha $ viscosities and disk lifetime. The
planets are initialised with co-planar and nearly-circular orbits with
the initial eccentricity of \num{e-7}. We stop the simulation at $
5\tau_{\rm disk}$ to ensure that the disk has sufficiently dispersed.

We adopt a constant $ \alpha $ model that is independent of time and
position. The initial value of $ \dot{M}_{\rm glob} $ is normalised by
the total disk mass:
\begin{equation}
	\label{eqn:cons_alpha_mdotglob_normalisation}
	\dot{M}^{\rm ini}_{\rm glob} = \frac{3}{2} \nu_{\rm 1au}
	\left(\frac{M^{\rm ini}_{\rm disk}}{\SI{1}{au^2}}\right)
	\left[{\left(\frac{R_{\rm out}}{\SI{1}{au}}\right)
		-\left(\frac{R_{\rm in}}{\SI{1}{au}}\right)}\right]^{-1},
\end{equation}
where $ M^{\rm ini}_{\rm disk} $ is the initial disk mass, $ \nu_{\rm
	1au} = \nu(R = \SI{1}{au}) $ is the kinematic viscosity at
\SI{1}{au}.

We assume the outer edge of the dust disk is the same as that of the
gas disk. The inner edge of the disk is often poorly resolved, so we
simply set the inner edge to be \SI{0}{au}, except for AA Tau, whose
inner edge is set to be the outer boundary of its inner cavity.

\section{Architecture of planetary systems at the end of the disk stage}
\label{sec:rslt_disk_stage}

In this section, we compute the evolution of planetary systems through
the disk-planet interaction, and present their architecture at the end
of the disk stage. The majority of planets has significantly migrated
inwards and/or increased the mass, and their final architecture varies
much when a different set of initial conditions and disk parameters
are employed. In the following subsections, we will compare the
resulting configurations of the planetary systems against the
observation, by emphasizing the dependence on the disk parameters and
initial planetary mass assignments.

\subsection{Overall results at the disk stage}

\label{sec:overall_rslts}
\begin{figure*}
	\centering
	\gridline{\fig{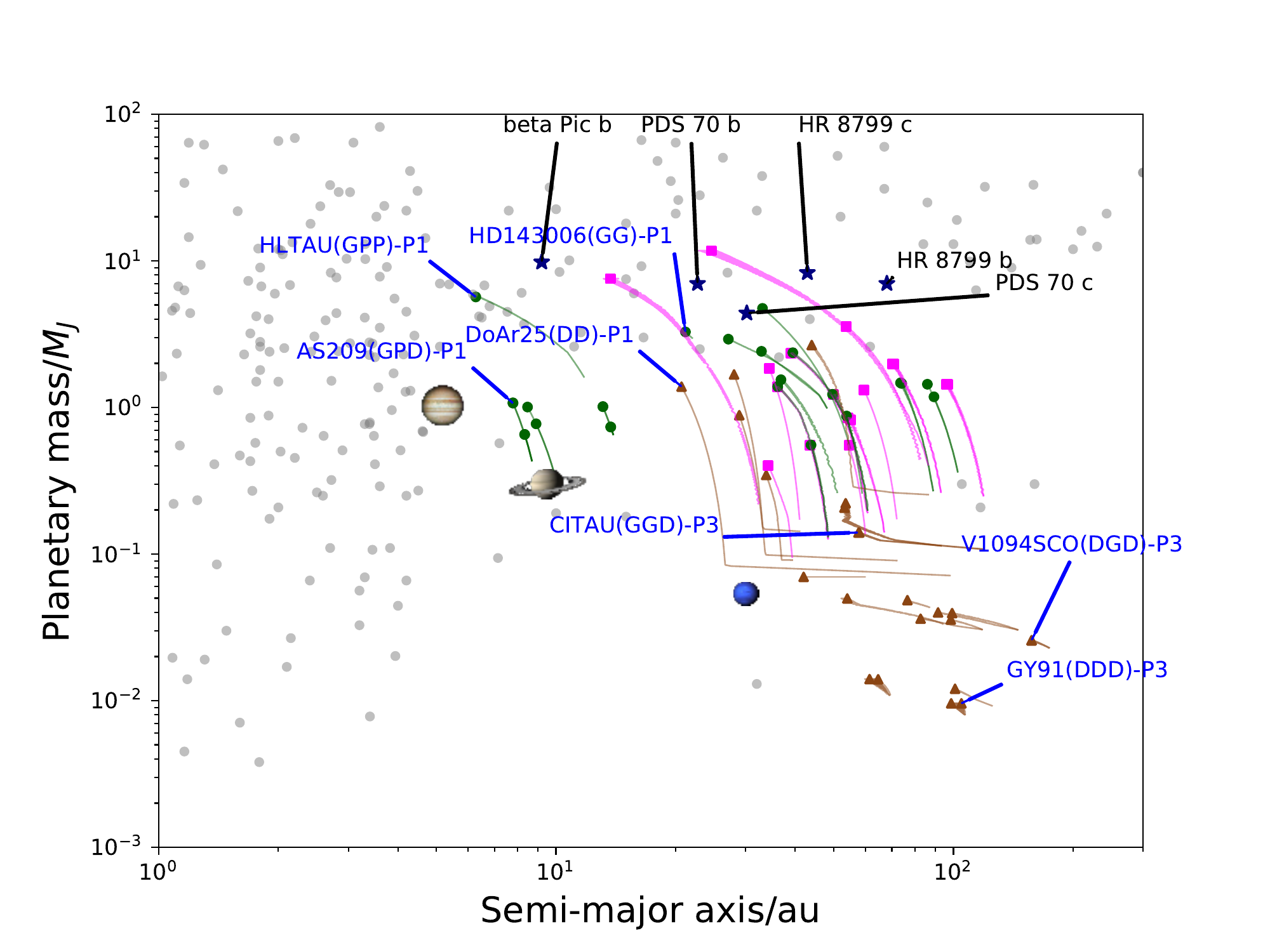}{0.65\linewidth}{(a)}
	  \fig{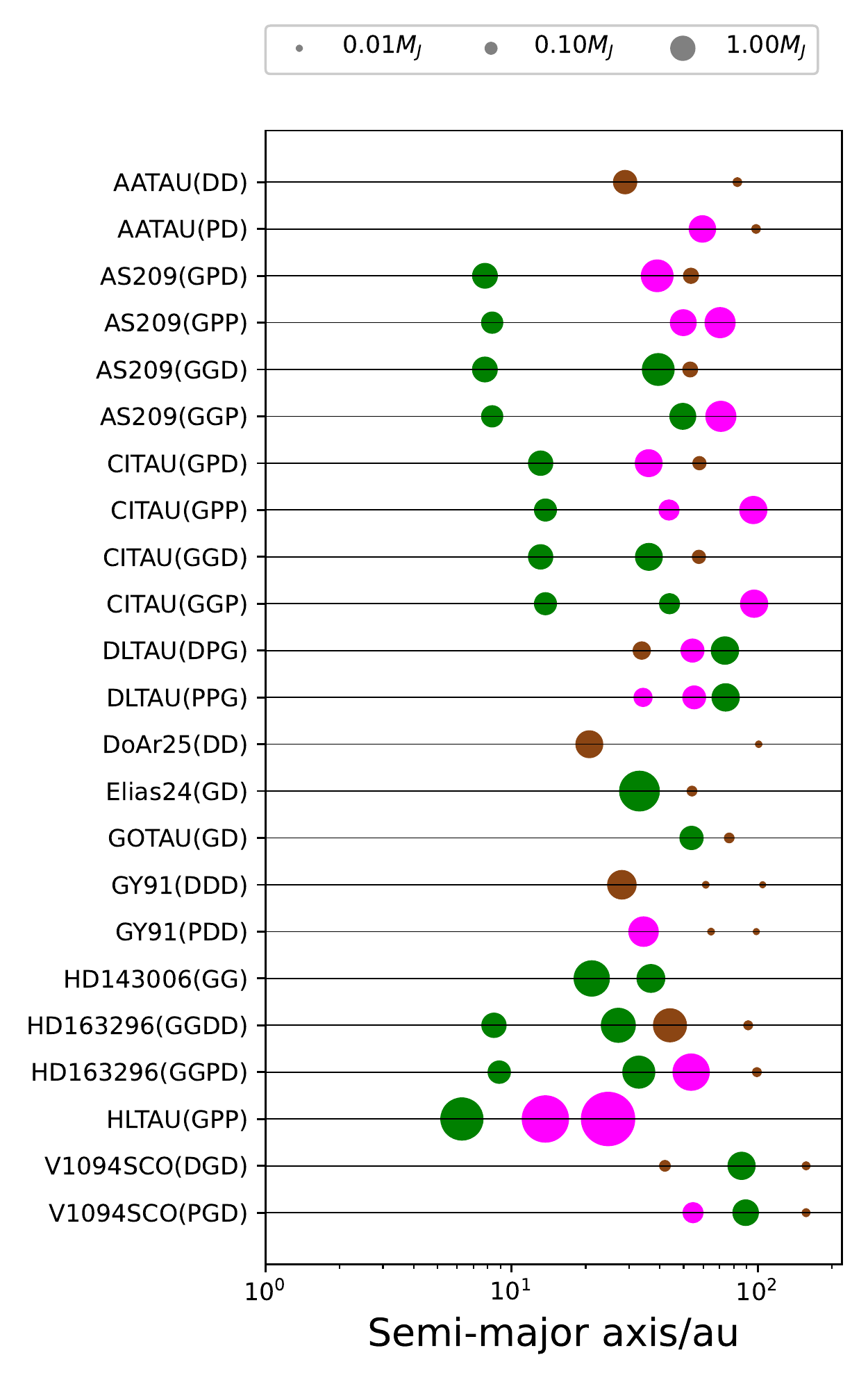}{0.33\linewidth}{(b)}}
        \caption{(a) Final mass and semi-major axis of the simulated
          planetary systems evolved with fiducial $ \alpha = \num{e-3}
          $, as well as the observed exoplanetary population in grey
          dots (data extracted from \texttt{exoplanet.eu},
          2021). Green, magenta, and brown color denote $ M_{\rm gas}
          $, $ M_{\rm iso} $, $ M_{\rm dust} $ adopted as initial
          planetary mass. The blue text tags some example planets. The
          respective color lines show the evolution track. Jupiter,
          Saturn, Uranus (cyan) and Neptune (dark blue) are also
          plotted. (b) Configuration of planetary systems at the end
          of disk dispersal at $ \alpha = \num{e-3}$. The marker size
          is proportional to $ M_{\rm p}^{2/3} $. The different
          colours denote the interpretation of the initial planetary
          mass assigned: brown indicates $ M_{\rm p, dust} $, magenta
          indicates $ M_{\rm iso} $ and green indicates $ M_{\rm
            p,gas} $.}
	\label{fig:comp_config_fidu}
\end{figure*}

Figure \ref{fig:comp_config_fidu} shows the evolution of mass and
semi-major axis of the planetary systems at the disk stage for the
fiducial disk parameters listed in Table \ref{tab:common_disk_para}.
In Figure \ref{fig:comp_config_fidu}(a), solid symbols indicate the
positions and masses of the simulated planets at the end of the disk
stage; different colours correspond to the type of initial planetary
mass (G: green circles, P: magenta squares, and D: brown triangles),
and lines associated with each symbol are evolution tracks of the
planets. For reference, the locations of the observed planets are
plotted in grey dots, as well as those of Jupiter, Saturn and Neptune
in the Solar system. Figure~\ref{fig:comp_config_fidu}(b) shows the
final mass and semi-major axis of planets for each system.

While most of the planets are initially located far away from the
observed population, during the disk-stage evolution the planets
migrate inwards and grow larger. Although the majority of the
resulting systems do not cover the range of observed planets, a
fraction of them turn out to be very close to the systems observed via
direct imaging, such as $ \beta $ Pic b, PDS 70b, PDS 70c, HR 8799b,
and HR 8799c. Those planets that are initialised with mass equal or
larger than $ M_{\rm iso} $ (magenta and green) undergo gas accretion
throughout the simulation, and their evolution tracks are
well-consistent with the predictions of \cite{Tanaka2019}
\citepalias[See figure 4 of][]{Wang2021}.

\begin{figure*}
	\centering
	\includegraphics[width=\linewidth]{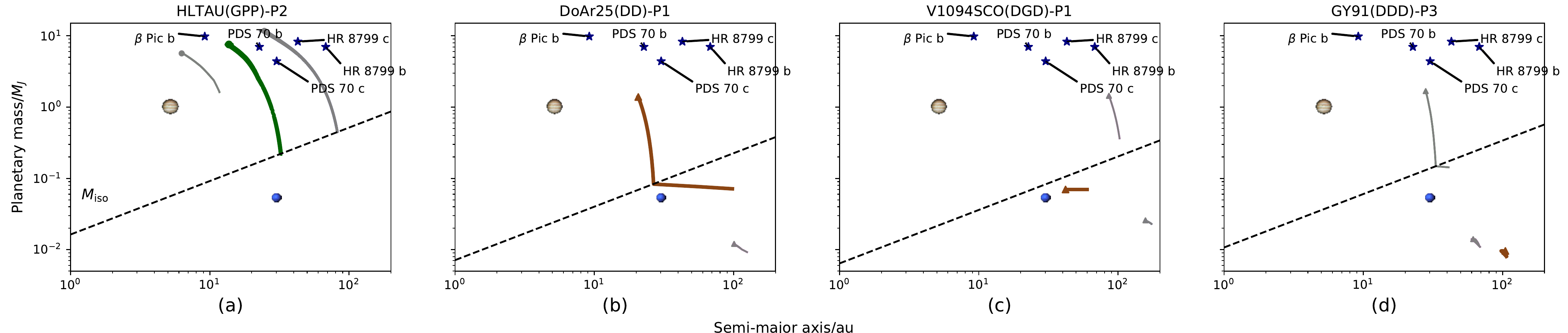}
	\caption{Example evolution of planets from four different groups. Evolution of other planets in the same system are also plotted in grey colour.}
	\label{fig:comp_config_zoom}
\end{figure*}

We divide the simulated planets roughly in four different groups whose
typical evolution tracks are plotted in Figure
\ref{fig:comp_config_zoom}. Although the planetary systems contain
multiple planets, the initial mass and location are still the most
important factors for the disk stage evolution as well as the final
configuration, rather than the presence of other planets. Hence, we
simply highlight the planet corresponding to the respective category,
while the rest of planets in the same system are plotted in grey
colour just for reference. In the following, we briefly summarise the
feature of planets in each category.

\begin{enumerate}
\item \textbf{Distant giant planets
  (Figure~\ref{fig:comp_config_zoom}a):} a fraction of our simulated
  planets including HL Tau(GPP), HD 143006(GG), and Elias 24 Planet
  1(G) end up with $ 0.5 - 3 $ Jupiter-mass planets or even more
  massive brown dwarfs with semi-major axis between \SI{30}{au} and
  \SI{100}{au}. Most of these planets start with gas accretion and
  fail to move to the inner region due to the slow migration. The most
  massive ones resemble those widely-separated systems discovered via
  direct imaging, HR 8799, $ \beta $ Pictoris and PDS 70 systems.

\item \textbf{Jupiter-like planets
  (Figure~\ref{fig:comp_config_zoom}b):} around 10 planets have final
  masses around $ 1-2 $ Jupiter masses with semi-major axis less than
  \SI{30}{au}. A few of them are initially far away ($ > \SI{50}{au}
  $) with mass slightly below $ M_{\rm iso} $, e.g., AA Tau Planet
  1(D), DoAr 25 Planet 1(D), GY 91 Planet 1(D). These planets first
  migrate significantly via the fast Type I migration, followed by the
  run-away gas accretion at the inner region of the disk, as
  illustrated by the L-shape brown tracks in
  Figure~\ref{fig:comp_config_fidu}(a). The rest of the planets (e.g.,
  AS 209 Planet 1(G), HD 163296 Planet 1(G)) are initially around
  \SI{10}{au}, which are close to the orbits of Jupiter and
  Saturn. Due to their large initial mass, they never experience
  strong migration nor accretion in the later stage.

\item \textbf{Neptune-like planets
  (Figure~\ref{fig:comp_config_zoom}c):} several planets (e.g. CI Tau
  Planet 3(D)) do not enter or barely enter the run-away gas accretion
  phase due to the combination of surface density drop and resonance
  block from the inner gas planet; they migrate very slowly without
  significant mass growth and stay at the region around
  \SI{50}{au}. V1094 SCO Planet 1(D) is close to Neptune in terms of
  both mass and semi-major axis.

\item \textbf{Distant small planets
  (Figure~\ref{fig:comp_config_zoom}d):} some low mass planets with a
  few $ \si{M_{\oplus}} $ (e.g. GY 91 Planet 3(D)) on the lower right
  area in Figure~\ref{fig:comp_config_fidu}(a) are too small and far
  away, so they do not evolve and stay at their initial locations
  without changing the mass.
\end{enumerate}

Incidentally, our simulated systems turned out not to reproduce the
observed population of Hot Jupiters. This conclusion is different from
the previous claim by \citet{Lodato2019}, who consider only the
migration of the planets without accretion. Indeed, the inward
migration of the gas giants is significantly slowed down once the
planets enter the run-away gas accretion phase, because the fast
growing planetary mass quickly depletes the gas gap. This effect has
been pointed out by \cite{Tanaka2019} in the context of a single,
isolated planet, and our simulation confirms that this is also valid
for multi-planet systems. Moreover, in a multi-planetary system, once
the inner planet becomes a gas giant, it blocks the outer planet from
migrating further inward via resonance. In order to become a Hot
Jupiter eventually, the embedded protoplanet should be less massive
and locate closer to the central star, so as to experience the fast
Type I migration to reach inner region followed by the substantial gas
accretion. This problem may point to several possibilities including
the observational bias and limitations of the current ALMA disks
and/or the additional perturbations from nearby objects, which will be
discussed in section \ref{subsec:implications}.

Note that when a planet is massive enough, the disk gas can be eccentric \citep{Kley2006}.
\cite{Tanaka2022} found that the disk is shifted to the eccentric phase when $K > 10^{4}$ (see Figure 16 of that paper).In our simulations, 7 out of 65 planets (e.g. planets of HL Tau) fall into this range (for $\alpha=10^{-3}$), and thus the gap eccentricity induced by the planet is rather low for the other planets. Even when the disk is eccentric, the eccentricity of the planet may be not significantly affected. According to the result of \cite{Papaloizou2001}, the eccentricity is boosted up to $0.2$ even the planet is as massive as $ 20 $ to $ \SI{30}{M_J} $. For those planets that are much less massive in our simulation, the eccentricity boosting due to this effect is not significant. Moreover, because most planet pairs are relatively widely separated in our simulation, $e = 0.1$ is not enough to cause the orbital crossing and the orbital stability of the planetary system is not strongly affected.

\subsection{Dependence on the different mass assignment for the embedded planets} \label{sec:mass_assign_evo}
\begin{figure*}
	\centering
        \gridline{\fig{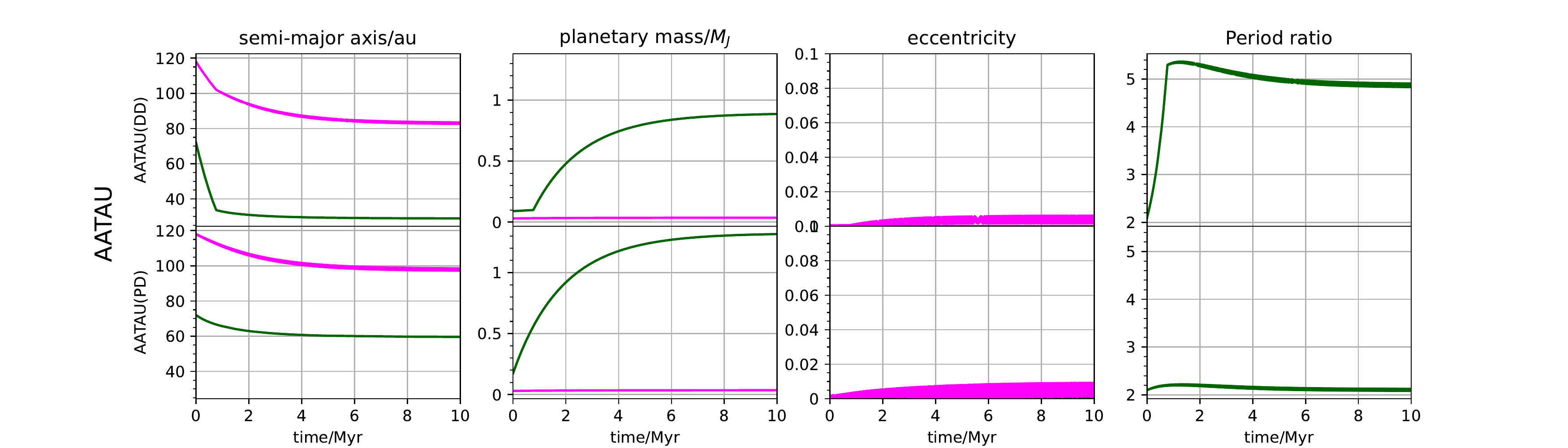}{1.0\linewidth}{(a) AA TAU}}
        \gridline{\fig{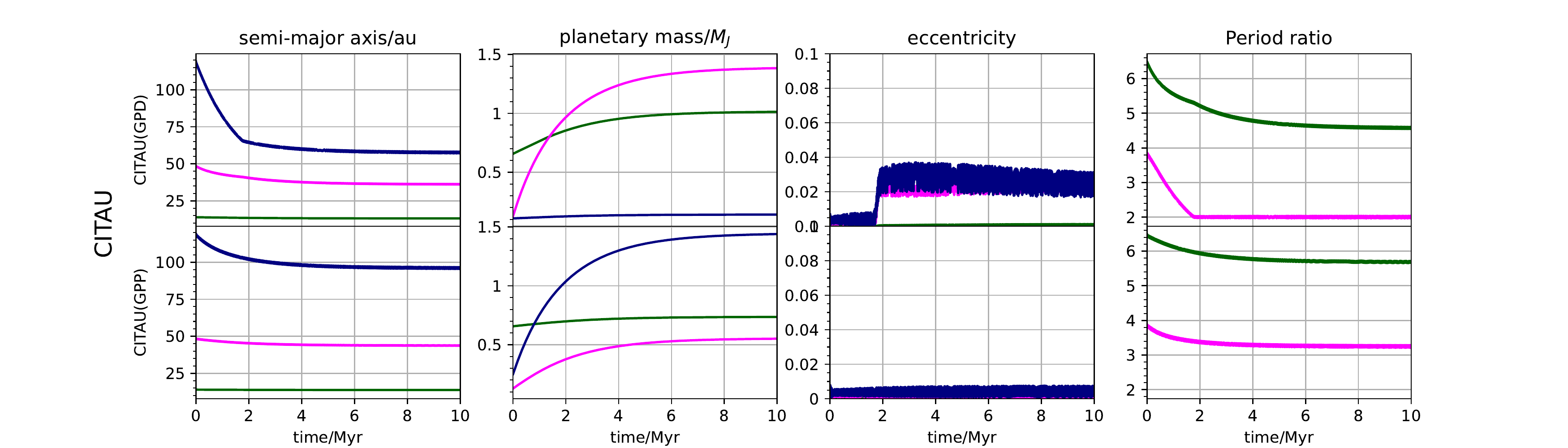}{1.0\linewidth}{(b) CI TAU}}
        \gridline{\fig{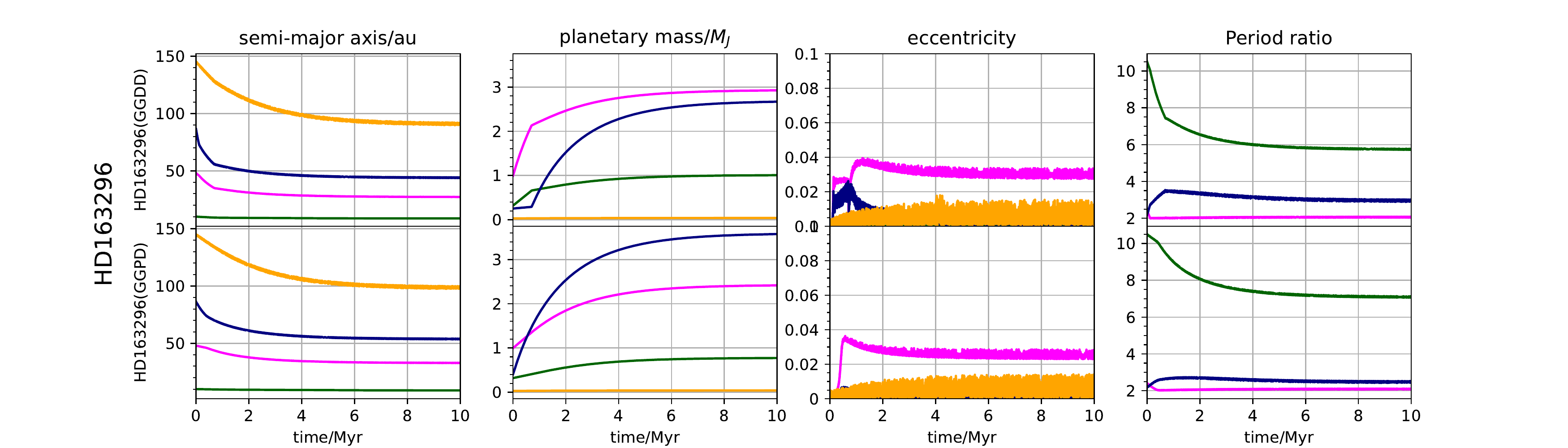}{1.0\linewidth}{(c)
            HD 163296}}
\caption{Example evolution of semi-major axis, planetary mass,
  eccentricity and period ratio of adjacent pair of disk (a) AS 209
  and (b) HD163296. Disk parameters are set to be the fiducial
  values.}
	\label{fig:evo_fidu}
\end{figure*}
We have shown in the previous subsection that the initial planetary
mass significantly affects the subsequent evolution, and thus the
final configuration of the system. To illustrate this dependence on
the initial mass, we select example systems with two planets (AA~Tau),
three planets (CI~Tau), and four planets (HD~163296). In each disk
system, we explore two different mass assignments for one of the
planets, considering theoretical and observational uncertainties of a
mass estimate, as mentioned in
Section~\ref{sec:ini_planetary_mass}. Other initial conditions, such
as locations of planets and disk parameters, are kept identical. Here
we focus on how the difference of the initial mass assignments affects
the evolution of planetary systems.

Our results show that initial planetary mass assignments lead to
different evolution outcomes, although the level of deviations depends
on the individual setup. As shown in Figure~\ref{fig:evo_fidu}~(a), in
AA Tau(DD), the inner and outer planets migrate from $ [72,118]\si{au}
$ to $ [28,83]\si{au} $, where in AA Tau(PD) they migrate from the
same initial locations to $ [60,98]\si{au} $. Similarly, Planets 1
(innermost), 2 (middle) and 3 (outermost) migrate from $
[14,48,119]\si{au} $ to $ [13,36,58]\si{au} $ in CI Tau(GPD) and to $
[14,44,96]\si{au} $ in CI Tau(GPP). Although the final position of the
Planet 1 is not strongly affected by the initial mass assignment, the
outer two planets of CI Tau(GPD) migrate further inward relative to
those of CI Tau(GPP).

CI Tau(GPD) experiences significantly stronger migration because its
Planet 3 has relatively small initial mass, and undergoes the fast
Type I migration at the beginning. Such a high migration speed was
sustained for around \SI{1}{Myr} due to the inefficient pebble
accretion. The fast approach between Planets 3 and 2 even excites the
eccentricity of Planet 2 at around \SI{2}{Myr}. When Planet 3 further
approaches Planet 2, its fast migration was hindered by both the
resonance block and drop of the gas surface density due to gas
accretion of of Planet 2. Consequently, both Planets 2 and 3
co-migrate with a period ratio very close to \num{2.0}. On the
contrary, in CI~Tau(GPP) Planet 3 starts from gas accretion, and that
lowers the surface density at the position of Planet 2. In effect,
both Planet 3 and Planet 2 suffer from slower migration, and their
final period ratio is larger than \num{3.0}.

The initial mass assignment also affects the mass evolution, and in
particular, results in distinct mass orders in CI Tau and HD 163296
systems. In CI Tau(GPD), Planet 3 undergoes pebble accretion
throughout, while both Planets 1 and 2 start with gas accretion. The
fast growing Planet 2 quenches the growth of Planet 1 by lowering the
surface density at the inner side (see equation
\ref{eqn:multip_sigmaunp}). It also prevents Planet 3 from migrating
further inward via 2:1 resonance. Therefore, Planet 3 can neither grow
efficiently nor migrate further inwards to trigger the gas accretion,
as shown by the period ratio evolution. In the end, Planet 2 becomes
the largest planet, and Planet 3 remains as a Saturn-size planet.

However, in CI Tau(GPP), Planet 3 initially starts with fast gas
accretion, and its large accretion rate quenches the growth of both
Planet 1 and 2. In effect, Planet 3 becomes a planet larger than
Jupiter at $ \SI{10}{Myr} $ epoch, while Planet 2 is the smallest
because of the reduced gas inflow due to the gas accretion of
Planet~3. The situation is similar in HD~163296(GGDD), except that
Planet 3 entered the gas accretion stage late at around \SI{1}{Myr},
instead of always staying at the pebble accretion stage. Such a delay
does not change the final mass qualitatively, but it is already
sufficient to reverse of the mass order of Planets 2 and 3.

The period ratio evolution shows that some planet pairs can enter
resonance via convergent migration. For example, the initial period
ratio of the outer pair in CI Tau(GPD) is around $ 4.0 $. Due to the
strong migration of Planet 3, the period ratio quickly decreases to $
2.0 $ within \SI{2}{Myr}, entering 2:1 resonance. Since Planet 3
undergoes Type I migration throughout the disk lifetime, its migration
speed can always match that of Planet 2 undergoing slower Type II
migration, which ensures the period ratio always sticks to $ 2.0 $. We
will discuss more about the overall period ratio distribution in
section \ref{sec:rslt_period_ratio} and section
\ref{sec:discussion_consistency} (also see Figure
\ref{fig:comp_instime})

To summarise, the difference of the initial mass assignments can
significantly change the orbital configuration the planetary system:
when the the initial planetary mass is below $ M_{\rm iso} $ (type
`D'), the planet migrates fast and its final position is much inward
than those planets assigned with masses equal or larger than $ M_{\rm
  iso} $. In general, the initial mass assignment has less significant
impact on the final planetary mass, because the fast migration of
planet initially below $ M_{\rm iso} $ can trigger the runaway gas
accretion in a short timescale, and thus the final masses are not
strongly affected. Nevertheless, in specific systems such as CI Tau
(Planet 3), qualitative difference exists between the final masses,
since the inner planet blocks the migration of the outer
planet. Therefore, it is important to take into account of the
planet-planet interaction and evolution feedback in such
multi-planetary systems.

\subsection{Sensitivity to $\alpha $ viscosity}
\label{sec:alpha_dependence}
\begin{figure*}
	\centering
	\gridline{\fig{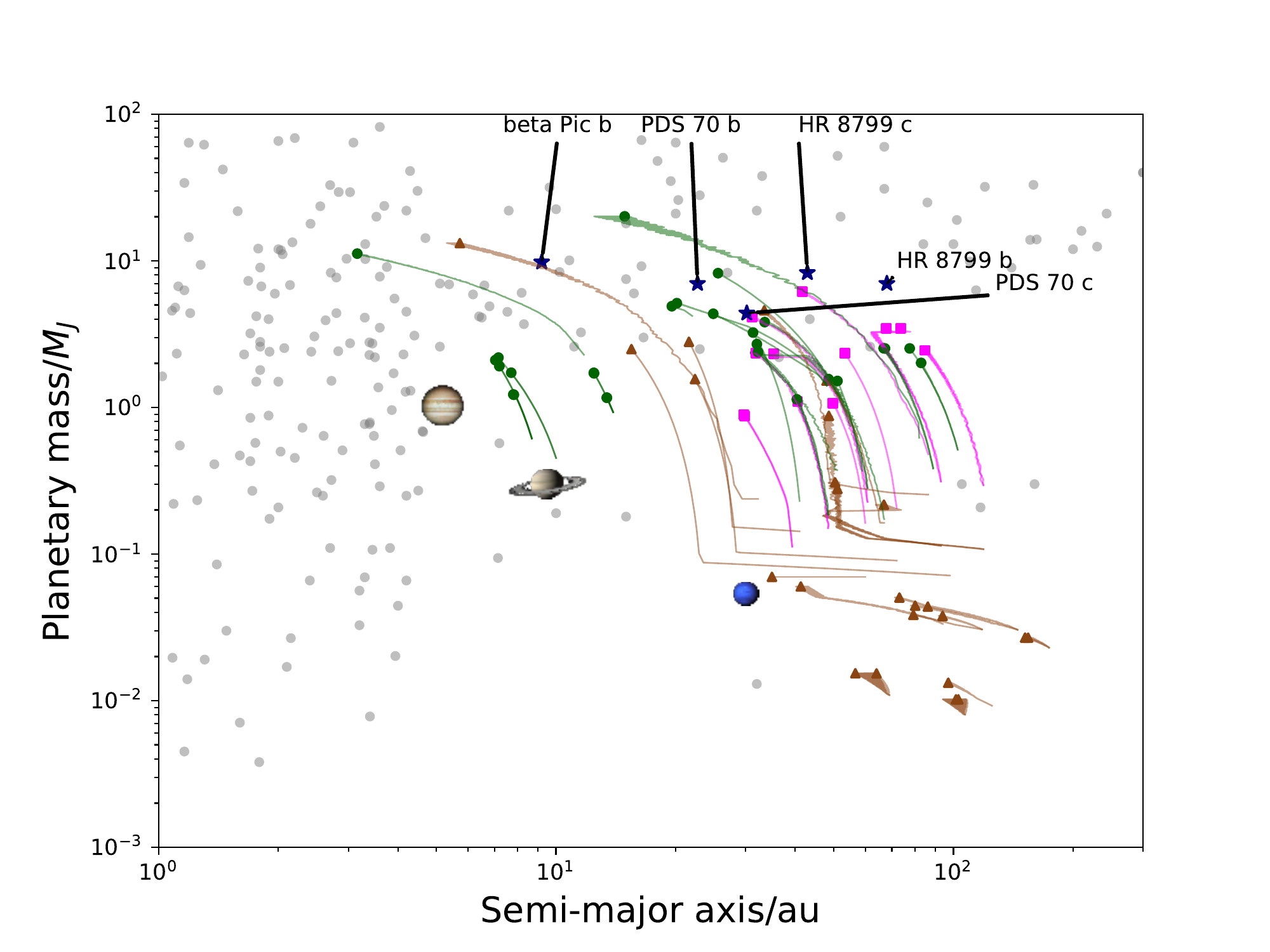}{0.65\linewidth}{(a)}
		      \fig{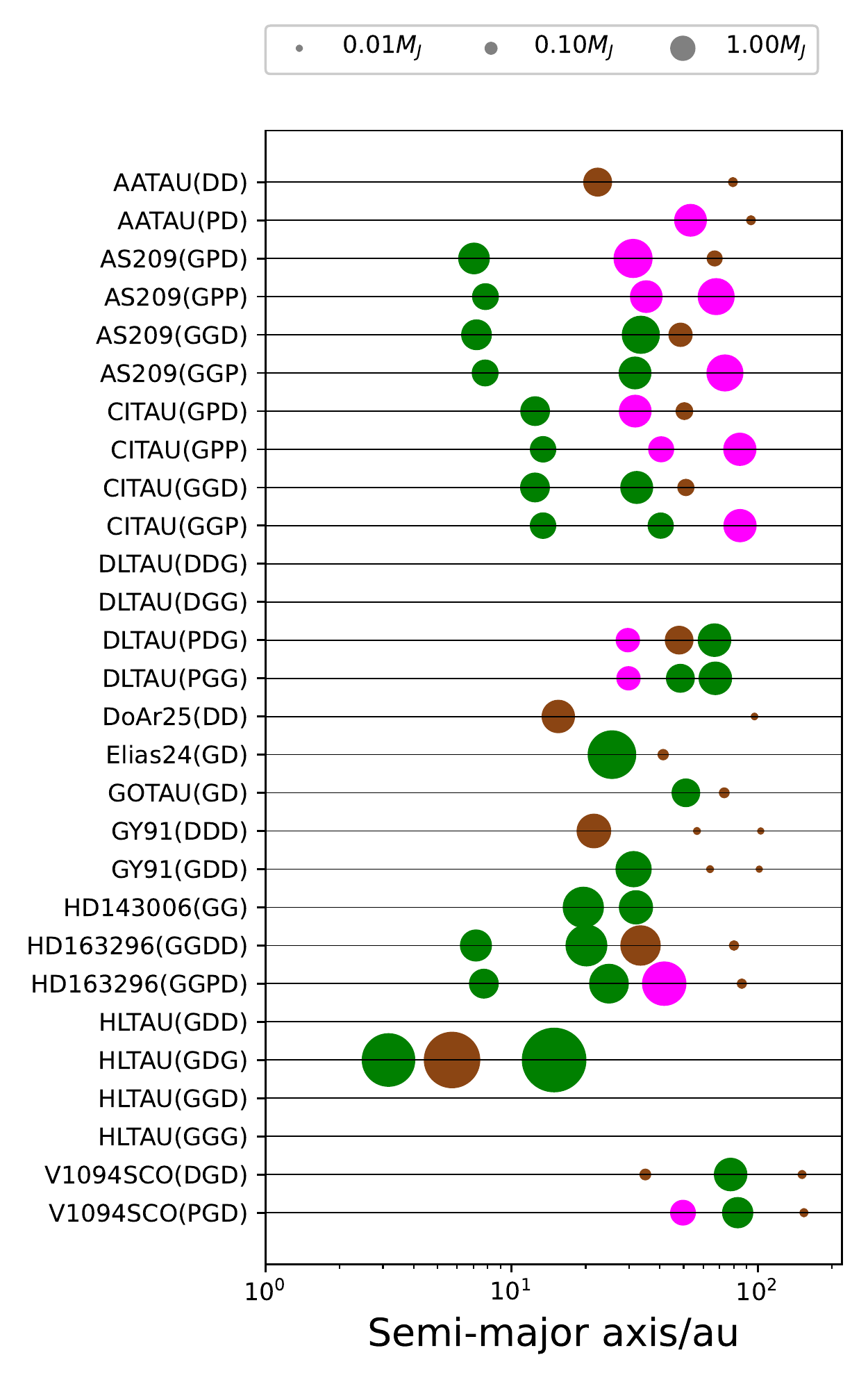}{0.33\linewidth}{(b)}}
	\gridline{\fig{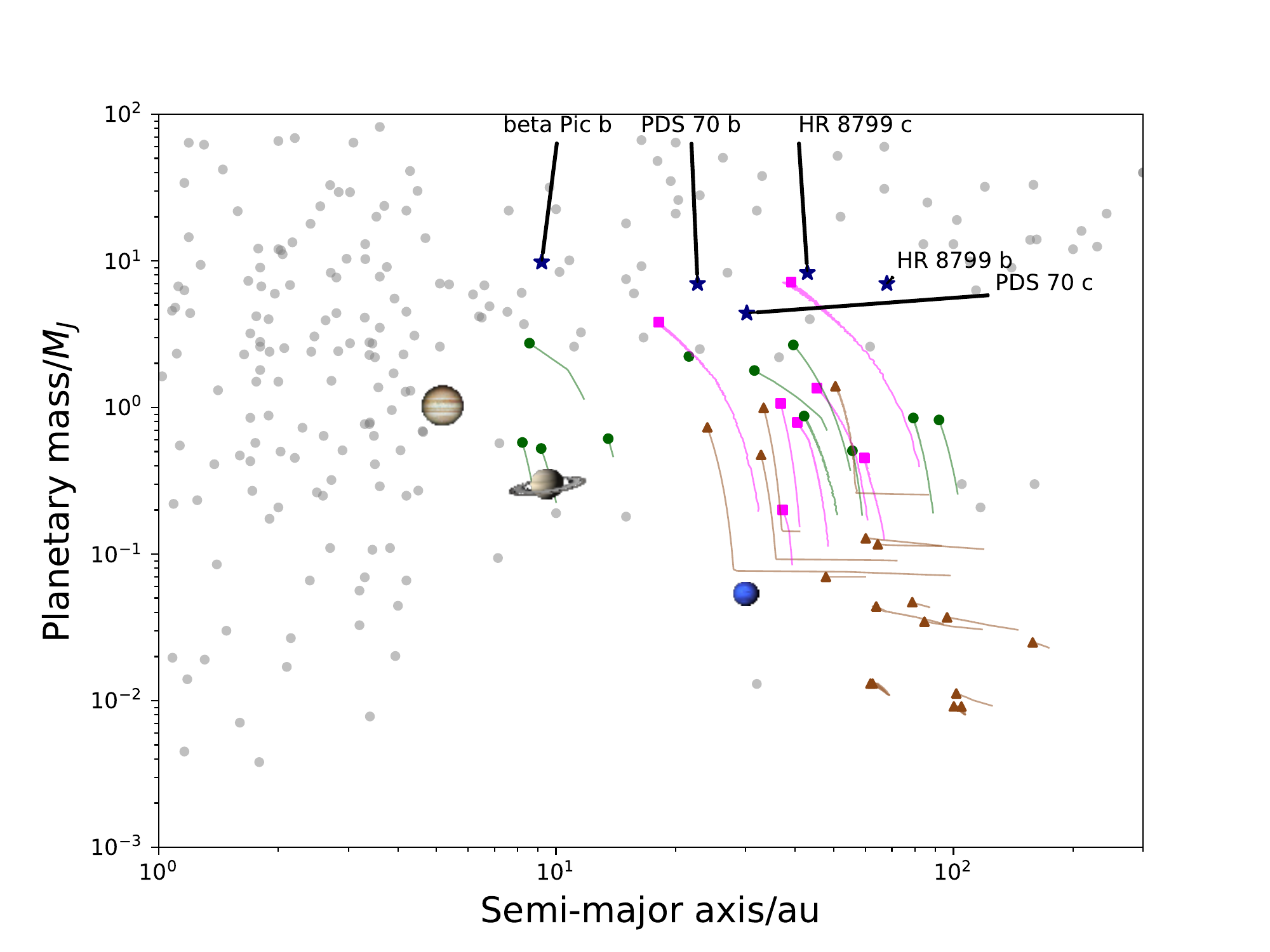}{0.65\linewidth}{(c)}
			  \fig{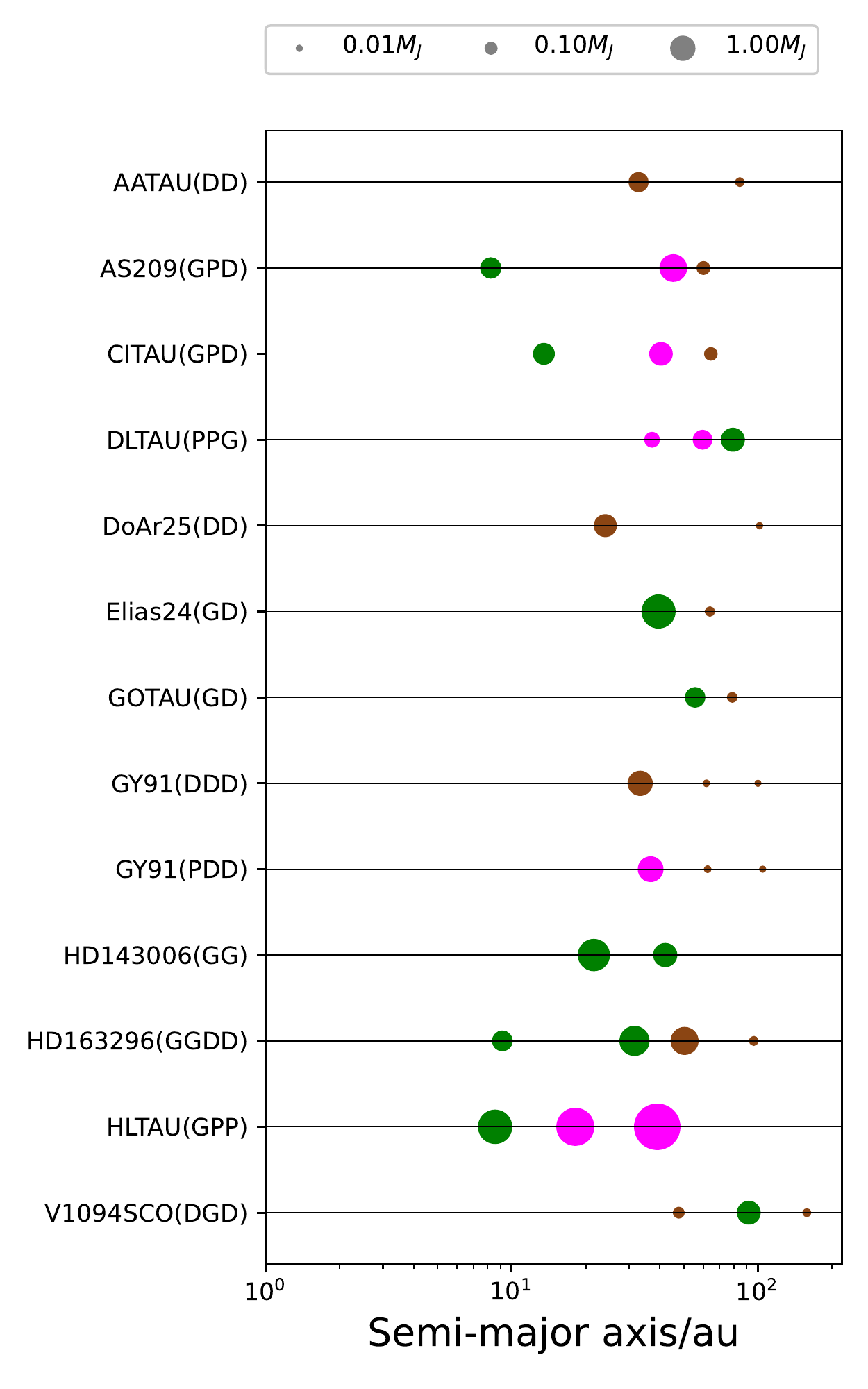}{0.33\linewidth}{(d)}}
\caption{Same as Figure \ref{fig:comp_config_fidu} but with different
  $ \alpha $ viscosities. (a)(b): $ \alpha = \num{2e-3} $. (c)(d): $
  \alpha = \num{5e-4} $. }
	\label{fig:comp_config_LSalp}
\end{figure*}

As mentioned in section~\ref{sec:ini_planetary_mass}, the $ \alpha $
parameter not only affects migration and accretion rates, but also
changes the initial conditions, including the classification of the
gaps and planetary mass estimation. To investigate the sensitivity to
the $ \alpha $ parameter, we also carried out simulations with $
\alpha = \num{2e-3} $ and $ \alpha = \num{5e-4} $, which differ by a
factor of two with respect to the fiducial $ \alpha(=\num{e-3})$. In
general, as $ \alpha $ increases, more gaps are classified as
indistinguishable gaps, and vice versa. For example, for $ \alpha =
\num{e-3} $, the HL Tau system only has one set of mass assignment,
but for $ \alpha = \num{2e-3} $, there are four different sets as both
Gap 2 and 3 can be interpreted as either gas gap or dust gap at larger
$ \alpha $. Moreover, both $ M_{\rm p,gas} $ and $ M_{\rm iso} $ are
scaled with $ \alpha $ parameters. Therefore, while comparing the same
system with different $ \alpha $ parameters, we focus on the overall
statistical trend rather than their specific case-by-case difference.

Figure \ref{fig:comp_config_LSalp} plots the final configuration of
the planetary systems at the end of the disk dispersal. Both migration
and accretion are enhanced when $ \alpha $ increases. For example,
while the innermost planet of HL Tau systems are located around
\SI{9}{au} with mass $ \sim \SI{2}{M_J} $ at $ \alpha = \num{5e-4} $,
they move further inward to around \SI{3}{au} with mass $ \sim
\SI{10}{M_J} $ at $ \alpha = \num{2e-3} $. It is because the gas depth
parameter $ K $ is inversely proportional to $ \alpha $ (see
equation~\ref{eq:def-K}), so increasing $ \alpha $ effectively
decreases the depth of the gas gap, therefore enhancing both rates of
migration and accretion that are proportional to the gas surface
density.

At $ \alpha = \num{2e-3} $, planets are both massive and closer to each other, and thus their resonance overlap and systems become unstable (eccentricity $ > 1 $): three out of four mass assignment sets of HL Tau become unstable during the disk stage, and similar instability also happens to two sets of DL Tau. Details about the instability time and evolution of the unstable cases are in Appendix \ref{app:unst_exp}. Since our migration and accretion models are only applicable to nearly circular orbits, we exclude these cases from the final configuration. It is worth to mention here that $ \alpha = \num{2e-3} $ is much larger than the fiducial $ \alpha $($ =\num{3e-4} $) of \citetalias{Wang2020} suggested by \cite{Pinte2015}, since the sharp edges of the HL Tau gap imply good dust settlement and thus weak turbulence level. Therefore, even though the majority of the HL Tau systems are unstable at $ \alpha = \num{2e-3} $, this should only be interpreted as extreme cases, and the majority of the HL Tau systems are still stable.

We also plot the semi-major axis and mass evolution of planets for $ \alpha = \num{2e-3}/\num{5e-4} $ in Figure \ref{fig:comp_config_LSalp}. Although the $ \alpha $ parameter in principle affects all accretion and migration mechanisms, the effect is only significant for planets entering the gas accretion stage, i.e., the distant giant planets and Jupiter-like planets. These planets initially above $ M_{\rm iso} $ migrate further inward and eventually become super-Jupiters or brown dwarfs when $ \alpha $ increases, exhibiting greater overlap with the observed planetary population.
	
On the other hand, some planetary systems consisting of the Neptune-like planets, such as V1094 SCO, DoAr 25 and GY 91, are relatively less sensitive to $ \alpha $ parameter since the majority of the planets always start from the pebble accretion stage. Some planets in these systems may enter gas accretion stage via inward migration, but by the time they reach $ M_{\rm iso} $, the disk surface density has already decayed. Therefore, in this case changing $ \alpha $ parameter does not affect their final configuration to a large extent. Finally, the evolution of distant low mass planets initially well below $ M_{\rm iso} $ is rather insensitive to the change of $ \alpha $. These sub-Neptunes remain far away from the star regardless of $ \alpha $, and have no counterpart in the observed population.

We can see that increasing $ \alpha $ generally results in more compact, massive and inner planetary systems, since both migration and accretion rates are enhanced. Also for some specific planets that are close to $ M_{\rm iso} $(e.g. AS 209 Planet 3), varying $ \alpha $ changes the time when the runaway gas accretion is triggered and thus can affect the final architecture to larger extent. The planetary systems tend to become less stable at larger $ \alpha $, and particularly at $\alpha = \num{2e-3}$, several sets become unstable during the disk phase. At $\alpha = \num{5e-4}$, there is no unstable cases and the evolution does not change qualitatively as compared with those systems evolved with $\alpha=10^{-3}$.

\subsection{Effect of disk lifetime}

\begin{figure*}[b]
	\centering
	\gridline{\fig{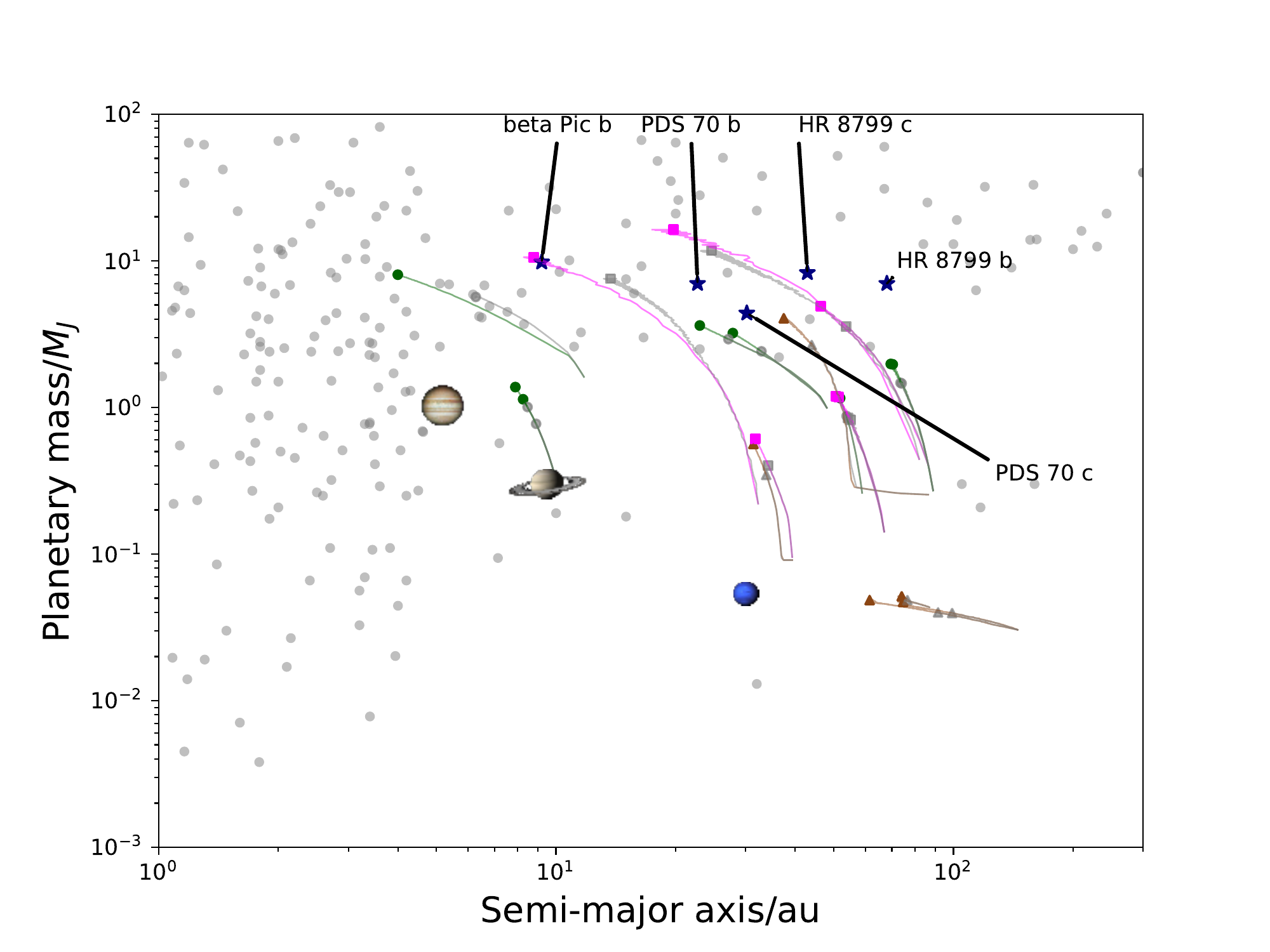}
          {0.6\linewidth}{}
	\fig{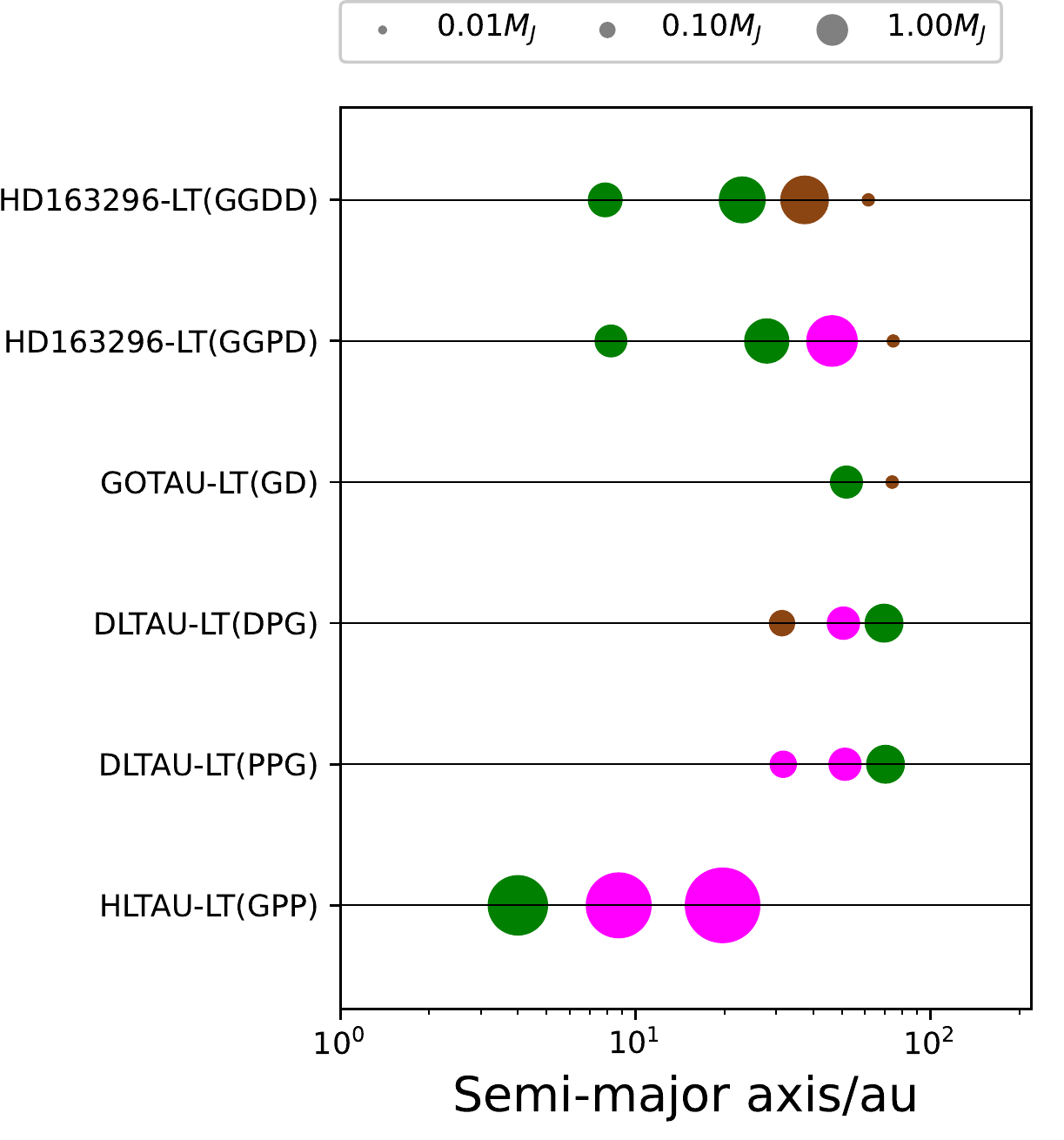}{0.40\linewidth}{}}
\caption{Same as Figure \ref{fig:comp_config_fidu} but with $
  \tau_{\rm disk} = \SI{3}{Myr}$. The total simulation time also
  changes to $ \SI{15}{Myr} $. The grey dots and tracks denote the
  same system evolved with fiducial $ \tau_{\rm disk} = \SI{2}{Myr}
  $. }
	\label{fig:comp_LT}
\end{figure*}

To investigate the impact of the disk lifetime, we select DL Tau, Go Tau, HD 163296 and HL Tau systems and re-run their simulation with longer disk lifetime $ \tau_{\rm disk} = \SI{3}{Myr} $. When the disk lifetime becomes longer, the disk surface density decays at a lower speed, therefore both the migration and accretion are expected to be enhanced, similar to increasing the $ \alpha $ parameter.

Figure \ref{fig:comp_LT} plots the evolution of planets with longer disk lifetime. For comparison, we plot the results of the same systems evolved with our fiducial value ($ \tau_{\rm disk} = \SI{2}{Myr} $) in grey colour. Indeed, all planets migrate to more inward region and become more massive. However, unlike increasing the $ \alpha $ viscosity, increasing the disk lifetime does cause any instabilities. We checked the evolution of the period ratios and found there is no significant difference as compared with the fiducial case, because the disk lifetime only extends the periods of the migration rather than increases the maximum migration speed. In this sense, although the disk lifetime can boost both the migration and accretion, it cannot strongly affect on the dynamical structure of the planetary systems.

\subsection{Period ratios of adjacent planets}
  \label{sec:rslt_period_ratio}

\begin{figure*}
	\centering
	\gridline{\fig{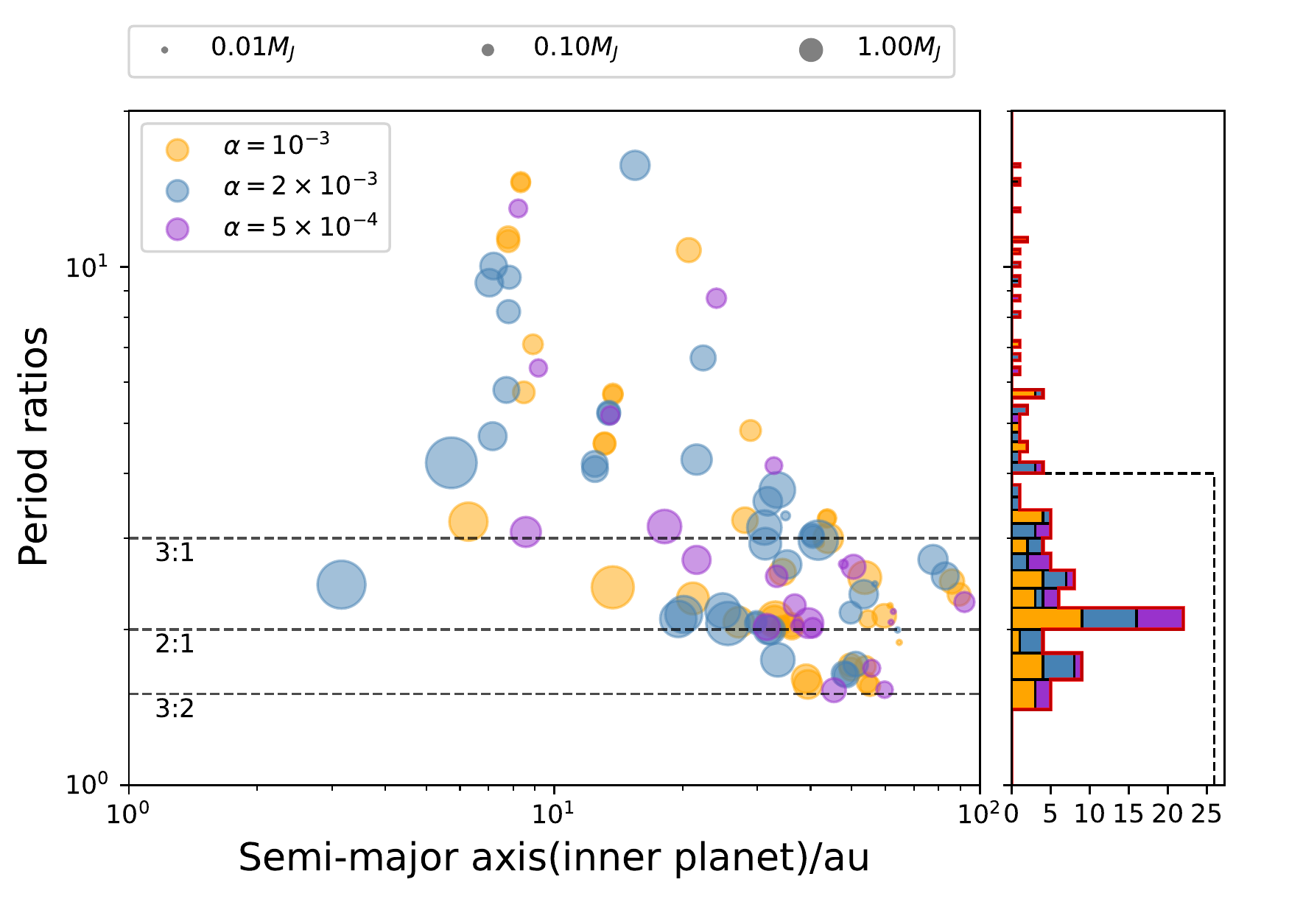}{0.7\linewidth}{(a)}}
	\gridline{\fig{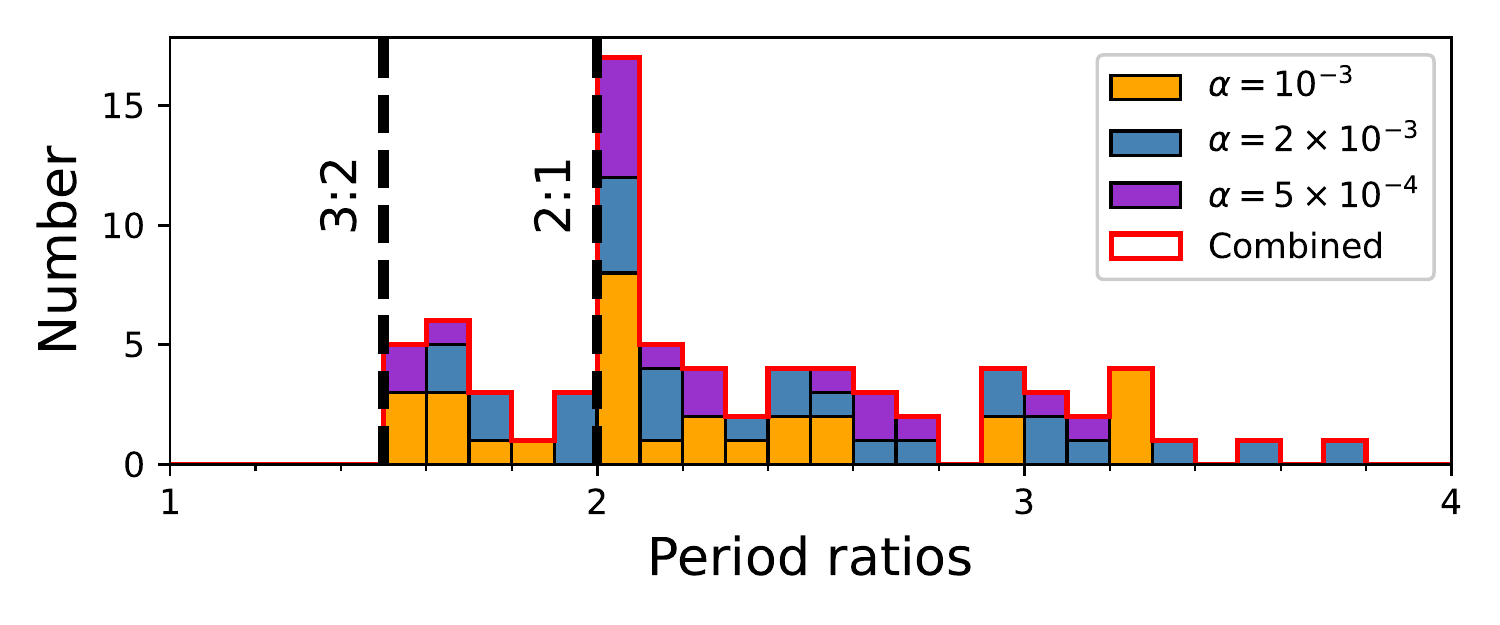}{0.5\linewidth}{(b)}}
\caption{(a)
  Period ratios of all adjacent planet pairs against the semi-major axis of the inner planets in the pairs. Colors represent results at different $ \alpha $, and size of the dot is proportional to $ 2/3 $ power of the mass of the inner planet. The histogram of the period ratios is plotted at bin-size = \num{0.2}. (b) Enlargement of the dotted
  region in (a), with bin-size = \num{0.1}.}
	\label{fig:period_hist}
\end{figure*}

Figure \ref{fig:evo_fidu} plots the evolution of period ratios for the three systems. In this subsection, we show the period ratios of all adjacent planet pairs against the semi-major axis of the inner planet at the end of the disk dispersal, together with the corresponding histograms in Figure \ref{fig:period_hist}~(a). The majority of period ratios lie within the range from \num{1.5} to \num{4}, with a long tail extending to nearly \num{14}. The inner planets of those widely-separated planet pairs (e.g. DoAr 25) have generally small mass. They are initially sub-$ M_{iso} $ planets that undergo fast inward migration at the beginning, while the outer planets are slow-migrating planets that are either very small (i.e., migration torque is weak) or very massive (i.e., gap is deep). Therefore, the outer planet cannot catch up with the inner planet, and the pair is gradually separated apart. As a result, the inner planets of these wide-separated pairs are located at inner region of $R_{\rm in} \lesssim 20$~au because of the significant inward migration.

Most planet pairs with relatively smaller period ratios ($<4$) are distributed in the outer region with $R_{\rm in} \simeq 50$~au. Figure \ref{fig:period_hist} (b) show that most of the rest of pairs have a period ratio $ < 2.8 $. There is a peak at \num{2.1}, showing many planet pairs are `trapped' outside of the strong 2:1 resonance. For those pairs initially with a sub-$ M_{\rm iso} $ planet outside and a massive gas-gap-opening planet inside, due to the fast Type I migration of the outer planet, most of them reach period ratios that are close to \num{2.0}. However, as the outer planet approaches the inner planet, the gas surface density drops because of the gas accretion of the inner planet (see equation \ref{eqn:multip_sigmaunp}), so the outer planet slows down. 

The situation for those pairs consisting of two massive gas-gap-opening planets is similar: when two planets are far apart, the gas surface density at the location of the outer planet is higher as its accretion quenches the gas inflow, so the migration speed of the outer planet is likely to be higher than that of the inner planet,
causing the outer planet to approach the inner one. As two planets get close, the difference between the gas surface density at their respective locations also decreases, so both the planets tend to co-migrate. In our simulations, we find no planet pair breaks into 2:1 resonance if the period ratio is initially above \num{2}.

Figure \ref{fig:period_hist} (b) also shows that around 20 planet pairs (including $ \alpha = \num{2e-3} \text{ and } \num{5e-4} $ cases) are inside the main resonance zone with period ratios less than 2.0, although their initial period ratios (e.g. outer pair of DL Tau) are below \num{2}. There is one obvious valley at \num{1.8}, which is mainly due to 3:2 and higher order resonances. Also such closely-packed planet pairs are absent in our previous investigation concerning only the HL Tau \citepalias[see Figure 8,][]{Wang2020}. There is no planet pair in our simulation breaking into $ 3:2 $ line, possibly because of the dense overlap of resonance zones near the $ 3:2 $ resonance, so planet pairs close to this range experience very chaotic evolution and thus quickly become unstable if they are closer (see Fig \ref{fig:unstable_examples}).

\begin{figure*}
	\gridline{\fig{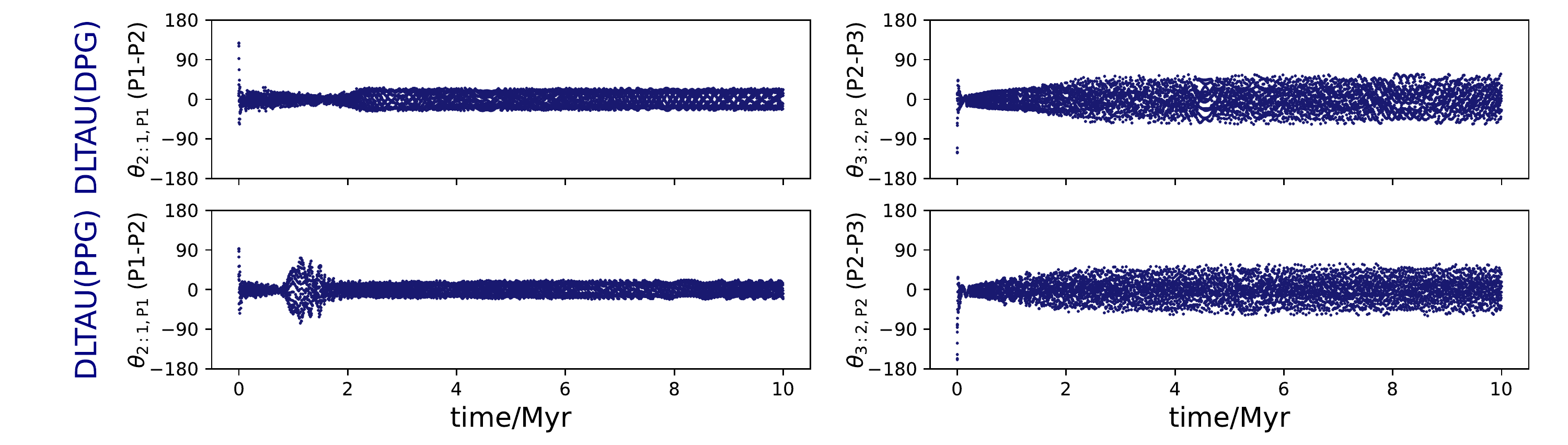}{1.0\linewidth}{(a)}}
	\gridline{\fig{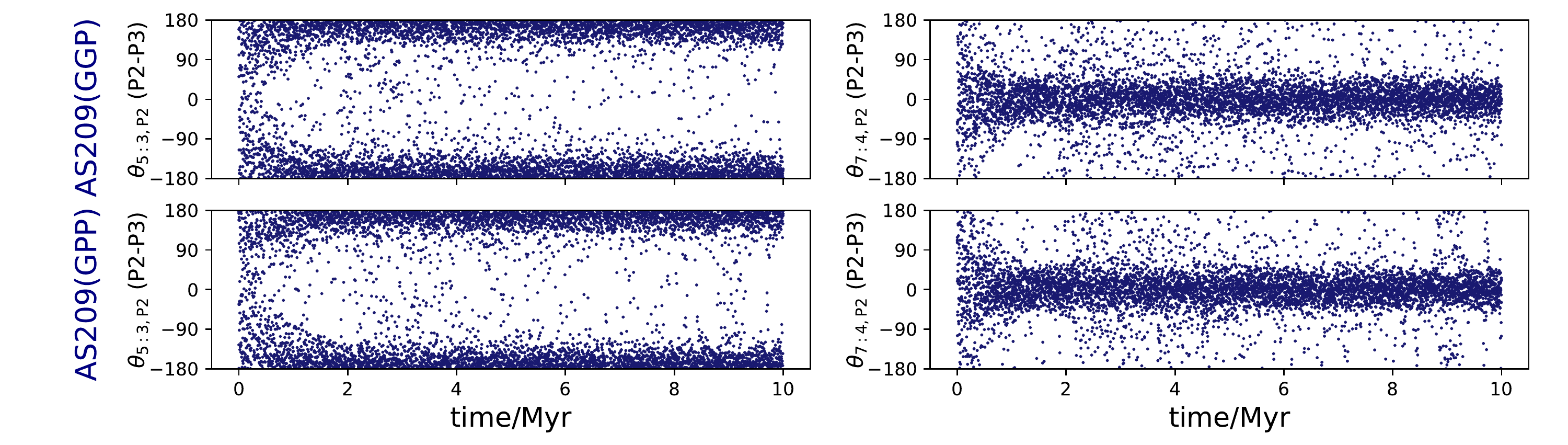}{1.0\linewidth}{(b)}}
	\caption{(a) 2:1 (of inner pair) and 3:2 (of outer pair) resonant
		angles of DL Tau system during the disk stage. (b) 5:3 and 7:4 resonant angles of Planet 2 in the outer pair in AS 209 sets GPP and GGP.}
	\label{fig:res_angle}
\end{figure*}

To better illustrate the MMR states of the planet pairs, we plot the resonant angle evolution ($ \alpha = \num{e-3} $) of the planet pairs in DL Tau and AS 209 systems in Figure \ref{fig:res_angle}. The resonant angle $ \theta_{p+q:q} $ for the $ p+q:p $ resonance is defined as :
\begin{equation}
	\label{eqn:def_res_ang}
	\theta_{p+q:q} = p\lambda_{\rm in} - (p+q)\lambda_{\rm out} + q\varpi
\end{equation}
where $ p $ and $ q $ are positive integers, the subscripts ``in'' and ``out'' are referring to the inner and outer planets that we considered in resonance. $ \lambda $ is the mean longitude and $ \varpi $ is the longitude of periastron of the inner or outer planet. Particularly, $ q $ is the order of the MMR.

Figure \ref{fig:res_angle} shows that both 2:1 and 3:2 resonant angles of DL Tau librate around $ \SI{0}{\degree} $ with small amplitudes, implying that the inner and outer pairs of DL Tau are in good 2:1 and 3:2 MMR, and all three planets are in a chain of resonance. For AS 209 sets GPP and GGP, the outer two planets are massive and close to each other with period ratio $ < 2.0 $, causing the resonance zone to overlap. As a result, the same pair oscillates between 5:3 and 7:4 resonant states, so most of the time the respective resonant angles are librating around $ \pm\SI{180}{\degree} $ and $ \SI{0}{\degree} $.

Overall, we see that the planet pairs become more closely packed after the disk stage evolution. Moreover, the convergent migration let some closely-packed planet pairs naturally enter MMR states that in turn prevents them from closer approach. As we will show in the next section, in planetary systems such as DL Tau(PPG), the chain of MMR stabilises the configuration of against small perturbations, and they can survive for at least \SI{10}{Gyr}.

\section{Evolution of planets after the disk dispersal under
   the stochastic perturbations 
  \label{sec:rslt_stochastic_force}}

\subsection{Perturbations from planetesimals}

After the disk dispersal, the disk-disk interaction is practically negligible, so we continue evolving the planetary systems at the end of disk phase (\S \ref{sec:rslt_disk_stage}) with their gravity alone in order to to examine the long-term orbital stability. This methodology has been applied in \citetalias{Wang2020}, which demonstrated that planetary architectures predicted from the observed HL Tau disk are mostly stable up to \SI{10}{Gyr}. In this paper, we further improve our previous methodology by incorporating the stochastic perturbative forces from numerous planetesimals surrounding each planet \citep[e.g.,][]{Nelson2004,Rein2009,Hands2014,Chatterjee2015}.

However, direct $N$-body simulations including star, planets and hundreds of planetesimals \citep[e.g.,][]{Chatterjee2015} are computationally expensive, and thus it is not feasible to calculate the long-term evolution using such a full treatment. Hence, we first examine the statistical nature of the perturbative force by carrying out a simple $N$-body simulation consisting of a planetary system with a planetesimal disk, and then formulate our simplified implementation calibrated by the result as described in details below.

We perform direct calculation of gravitational force from the planetesimals as follows in order to calibrate the strength of the perturbative forces, mainly following \cite{Chatterjee2015}. We pick up the HD~143006 system as a representative system, which consists of two giant planets initially close to 2:1 period ratio. The initial semi-major axis (\SI{21}{au}, \SI{37}{au}) and planetary masses ($ \SI{3.2}{M_J}, \SI{1.5}{M_J}$) are the same as those at the end of the disk stage. We introduce a planetesimal disk containing \num{1000} planetesimal particles. The planetesimal disk extends from $a_{\rm min}$ = 1:3 period orbit inside of the inner planet to $a_{\rm max}$ = 3:1 period ratio outside of the outer planet, so as to avoid the edge effect \citep{Chatterjee2015}.

\begin{figure*}
	\begin{minipage}{.4\linewidth}\vspace{0pt}
		\includegraphics[width=\linewidth]{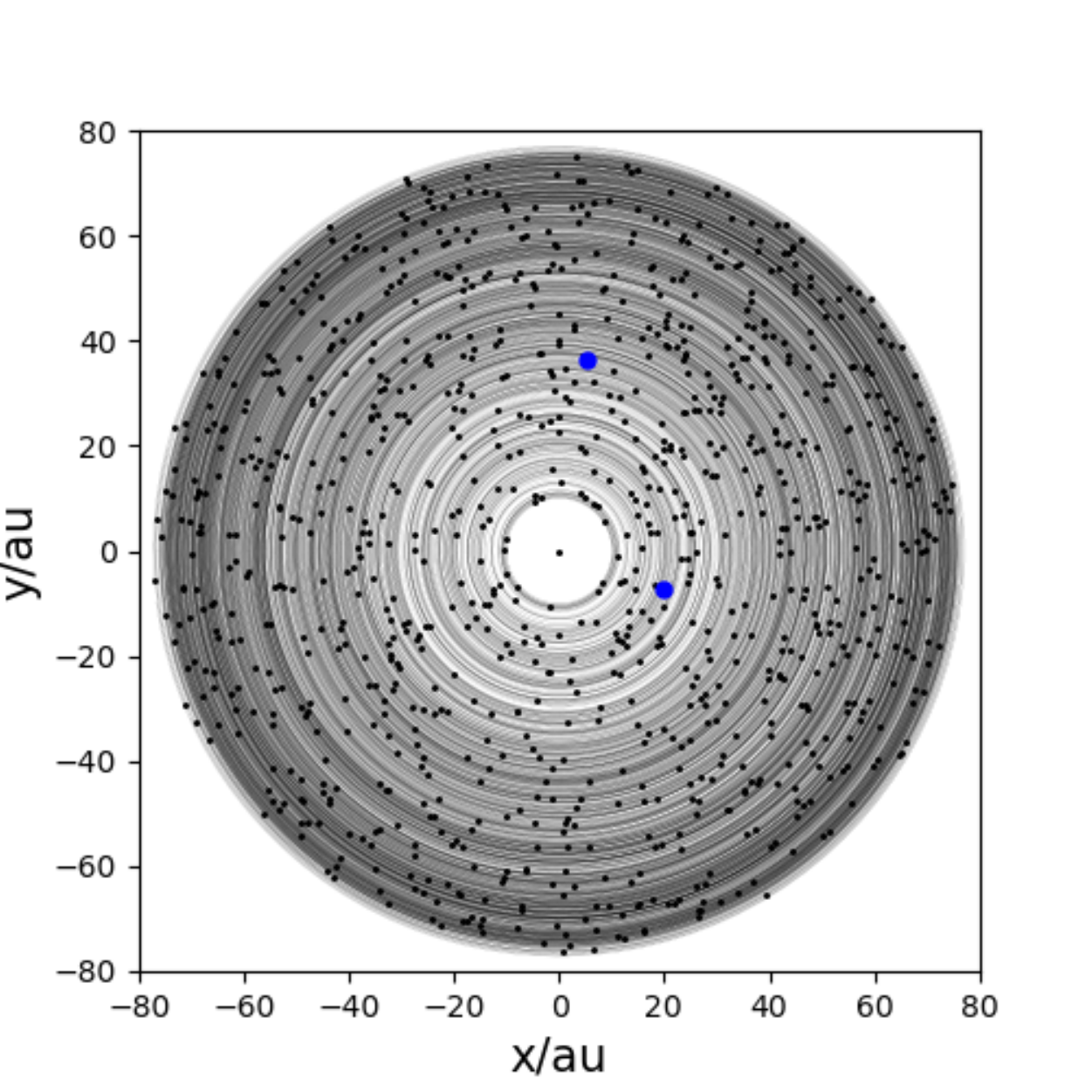}
	\end{minipage}
	\hfill
	\begin{minipage}{.6\linewidth}\vspace{0pt}\raggedright
		\begin{minipage}[t]{\linewidth}\vspace{0pt}\raggedright
			\includegraphics[width=\linewidth]{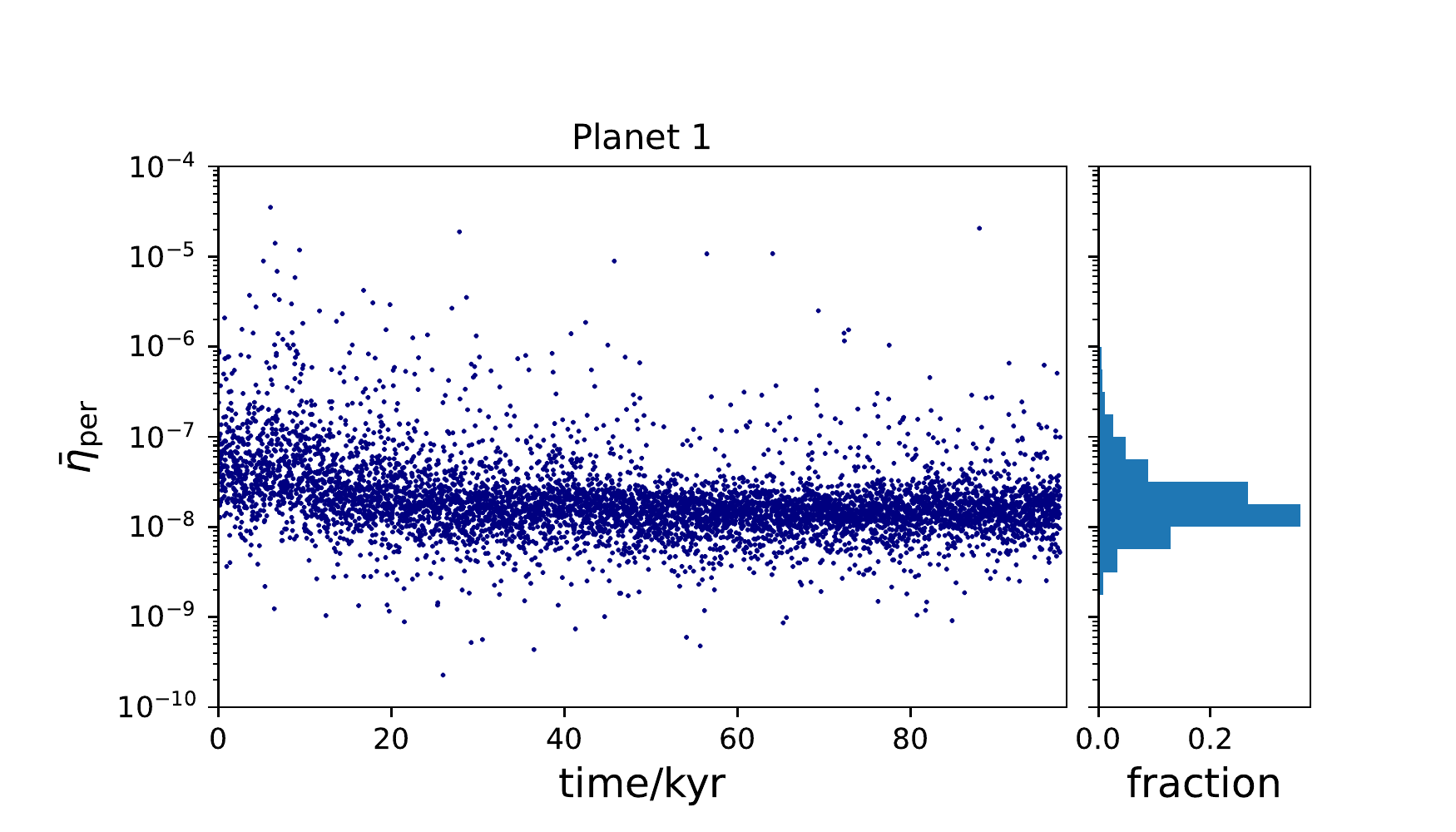}
		\end{minipage}
		\begin{minipage}[t]{\linewidth}\vspace{0pt}\raggedright
			\includegraphics[width=\linewidth]{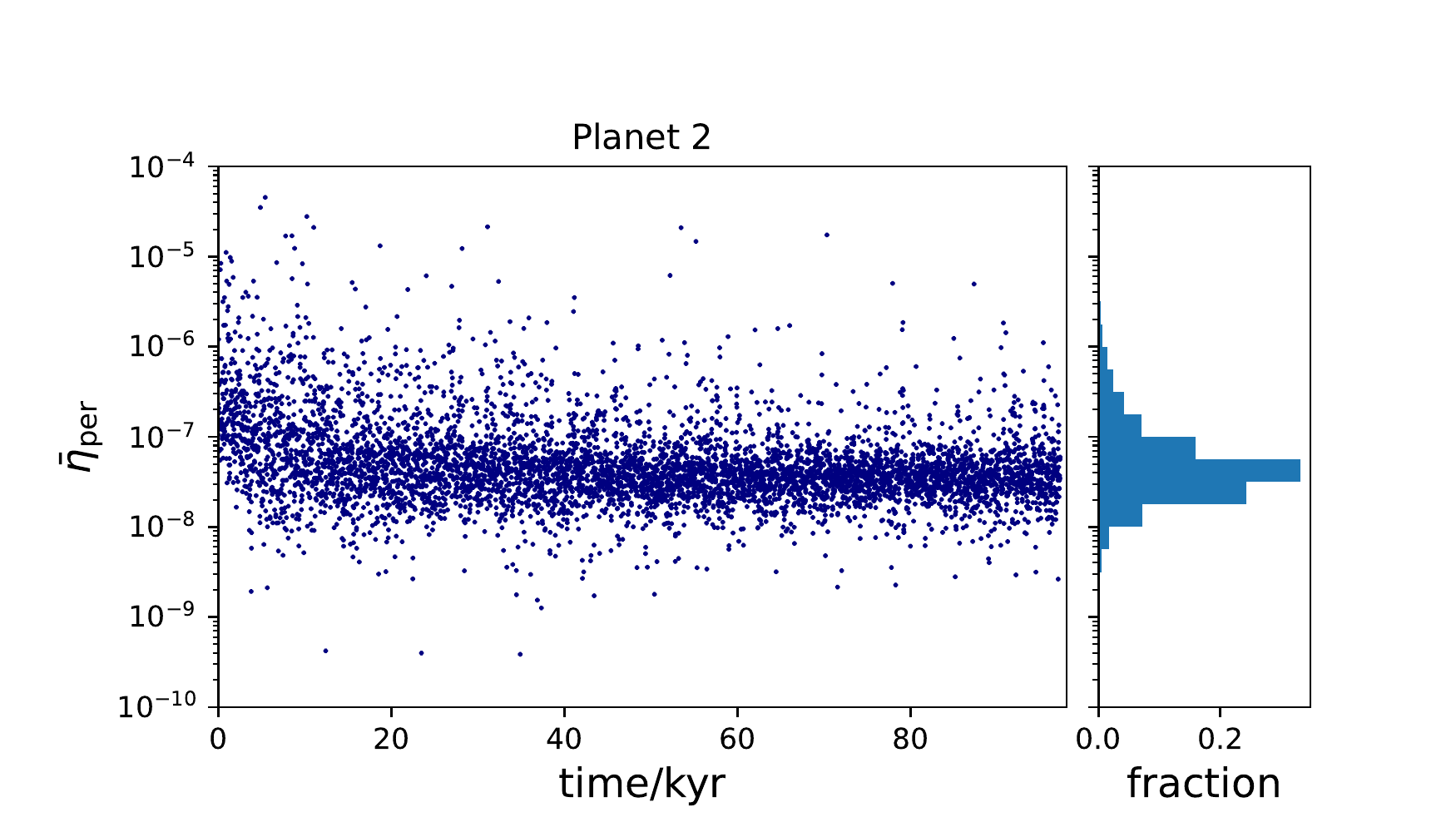}
		\end{minipage}
	\end{minipage}
\caption{Left: Initial setup of the planets with a planetesimal
  disks. Black dots are planetesimals, and blue dots are the
  planets. The initially orbits of the planetesimals are also
  plotted. Right: Time-averaged $ \eta_{\rm per} $ against the
  simulation time for both planet 1 and planet 2. The perturbative
  force is averaged over $ \SI{0.16}{yr} $ to be consistent with our
  random force implementation in the previous section.}
\label{fig:avg_perturb_force}
\end{figure*}

The left panel of Figure \ref{fig:avg_perturb_force} shows the initial setup. All planetesimal particles have identical mass, and their initial semi-major axis follow $ \propto R^{-1} $ power law distribution. The total mass of the disk is given by $\SI{3.3e-3}{M_{\oplus}}(a_{\rm max}/\si{au}-a_{\rm min}/\si{au})$, so the mass of each particle is at the order of $ \sim\SI{e-4}{M_\oplus} $, or roughly the mass of Ceres. In the simulation, we treat all the planetesimal particles as semi-test particles, which means they can mutually interact with the star and planets however not among themselves. The same $N$-body code \texttt{REBOUND} and the hybrid integrator \texttt{HERMES} are used with a typical time step equal to $ \SI{0.016}{yr} $\footnote{Our simulation uses natural units, i.e., $ \si{M_\odot} $ and \si{au}, and the gravitational constant is set to be unity. Therefore, the unit time corresponds to $ {1/2\pi} $ \si{year} $\sim \SI{0.16}{year}$.}, and the integration finishes at \num{1000} orbital periods of the inner planet.

The perturbative force per unit mass of the $i$-th planet is computed by summing up a contribution from the $k$-th planetesimal at time $ t $:
\begin{align}\label{eqn:sum_perturb_force}
	\boldsymbol{f}_{{\rm per}, i}(t) = -
	\sum\limits_{k \in \rm planetesimals}
	\frac{\mathcal{G}M_k}{|\mathbf{R}_k(t)
	  -\mathbf{R}_i(t)|^3}\left[\mathbf{R}_k(t)-\mathbf{R}_i(t)\right], 
\end{align}
where $M_{k}$ and $ \mathbf{R}_k $ denote the mass and the position vector of the $k$-th planetesimal with $ \mathbf{R}_i$ being the position vector of the $i$-th planet.

The perturbative force is then normalised by the stellar gravity and averaged over each time interval $ \Delta T = \SI{0.16}{yr}$ to give the dimensionless, time-averaged strength $ \bar{\eta}_{per}(t) $ as
\begin{align}
\label{eqn:t_avg_perturb_strength}
\bar{\eta}_{per,i}(t)  =
\left|\frac{1}{\Delta T}\int_{t-\Delta T}^{t} \boldsymbol{\eta}_{per,i}(t')dt'\right|
\end{align}
where
\begin{align}
  \boldsymbol{\eta}_{per,i}(t) =
  \dfrac{\boldsymbol{f}_{{\rm per}, i}(t)}{f_{*,i}}
=  \dfrac{\boldsymbol{f}_{{\rm per}, i}(t)}{\mathcal{G}{M_*}R_i^{-2}(t)}.
\end{align}

The right panels of Figure \ref{fig:avg_perturb_force} plot the averaged dimensionless strength of the perturbative force against the time, with upper and lower panels correspond to Planets 1 and 2, respectively. Initially, $ \eta_{\rm per} $ fluctuates with a relatively large amplitude. After a few hundreds orbits, the system approaches equilibrium and the fluctuation level decreases. While the total range of $ \bar{\eta}_{per,i} $ covers nearly three orders of magnitude, histograms show that the majority of the data points are in the range between \num{e-7} and \num{e-8}.  The histograms also indicate that the peak of $ \eta_{\rm per} $ distribution is a few times larger for Planet 2 relative to Planet 1, because the stellar gravity acting on outer Planet 2 is weaker. This difference is not significant in terms of the order of magnitude.

We also investigate the dependence of $ \bar{\eta}_{per} $ on the total mass of the planetesimal disk and the particle number. The peak value of $ \bar{\eta}_{per} $ is found to scale linearly with the total mass of the planetesimal disk but it is not sensitive to the number of planetesimal particles as long as the total disk mass is the same (see Appendix \ref{app:eta_mass_number_dep}). Therefore, even considering the uncertainty of the total mass of the planetesimal disk and the variation of the semi-major axis, $ \bar{\eta}_{per} $ is expected to be less than $ \sim \num{e-6}$. We will consider next if the planetary systems are stable against the mutual gravity and the stochastic perturbations at this level.

\subsection{Numerical method to test the long-term stability}
\label{subsec:pertative_force}

On the basis of the simulation run in the previous subsection, we consider a simplified model incorporating the perturbative force in the post-disk stage, and evolve the systems up to $ \SI{10}{Gyr} $.  The $ i $-th planet now obeys the following stochastic equation of motion:
\begin{equation}
	\label{eqn:postdisk_eom}
	\boldsymbol{\ddot{R}}_i = \boldsymbol{f}_{\rm{grav},i}
	+\boldsymbol{f}_{\rm{per},i},
\end{equation}
where $\boldsymbol{f}_{\rm per,i} $ represents the perturbative force
per unit mass acting on the $ i $-th planet.

We assume that the magnitude of $\boldsymbol{f}_{\rm per} $ is given by $ f_{{\rm per},i} = \eta_{\rm per} f_{*,i}, $ with $\eta_{\rm per} $ being the dimensionless strength factor. In each simulation run, we fix the value of $\eta_{\rm per} $, but repeat many runs by systematically varying the value.  The perturbative force is assumed to be uncorrelated beyond each time step: the direction of the perturbative force acting on each planet is uniformly random between $0^\circ$ and $\SI{360}{\degree}$ at each time step $ \Delta t = \SI{0.16}{yr} $.

In this post-disk stage, we integrate the system using the \texttt{WHFAST} integrator \citep{Rein2015}. We reset the simulation epoch as $t=0$ at the beginning of this stage. Systems are integrated by varying $ \eta_{\rm per} $ from $ \num{e-1} $ to $ \num{e-8} $. When one of the planets is ejected from the system (semi-major axis $ a_p > \SI{e3}{au} $), collides with either its hosting star ($ a_p < \SI{0.01}{au} $) or the other planets (mutual distance $ < \SI{2}{R_J} $), we regard this system as unstable and stop the simulation. Otherwise, the system is integrated up to \SI{10}{Gyr}.

\section{Stability of the resulting multi-planetary systems
  \label{sec:result-stability}}

\subsection{Stability under the presence of perturbative forces}

Based on the model we described in section~\ref{sec:rslt_stochastic_force}, we examined the instability time of the multi-planetary system with random perturbative force in various magnitudes ($10^{-8} \leq \eta_{\rm per} \leq 10^{-1}$). We eliminate those systems that has already become unstable during the disk phase and only carry out stability check for the remaining systems.

\begin{figure}
	\centering
	\gridline{\fig{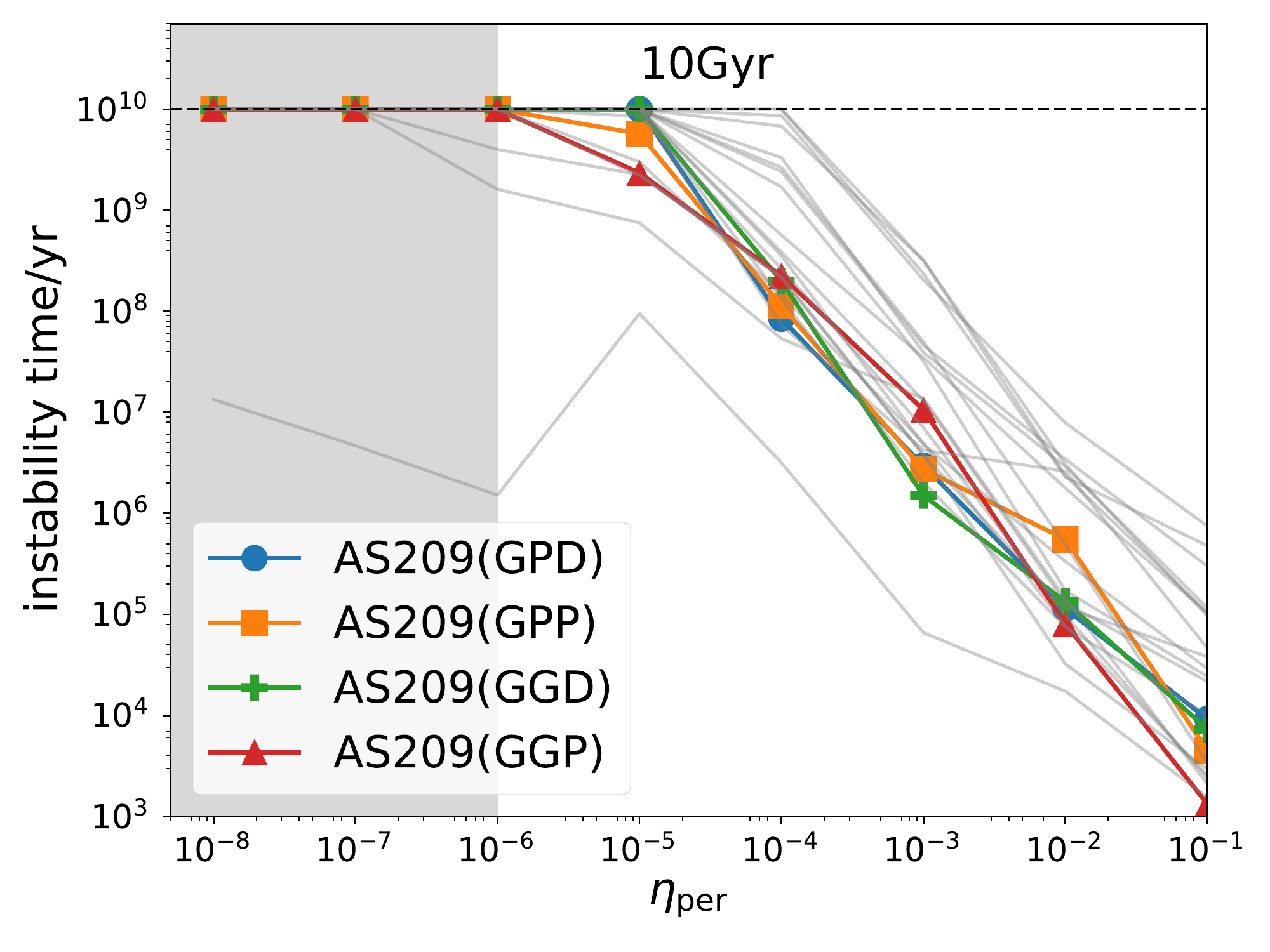}{0.33\linewidth}{(a) AS 209}
		\fig{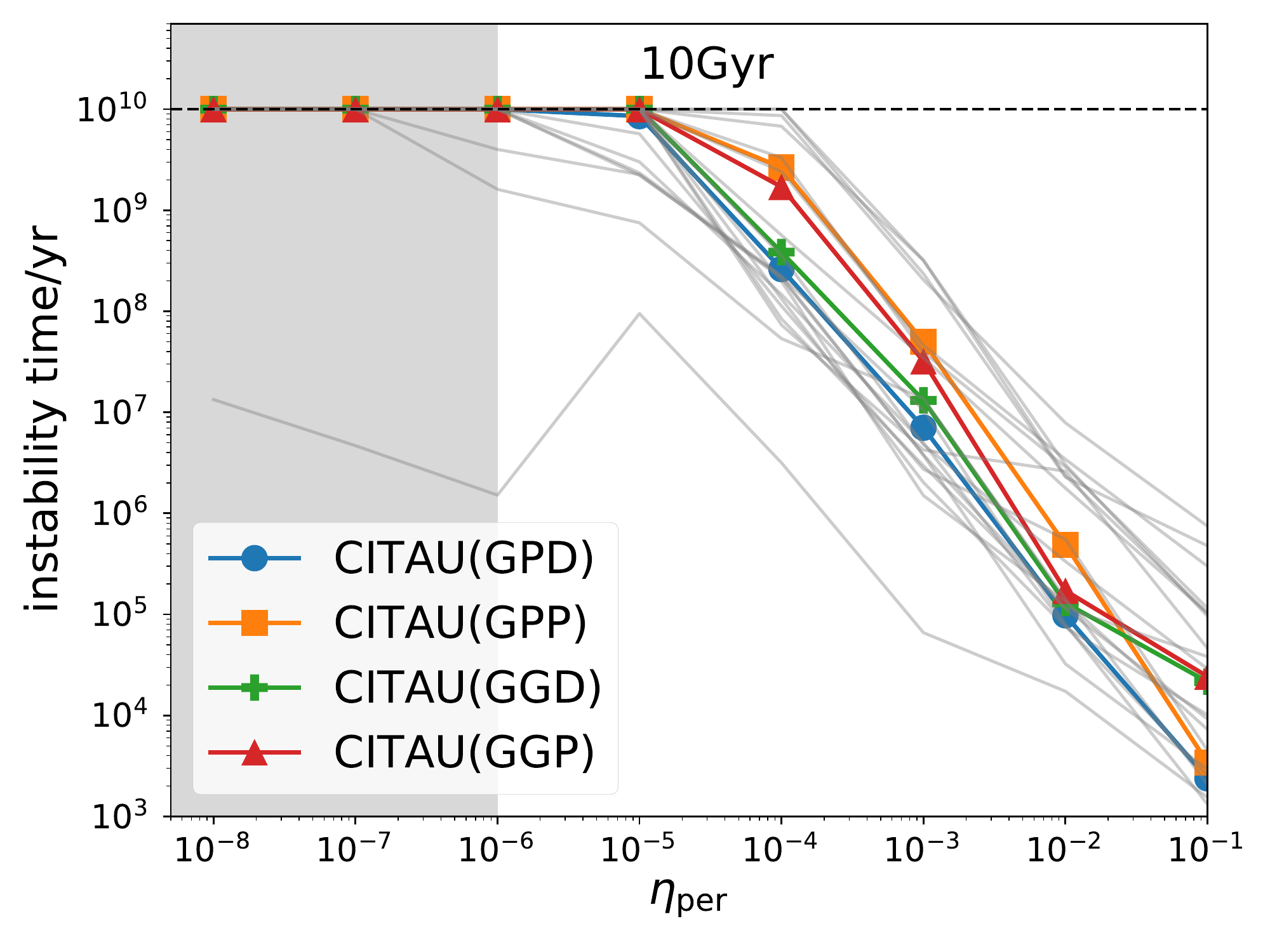}{0.33\linewidth}{(b) CI Tau}
		\fig{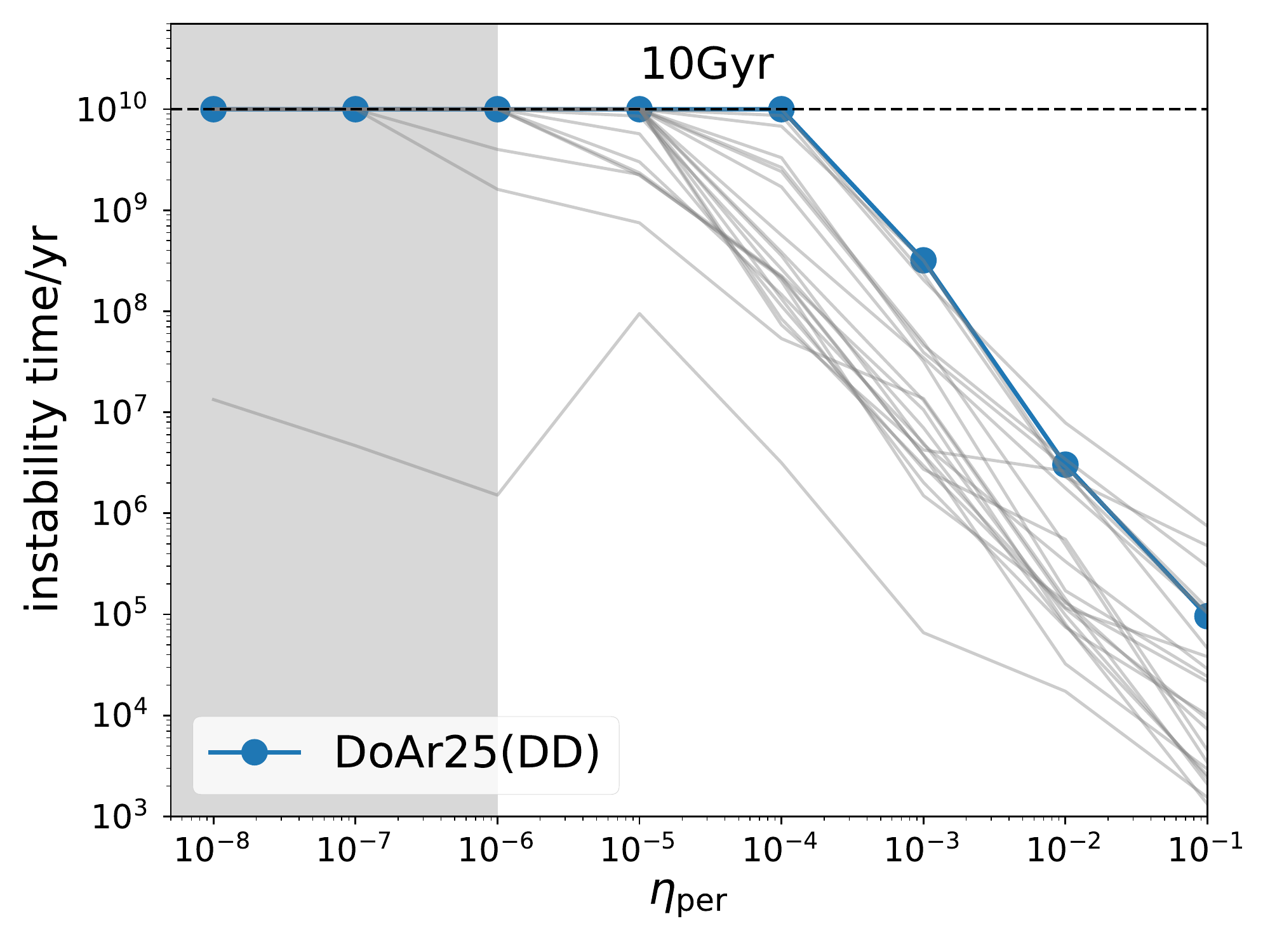}{0.33\linewidth}{(c) DoAr 25}}
	\gridline{\fig{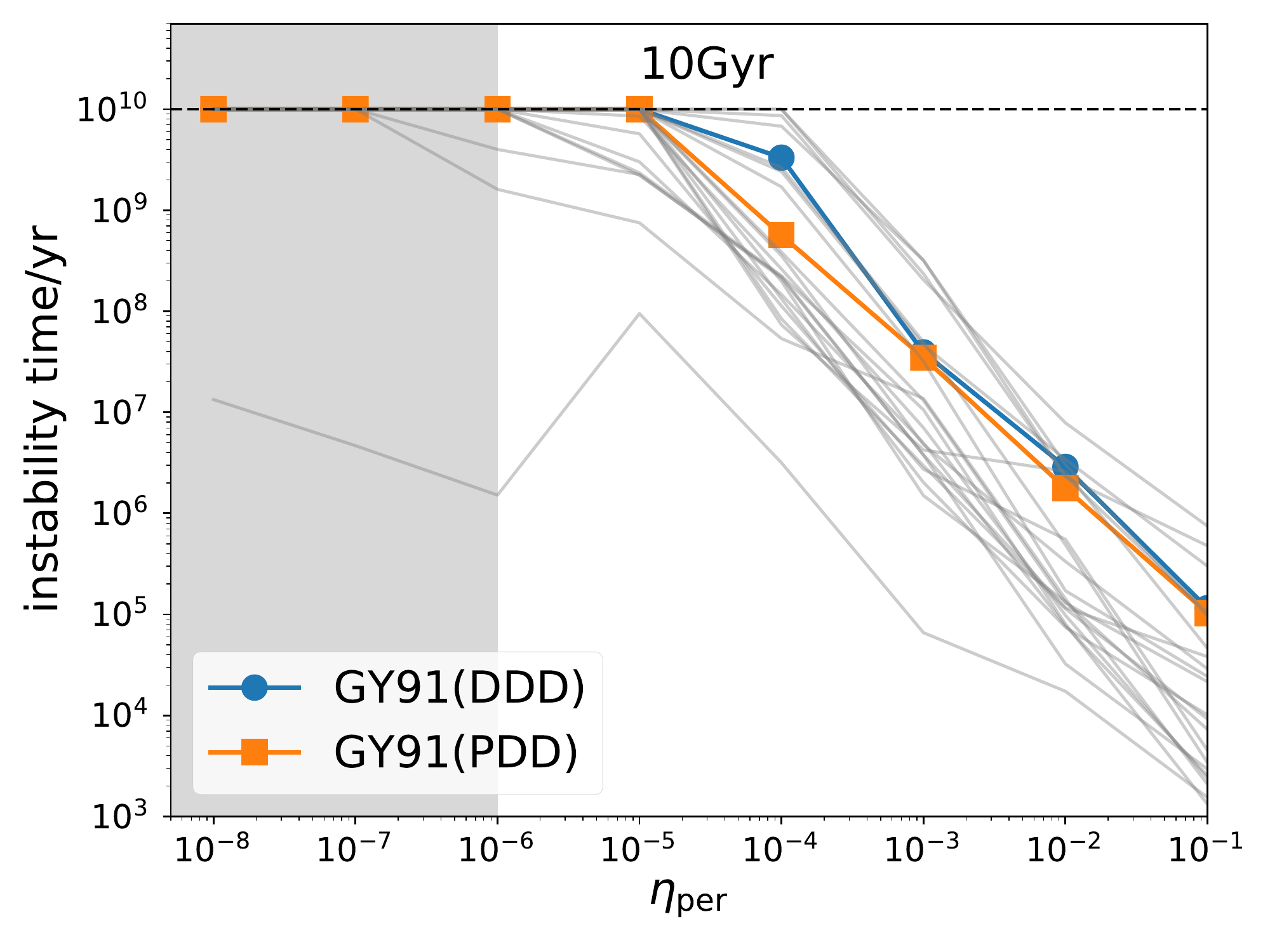}{0.33\linewidth}{(d) GY 91}
		\fig{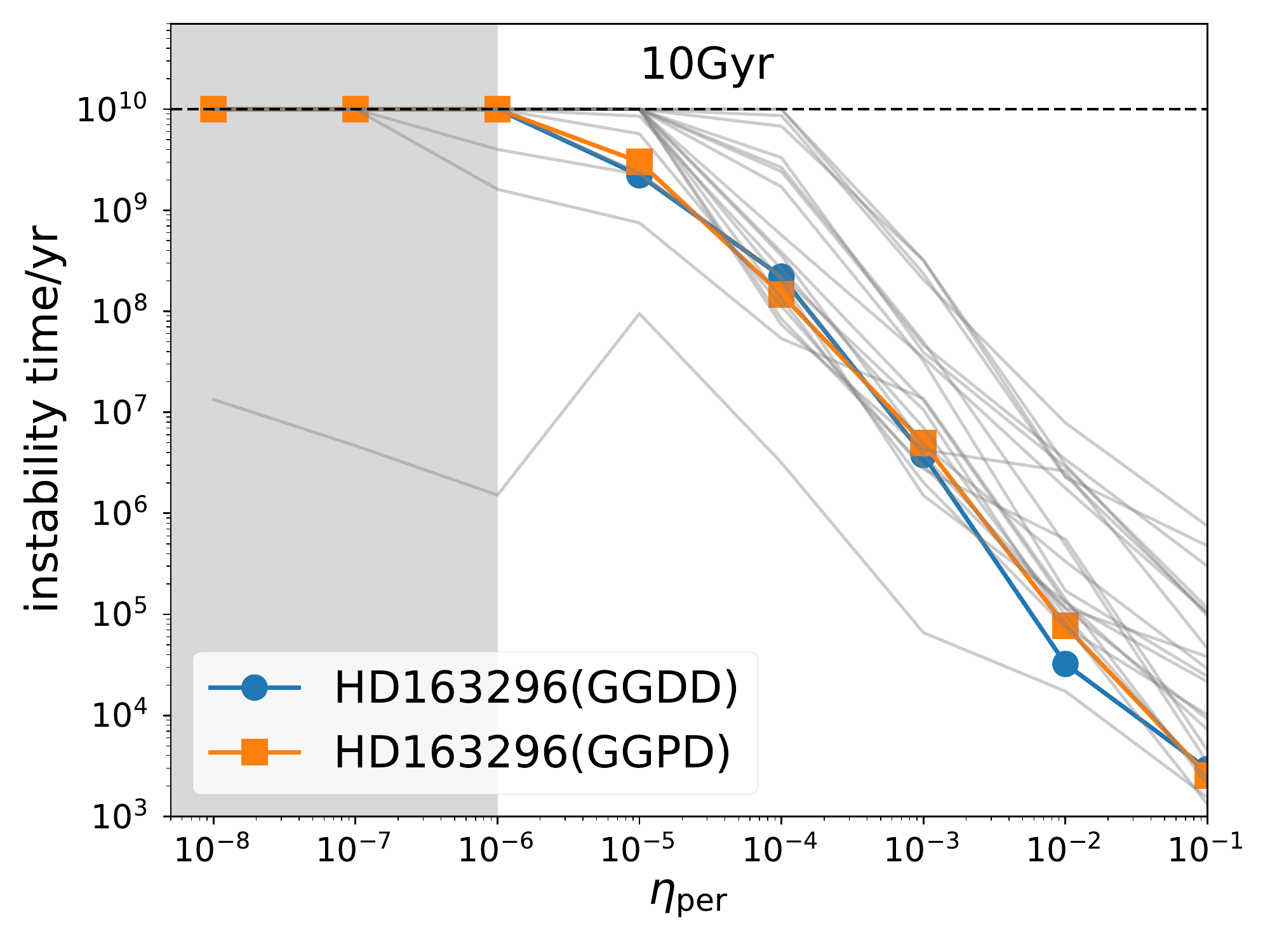}{0.33\linewidth}{(e) HD 163296}
		\fig{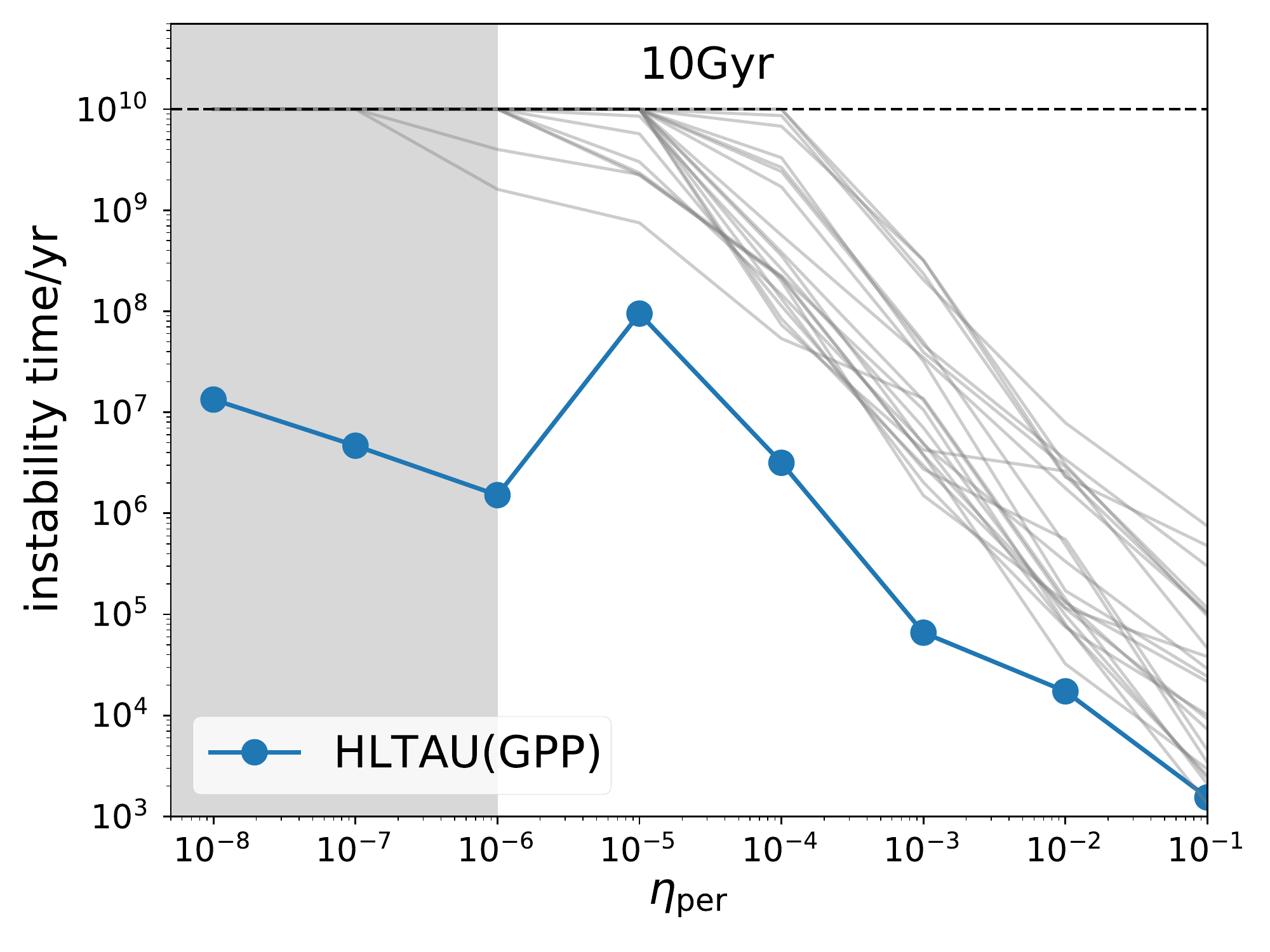}{0.33\linewidth}{(f) HL Tau}}
\caption{Instability time against the dimensionless magnitude of the
    stochastic perturbative force $ \eta_{\rm per}$. Grey lines
    correspond to the 23 different realizations from the twelve systems
    (Table \ref{tab:fidu_mass_table} for $\alpha=10^{-3}$).  The
    shaded region ($\eta_{\rm per}<10^{-6}$) is the realistic range
    for the planetesimals. Six different panels highlight the behaviour
    of the selected systems, together with different initial mass
    assignments of the planets.}
	\label{fig:instfratio}
\end{figure}

Figure \ref{fig:instfratio} plots the instability time against $ \eta_{\rm per} $ for all the 23 realisations from the 12 disk systems (Table \ref{tab:fidu_mass_table}) of fiducial parameters in thin grey lines.  Different panels depict six disk systems, respectively, so as to highlight their behaviour.  The amplitude of the instability time is dependent on the specific architecture of the systems and varies within two orders of magnitude, but is roughly proportional to $\eta_{\rm per}^{1.5}$ for $\eta_{\rm per}>10^{-6}$, except HL Tau. Note that we stop the run at $t=\SI{10}{Gyr}$, and most of the systems survive for at least \SI{10}{Gyr} for the realistic range of the perturbative forces, {\it i.e.,} $\eta_{\rm per} \leq \num{e-6}$.

As expected, systems become destabilised in general as the number of planets increases. For the same system, different initial mass assignments for planets do not significantly change the instability time that varies just within one order of magnitude. The result also shows that the minimum separation between planets plays a decisive role in the instability: sets GPP and GGP of CI Tau exhibit stronger resistance to the perturbation than their counterparts, as the minimum planet-planet separation in these two sets are larger than those of rest. As another evidence, both GY 91 and AS 209 have three planets, yet GY 91 is more stable than AS 209 because the closest pair in AS 209 (Planet 2 and 3) has smaller separation, and the planets are also more massive.

\begin{figure*}
	\gridline{\fig{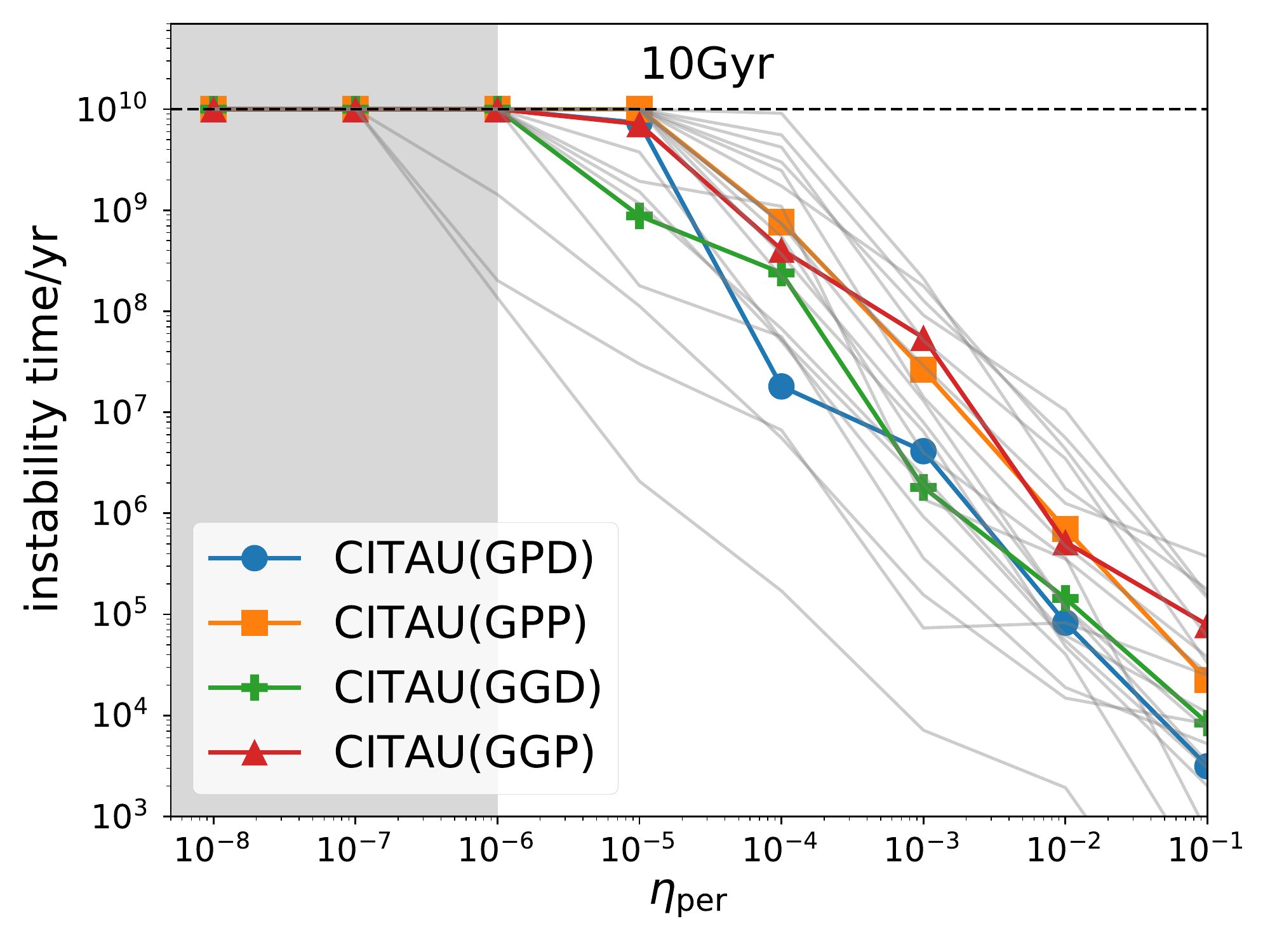}{0.33\linewidth}{(a)}
			  \fig{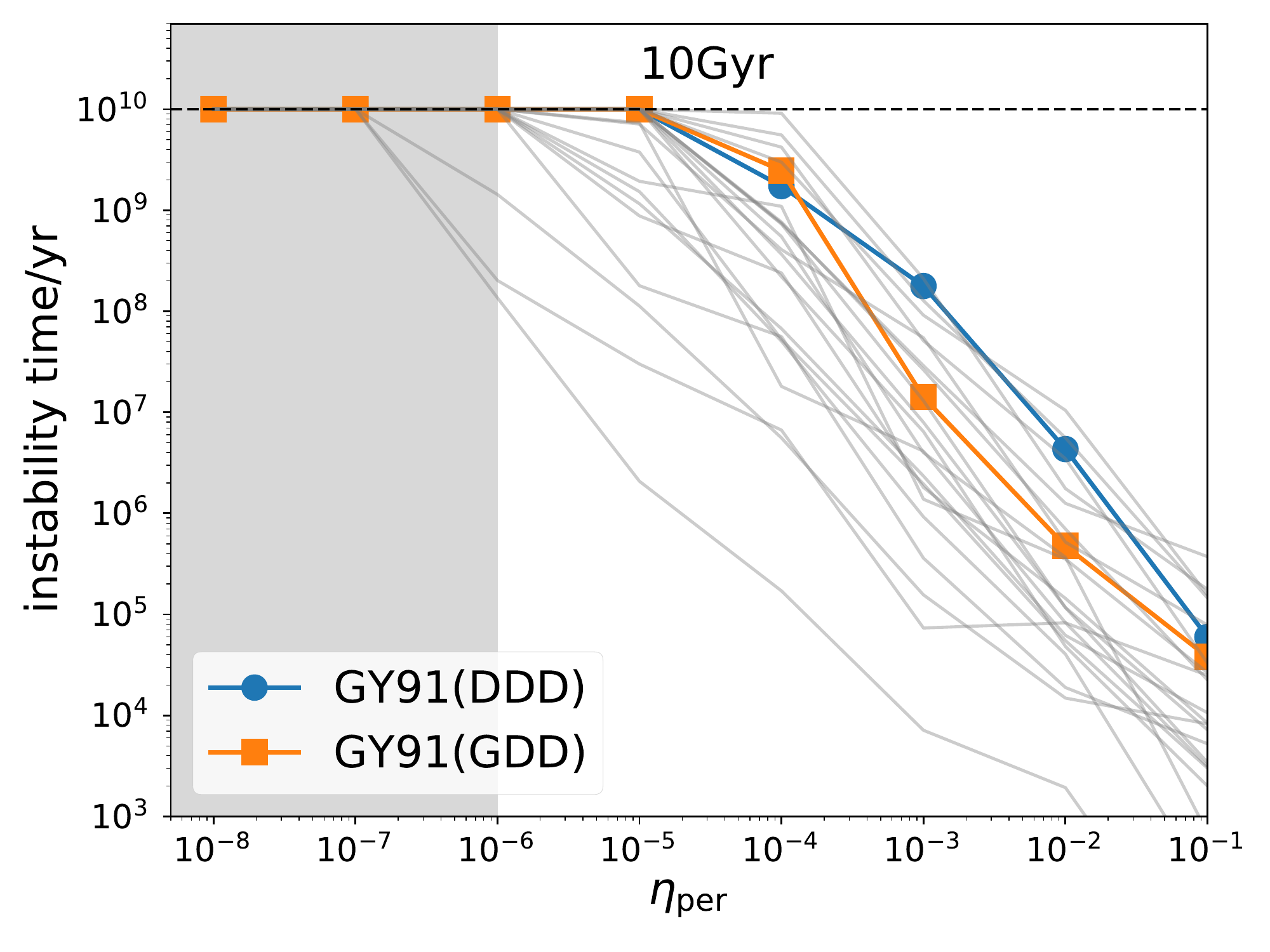}{0.33\linewidth}{(b)}
			  \fig{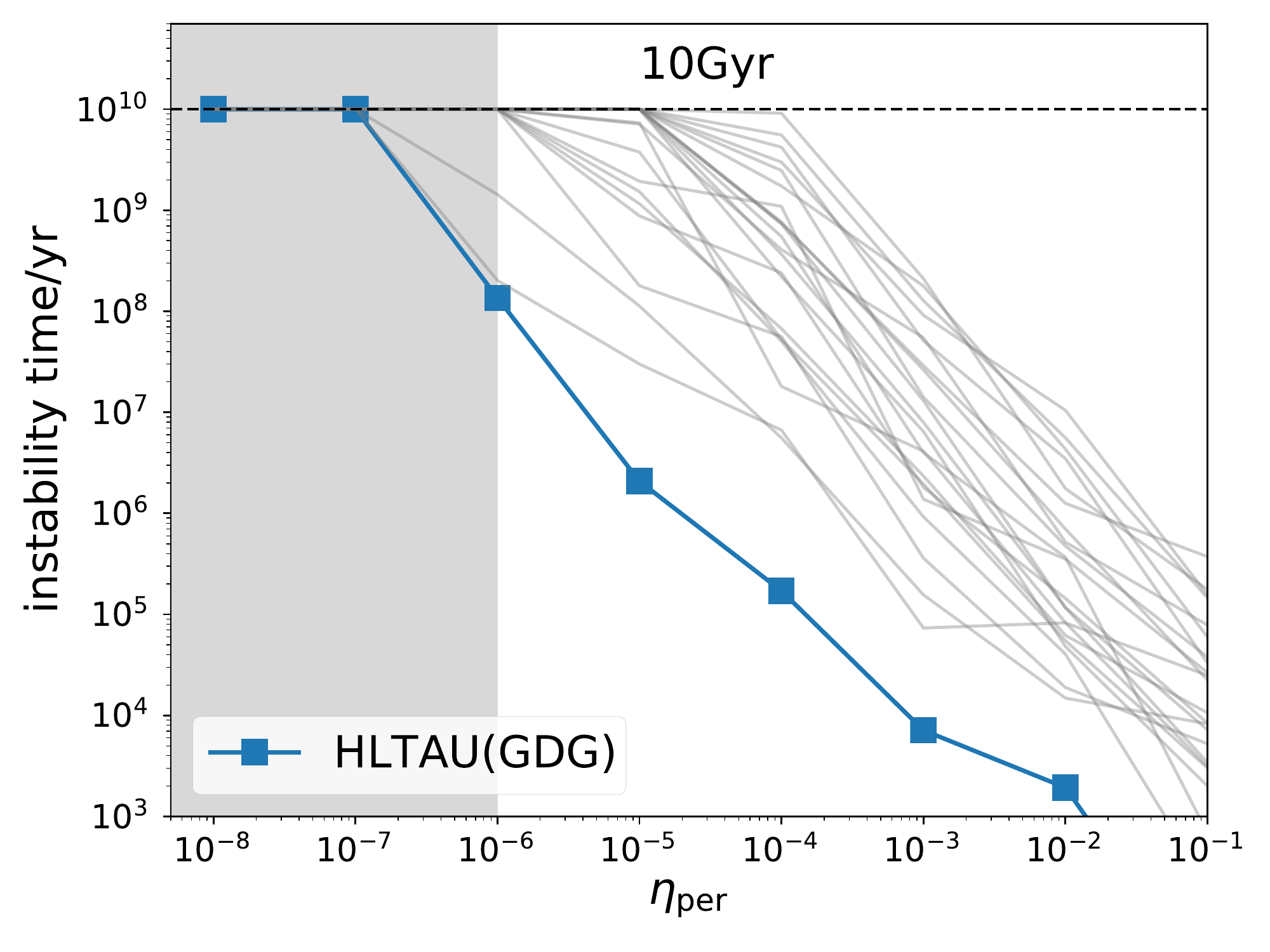}{0.33\linewidth}{(c)}}
	\gridline{\fig{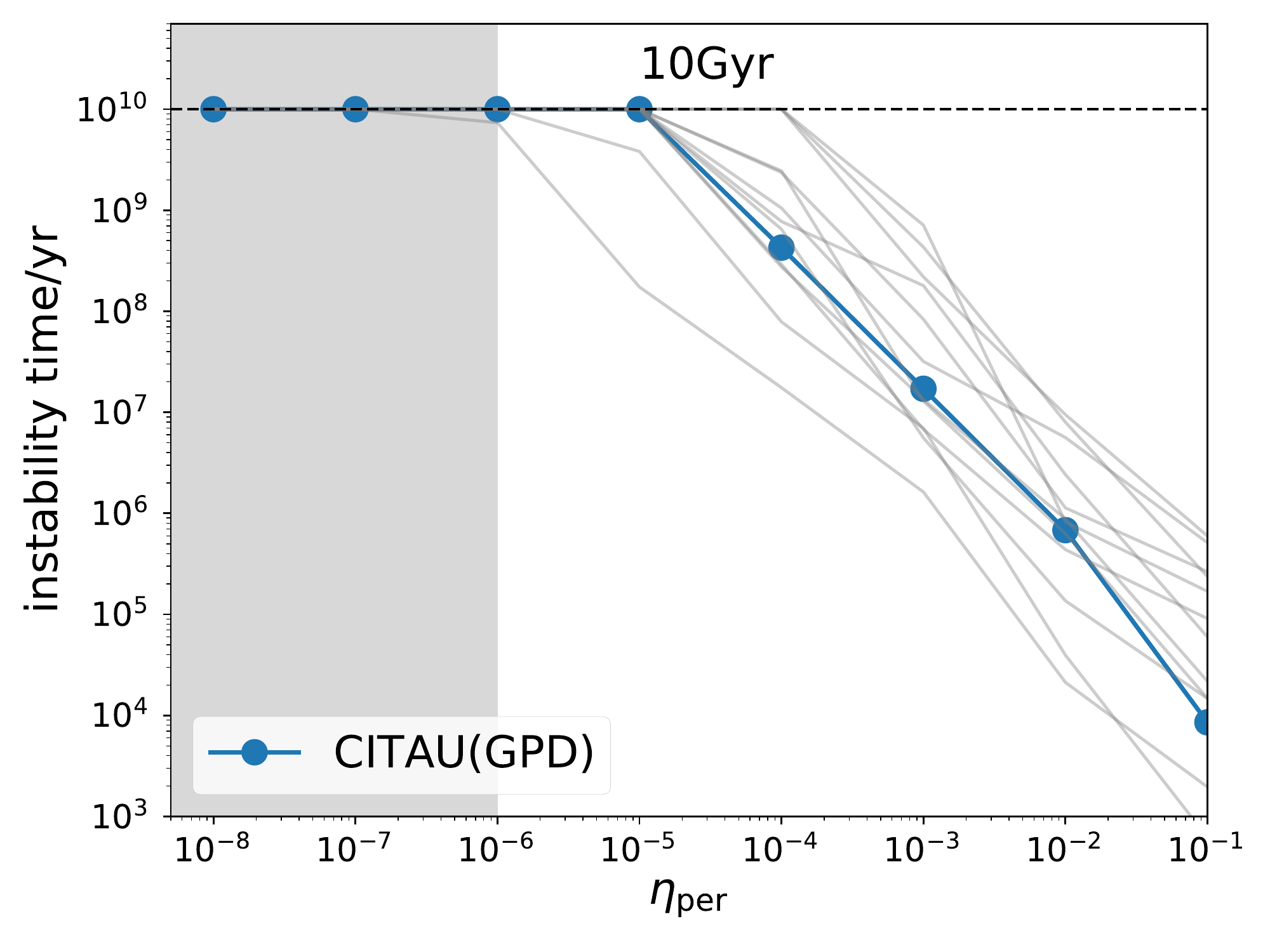}{0.33\linewidth}{(d)}
		\fig{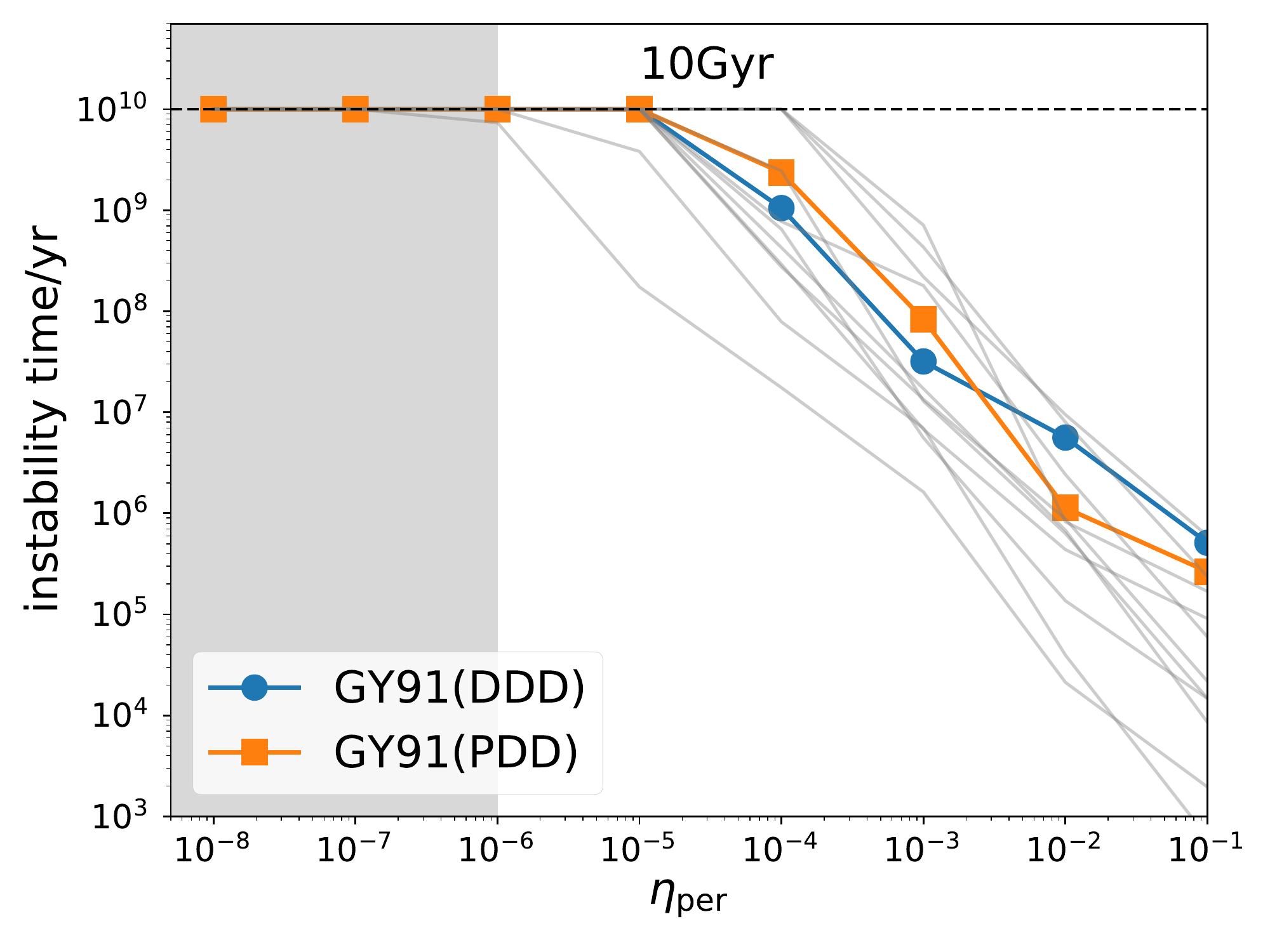}{0.33\linewidth}{(e)}
		\fig{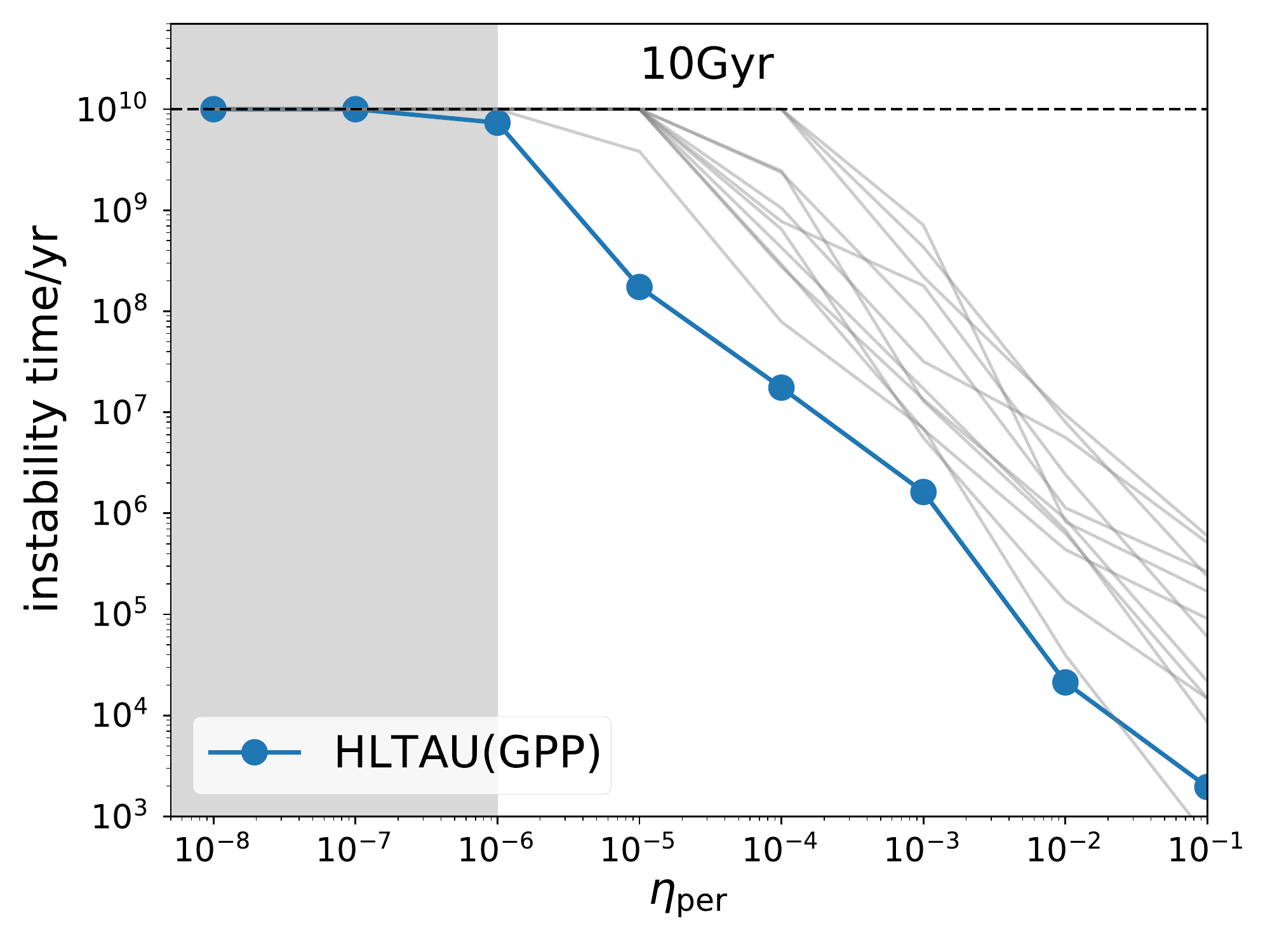}{0.33\linewidth}{(f)}}
\caption{Same as Figure \ref{fig:instfratio} but with different $
  \alpha $. (a)(b)(c): Instability time of CI Tau, GY 91 and HL Tau
  systems evolved with $ \alpha = \num{2e-3} $. (d)(e)(f): Instability
  time of CI Tau, GY 91 and HL Tau systems evolved with $ \alpha =
  \num{5e-4} $.}
	\label{fig:instfratio_LSalp}
\end{figure*}

We also examine the instability time versus the perturbative strength $ \eta_{\rm per} $ for planetary systems evolved with large $ \alpha = \num{2e-3} $ and small $ \alpha = \num{5e-4} $, as plotted in Figure \ref{fig:instfratio_LSalp}. In Figure \ref{fig:instfratio_LSalp}, we select three systems, CI Tau, GY 91 and HL Tau, as examples to show the $ \alpha $-dependence of the instability curve. In section \ref{sec:alpha_dependence} we have seen that $ \alpha = \num{2e-3} $ generally boosts both the migration and accretion rates during the disk stage, therefore the planetary systems are closer to the star and become more massive. In the case of HL Tau, its stability curve moves down as $ \alpha $ increases, showing greater instability. However, such a trend is not obvious for the rest of the systems. For GY 91, the outer two planets are dust planets, therefore their evolution are not sensitive to change of $ \alpha $. For CI Tau, although the planetary masses increases with $ \alpha $,
the separation between the planets do not change much due to the convergent migration, therefore the stability of the CI Tau system is not strongly affected within \SI{10}{Gyr} range. At $ \alpha = \num{5e-4} $, the majority of the systems stay stable for \SI{10}{Gyr} when $ \eta_{\rm per} \leq \num{e-5} $. Even the least stable HL systems cut-off at $ \eta_{\rm per} = \num{e-7} $.

Clearly, the HL Tau system with $ \alpha = \num{e-3} $ seems to be much less stable than the other systems: it becomes unstable at around \SI{10}{Myr} after the disk dispersal regardless of the $ \eta_{\rm per} $ down to \num{e-8}, indicating that its final configuration is sensitive to even minor perturbations. Compared with the conclusion of \citetalias{Wang2020} that the HL Tau systems are stable in general after the disk stage, this different result is due to the $\alpha$ adopted: \citetalias{Wang2020} adopted $2\times 10^{-4} \leq \alpha \leq 6\times 10^{-4}$, which are much smaller than the fiducial value $\alpha=10^{-3}$ in the present paper. As we will later discussed in section \ref{sec:discussion_consistency}, the special stability behaviour of the HL Tau system ($ \alpha = \num{e-3} $) is because a scattering event happening in the last \si{Myr} drives all three planets out of the resonance, leaving the system in a marginally stable states. Since no similar behaviour has been observed in other systems, this marginally stable HL Tau system can be regarded as only a serendipitous case. Nevertheless, this may be a good example to show how the stability can be affected by the resonant states.

Moreover, HL Tau systems are less stable because its initial disk mass ($ \SI{0.105}{M_\odot} $) is the largest among all disks, and therefore the resulting planets in HL Tau are the most massive and closest to the star in all systems. Theoretically, the semi-major axis $ a_p $ of a single, isolate planet under such stochastic perturbative forces is expected to undergo random walk with the root mean square deviation $ \sqrt{\left\langle \Delta a_p^2 \right\rangle} \propto \sqrt{Dt} $, where $ t $ is the simulation time and $ D \propto \eta_{\rm per}^2M_{*}^2R_p^{-4} $ is the diffusion constant \citep[equation 46, ][]{Rein2009}. Therefore, under the perturbative force with the same $ \eta_{\rm per} $, the inner planets deviate faster from their original semi-major axis and have shorter instability time, as the diffusion constant $ D $ increases with decreasing $ R $.

Our analysis shows that given the fiducial parameters that we adopt, the perturbation from the planetesimals can only marginally affect the most massive and unstable systems. In order for the majority of the systems to be unstable, either the mass of the planetesimals have to be much larger, or additional source of perturbation (e.g., close stellar flyby) is required to significantly destabilise the configuration reached at the end of the disk dispersal. We discuss the implications in section \ref{subsec:implications}.

\subsection{Consistency with previous stability criteria }
\label{sec:discussion_consistency}

\begin{figure*}
\centering
\includegraphics[width =0.5\linewidth]{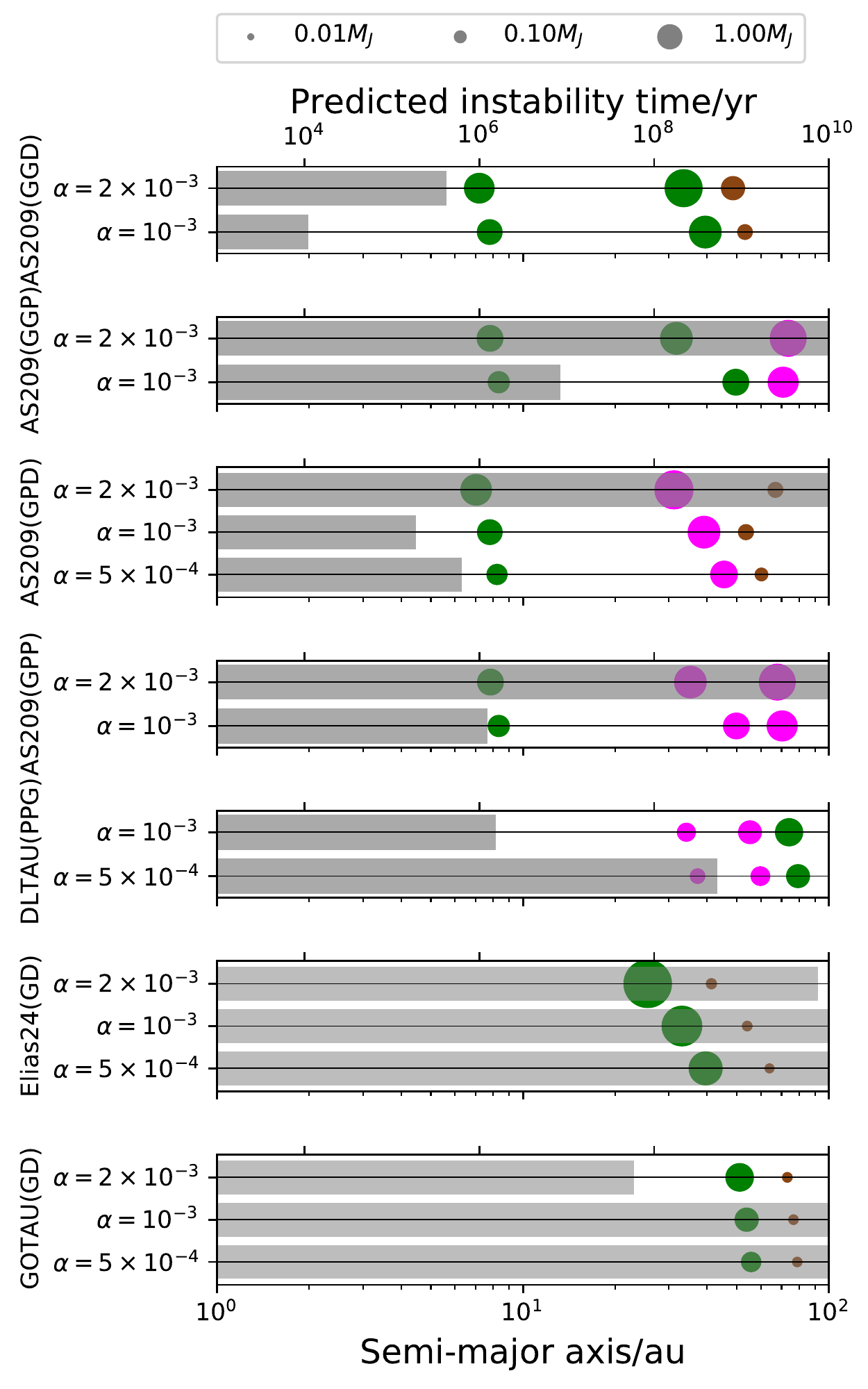}
\caption{Instability time predicted by \cite{Morrison2016} for
  each system based on the configuration at the end of the disk stage
  with different $ \alpha $ values.}
\label{fig:comp_instime}
\end{figure*}

We have shown that most of the systems are stable for at least \SI{10}{Gyr} even under the presence of stochastic perturbative forces due to planetesimals. It is interesting to check the consistency of these results with the theoretical prediction. We compute the instability time from the configuration at the end of the disk stage using the same formula \cite{Morrison2016} that we adopted in our previous paper \citepalias[section 5.1,][]{Wang2020}. It is worth to mention that the planetary configurations out of our simulation have a wide distribution over mass and semi-major axis; they are different from the theoretical initial conditions considered in \cite{Morrison2016}, who concern the general stability criterion of equal-mass multi-planetary systems. Therefore, applying the formula of \cite{Morrison2016} to our systems should be interpreted as an extrapolation of their original result, and deviations are expected if the system configuration is largely different from those valid in the original context. We would also like to stress here that the instability time predicted by \cite{Morrison2016} neglects the possible effect of the perturbative force.

Having said so, however, we found that most of the systems are consistent with the prediction of \cite{Morrison2016} at the level of \SI{10}{Gyr}. Thus we focus on all the configurations that are predicted to be unstable according to the formula of \cite{Morrison2016}.  Figure \ref{fig:comp_instime} indicates the instability time of those systems (grey bars in the upper axis) together with their planetary configurations (filled circles in the same manner of the right panel of Figure \ref{fig:comp_config_fidu}). For each set, we also plot the same set initialised and evolved with
different $ \alpha $ at the disk stage whenever exist. In the same figure, the configurations at the end of the disk systems evolved with the respective $ \alpha $ are also plotted for reference. Note that we exclude the five cases that have already become unstable at the disk phase for $\alpha = \num{2e-3}$ case.

Figure \ref{fig:comp_instime} show that all the realisations that are predicted to be unstable by \cite{Morrison2016} have an outer planet pair with a period ratio close to or less than \num{2.0}. Also in most of the cases, the third planet is relatively far away from the outer pair. For example, the outer pair in AS 209(GPD) and AS 209(GGD) with $ \alpha = \num{e-3} $ consists of a super-Jupiter and a sub-Saturn in the outer region, while the innermost planet is around \SI{10}{au}. Since the instability time is computed from the minimum orbital separation of all planet pairs within planetary system, such configurations are predicted to be unstable for less than \SI{1}{Myr}. The predicted instability time for GO Tau(GD), Elias 24(GD), DL Tau(PPG) decreases with increasing $ \alpha $, because a large $ \alpha $ value promotes both migration and accretion and results in a more compact systems. However, for AS 209 systems evolved with $ \alpha = \num{2e-3} $, the scattering
between Planet 2 and 3 increases the separation between them, therefore the final configurations at $ \alpha = \num{2e-3} $ is more stable than those at $ \alpha =\num{e-3} $.

However, this configuration is largely different from the theoretical setup investigated by \cite{Morrison2016} in at least two aspects. Firstly, the planetary masses in the pair are not equal in our simulation; for AS 209(GPD), the mass difference is more than $ 10 $ times. The unequal mass affects the metric that is used to normalise the orbital separation and therefore the predicted instability time. Secondly, while \cite{Morrison2016} consider multi-planets with equal normalised orbital separation, in our simulation the separations of planet pairs may differ by a large extent. As a result, the closest planet pair experiences much weaker influence from the third planet as compared with that investigated by \cite{Morrison2016}. Therefore, these configurations are not in the applicable range of the prediction formula.

Besides the difference in configuration, we think the MMR may be another reason to cause the underestimate of the instability time for some systems such as DL Tau(PPG). As we show in Figure \ref{fig:res_angle}, all two planet pairs of DL Tau are in either 2:1 or 3:2 MMR.  Since the planet pairs are locked in a chain of resonance, their configurations are more resistant to the small perturbations and thus can be stabilised over long time. This argument is also consistent with our previous findings in \citetalias{Wang2020} that the instability time prediction based on the mutual separation can underestimate the stability when the planets are in a chain of resonance.

\begin{figure}[t]
	\gridline{\fig{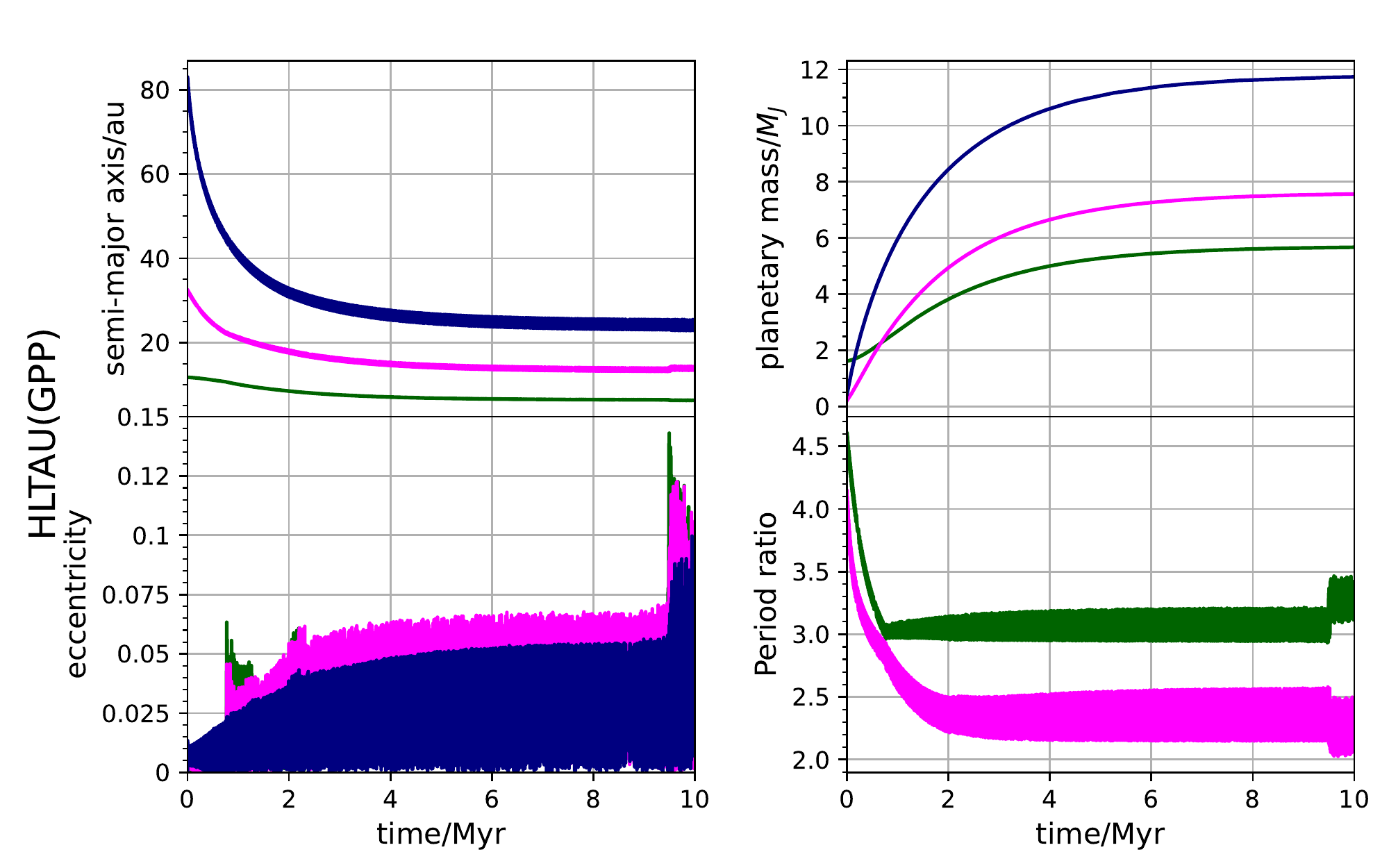}{0.75\linewidth}{(a)}}
	\gridline{\fig{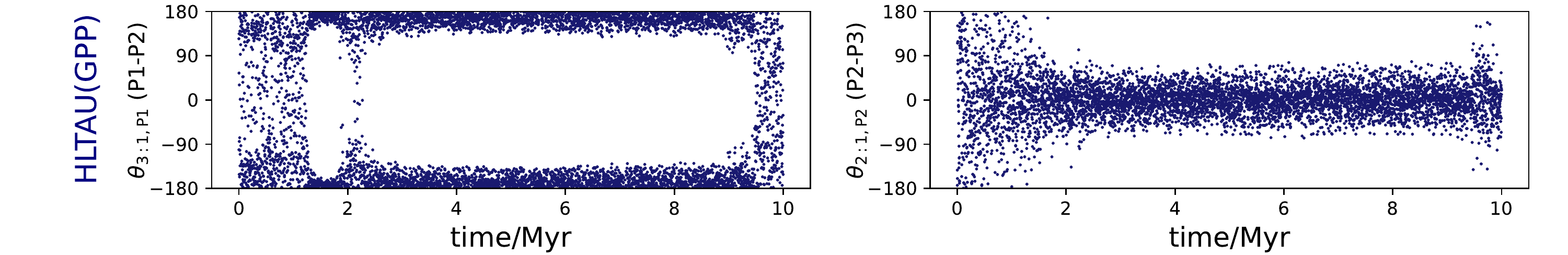}{1.0\linewidth}{(b)}}
\caption{(a) Similar to Figure \ref{fig:evo_fidu} but the evolution of
  the HL Tau system at $ \alpha =\num{e-3} $. (b) 3:1 resonant angle
  of the inner pair and 2:1 resonant angle of the outer pair. Both
  angles start to circulate within the last \SI{1}{Myr} due to a
  scattering event that excites the eccentricity.}
	\label{fig:evo_HLTAU}
\end{figure}

We noticed that the fiducial set of HL Tau is the only case whose stability is overestimated. Considering three out of four sets in the $\alpha = \num{2e-3}$ cases do not even survive the disk stage, this special set is in marginal stability that is vulnerable to even minor perturbations. Figure \ref{fig:evo_HLTAU} plots the evolution of HL Tau at the disk stage. At around \SI{9.5}{Myr}, there is a scattering event that changes both the semi-major axis and boosts the eccentricity to the level as high as \num{0.1}. Although the change of the semi-major axis is not significant, this scattering significantly changes the resonant states, as shown by the sudden change of the period ratios as well as the resonant angles. Before \SI{9.5}{Myr}, both 3:1 resonant angle of the inner pair and 2:1 resonant angle of the outer pair librate; however, after the scattering takes place, both resonant angles start to circulate. Since all three planets are massive, their resonant zones may overlap and once the planet is out of the resonant lock, the system becomes chaotic and unstable.

\section{Discussion\label{sec:discussion}}

\subsection{Comparison with the previous result concerning the HL Tau system}

Our previous study \citetalias{Wang2020} investigates the evolution of HL Tau system at the disk stage, assuming that all the three planets have opened gas gaps. Although the fiducial disk parameters and adopted initial masses are different, in both studies the HL Tau system evolve to become a widely-separated planetary system consisting of three super-Jupiters at the end of the disk stage.  However, due to the relatively larger viscosity adopted here ($\alpha=10^{-3}$ in contrast to $2\times 10^{-4} \leq \alpha \leq 6\times 10^{-4}$ in \citetalias{Wang2020}), the final masses of the HL Tau planets are a few times larger. The resulting HL Tau system at $ \alpha = \num{e-3} $ is only marginally stable. If we adopt $ \alpha = \num{2e-3}$, three out of four initial mass assignment sets become unstable during the disk stage. We note that only one out of 75 sets in \citetalias{Wang2020} becomes unstable during the disk stage.

For the rest of the ALMA disks, overall the final architecture exhibits greater diversity in terms of semi-major axis and masses, partially due to their diverse initial conditions and different gap interpretations. The semi-major axis of the planets ranges from around \SI{10}{au} to over \SI{100}{au}, and some planet pairs are in 3:2, 5:3 and 7:4 MMR with period ratios less than \num{2.0}, which are absent in previous HL Tau study. While the final planetary masses of some systems (e.g., HD 163296, CI Tau, AS 209) are similar to HL Tau, some planets are strongly quenched in mass growth due to both inefficient pebble accretion at the outer disk and failure to trigger the run-away gas accretion. These planets remain as Neptune-sized planets and co-migrate with the inner planets until the end of the disk dispersal.

Despite the diversities of the architecture, the dynamical structure of the systems are consistent with what we found in \citetalias{Wang2020}. Most systems naturally enter good MMR states because of the convergent migration, and under the resonance lock they can stay stable for at least \SI{10}{Gyr} with even minor perturbations. Our conclusion in \citetalias{Wang2020} still holds for other ALMA disks that the final configuration of the planetary systems is significantly stabilised by the planet-disk interaction. Once the systems survive the disk stage without being gravitationally excited, it is hard to induce instabilities afterwards unless significant level of perturbation ($\eta_{\rm per} > \num{e-6}$) is present.

\subsection{Implications for future observation and possible
  scenarios of close-in planet formation}
\label{subsec:implications}

The evolution outcomes of the planetary systems that we obtained may have several implications on the future observation. First of all, the similarities between the planetary systems discovered via direct imaging and our synthesised giant planetary systems may imply potential connections between the two. The similarity is not only limited to the mass and semi-major axis, but also the dynamical structure: \cite{Gozdziewski2014,Gozdziewski2018,Gozdziewski2020} found the planet pairs in HR 8799 are in good 2:1 resonant states, and such orbital resonance can stabilise the system for \si{Gyr} timescale. As a natural consequence of the convergent migration, the 2:1 resonance appears in many of our planet pairs, without any fine tuned planet or disk parameters. Furthermore, in our past study \citetalias{Wang2020}, we have found the HL Tau disk has the potential to produce such systems. In this study, we found not only the HL Tau system, but also other ALMA disks such as HD 163296 and DL Tau,
can evolve to systems alike. The relative abundant sample of the disks implies that multiple giant planet systems in wide separation like HR 8799 may not be rare, and they may be detected as the observation technique advances in the future such as the James Webb Space Telescope (\textit{JWST}).

We have also found that there is not much overlap between the short-period exoplanets observed by Kepler and TESS and the planets we produced from the observed PPDs, even after varying different parameters. Although some planets have demonstrated strong inward migration, their final locations are mostly around \SI{10}{au}, partially because their initial locations are too far away, and the migration is strongly quenched once the planet enters run-away gas accretion stage. In order to produce close-in planetary systems, planetary cores have to be formed closer to the star at the first place. Moreover, planets initially much below the pebble isolation mass can barely migrate and accrete, and so far no such exoplanets have been observed. The above mismatch between the synthesised planetary population and the observed population can be a motivation for both the disk and exoplanet observation: the lack of discovered planetary signatures at the inner region of the disk encourage future observation such as the Next-Generation Very Large Array (\textit{ngVLA}) to explore the inner disk with higher resolving power, and the abundance of planets far away from the star implies there are potentially more exoplanets to be observed, even in exoplanetary systems that we have discovered.

Finally, our results show that those planetary systems predicted from the disks are likely to be stable. The stable configuration due to the convergent migration may indicate that the production of Hot Jupiters or planets with large spin-orbit misalignment via instability is not efficient. However, we see that given large viscosity at the disk stage and/or strong enough perturbation $ \eta_{\rm per} > \num{e-5} $ after the disk dispersal, the system can be disturbed and become unstable in short time scale, which may eventually lead to the formation of Hot Jupiters via tidal circularisation of eccentric orbit caused by planet-planet scattering \citep[e.g.,][]{Rasio1996,Nagasawa2008}. Due to the relatively weak perturbations from the planetesimals, an external source of strong perturbation may be critical to generate short-period giant planets. Such perturbative source might be realised by stellar flyby \citep[e.g.,][]{Rodet2021}, as recent observations have revealed that stars in a cluster are likely to have a short period planets including Hot Jupiters, as compared to field stars \citep{Winter2020}. Other mechanisms, such as interaction with free floating planets \citep{Varvoglis2012,Goulinski2018}, can also result in loosely bounded planetary orbits at high eccentricity. These possibilities motivate future observation to better constrain the viscosity in a PPD, possible sources as well as the realistic ranges of the possible perturbations.

\subsection{Caveats}
\label{sec:caveats}
There are several caveats involved in this research. First, our gas
gap depth, migration and accretion models are originally derived for a
single planet. While implementing these effects onto multiple planets,
we assume each planet can still be treated independently, even they
are coupled by the gas surface density profile. This assumption does
not always hold as when the planets are both massive and close to each
other, their gas gaps may overlap and combine to become a common gap
\citep[e.g.,][]{Pierens2008,Podlewska2009,Duffell2015a,Cimerman2018},
which will revise the migration and accretion model. It is largely
unknown that how the common gap will affect the migration and
accretion processes, and careful consideration of this effect is
necessary for future studies concerning evolution of compact and
massive planetary systems.

Second, our evolution model assumes the planets are co-planar and in
nearly circular orbits. It is a good approximation in the disk stage
as the disk can effectively damp the eccentricity as well as the
inclination of the planets. However, in some extreme cases, two
planets may be very close to each other due to strong migration, and
the planet-planet scattering event can happen afterwards, which
excites the inclination or eccentricity to high level. Our current
numerical model cannot accommodate such a scattering event and thus we
cannot predict the outcome once instability occurs at the disk
stage. However, although most our systems are stable in the disk
stage, planet-planet scattering events between planets occur at $
\alpha = \num{2e-3} $, and the eccentricity is excited. The
eccentric/inclined orbits induced by the disk-stage instability are
prerequisites for tidal circularisation to take effects and eventually
produce close-in Hot Jupiters \citep[e.g.,][]{Nagasawa2008}. They are
also important in producing planets with large spin-orbit misalignment
at later stage via Lidov-Kozai oscillation \citep[e.g.,][]{Pu2021}. In
addition, large initial eccentricity is also required for Tidal
circularization to take effects. It is therefore necessary for future
work to incorporate the planet-disk interaction model at high
eccentricities to further investigate the outcome of such unstable
cases.

Third, in our model, we adopt a one-to-one correspondence for the
planet and the gap, and do not include any planets unseen to
ALMA. There are may be undetected gaps in the the poorly resolved
inner disk \citep[e.g.,][]{Jennings2021}, and planets from which may
change the dynamical structure and the evolution of the outer
planets. We also do not consider the possibility of the new planet
formation that may happen at the dust-rich region, e.g., the dust ring
outside of the gas gap \citep[e.g.,][]{Carrera2021}. Due to the
suppression of growth from the existing giant planets, the new-born
planets are unlikely to significantly change the configuration of the
system. However, they may contribute more to the rocky planetary
population at the inner disk, which are missing in our simulation
while comparing with the observed population.  Moreover, when the
turbulent viscosity is very low, multiple gaps can be formed by a
single planet \citep{Dong2017,Dong2018,Bae2017}. In this case, our
assumption that the planet and gap follow one-to-one correspondence
overestimates the number of the planets.  Alternatively, we may also
overestimate the number of the planets in the disks with the low
viscosity, where an outer ring can be left behind of a migrating
planet \citep{Kanagawa2021a}.

Fourth, we should note uncertainty of the interval between the onset
of the runaway gas accretion and the pebble isolation. For simplicity,
we assumed that the runaway gas accretion occurs quickly after the
pebble-isolation of the planet, as compared with the migration
timescale of the planet.  However, the interval between the two
accretion regimes may increase with the dust opacity of the planetary
atmosphere \citep{HUBICKYJ2005,Lambrechts2019}. If this interval is
comparable with the timescale of planetary migration, the planet would
experience more inward migration than that in our simulation. For the
substantial number of planets that are initialised with $ M_{\rm iso}
$, they can migrate further inward by staying longer at around $
M_{\rm iso} $ if the run-away gas accretion is delayed. Such an effect
should be investigated and carefully treated in future works.

Finally, we should discuss the effect of the uncertainly of the disk mass.
Since we inferred the disk mass from the millimetre luminosity by assuming the disk is optically thin (see equation~\ref{eqn:disk_mass_from_flux}), we may underestimate the mass of the gas if the disk is optically thick, especially in the opaque centre region of the disk \citep{Zhu2019,Liu2019}. We also assume that the gas-to-dust ratio is 100, and the uncertainty of this ratio can also affect the estimated disk mass. Having said so, a very massive disk is not stable due to the gravitational instability, and so far there is no obvious evidence of such an instability in the observed disks. Therefore, considering the limit of the gravitational instability, the disk mass may has an upper limit up to $\sim 0.1 M_{\odot}$, which is about two times of our current estimate in the most cases (see Table~\ref{tab:fidu_mass_table}). At this upper limit, planets may experience stronger migration and accretion and eventually become more inward and massive, similar to the case of HL Tau.

We should also note that the stellar accretion rates given by our model with $ \alpha = 10^{-3} $ are smaller than those estimated from the observed accretion luminosity in several disks (e.g. HD~169243), even before considering the reduction of the stellar accretion rate due to the planetary accretion \citep[see Table 1 of ][]{Andrews2018}. A higher viscosity $\alpha$ may explain the observed accretion rate, but it is not observationally favoured as the recent study \citep[e.g.,][]{Flaherty2015} suggests that the disk viscosity may be rather low . This inconsistency is mainly due to our simplified viscous model that assumes a constant $ \alpha $ viscosity, and more sophisticated viscous accretion \citep{Delage2022} or disk-wind accretion \citep{Bai2016,Suzuki2016,Ida2018} models are required to address this inconsistency.

\section{Summary and Conclusion\label{sec:summary}}

Despite numerous important and interesting progresses in theoretical and observational understanding of exoplanetary systems over the last couple of decades, there is no accepted formation scenario that explains their universality and diversities simultaneously. One of the main difficulties that hinder establishing the successful scenario is the lack of the realistic initial conditions at the protoplanetary disk phases. Most of past theoretical studies have tried to construct exoplanetary systems simply by adopting somewhat artificial initial conditions. Given a lot of uncertainties in the physical processes in the formation and evolution of protoplanetary disks, the limitation due to the approach is inevitably large.

This difficulty is expected to be less serious thanks to recent discoveries of ring and gap substructures in many dust disks by ALMA. Following a conventional planetary interpretation for the substructures, one can bypass the uncertainties of the initial conditions, perform numerical simulations, and confront the resulting architecture of the planetary systems.

In this paper, we perform two-stage N-body simulations, and investigate the outcomes of 12 ALMA disks, the initial conditions of which are based on the orbital and mass prediction in  \citetalias{Wang2021} (section \ref{sec:ini_planetary_mass}). During the disk stage, we include the planetary migration (section \ref{sec:method_mig}) as well as pebble/gas accretion (section \ref{sec:method_acc}) schemes to mimic the interaction between the planet and the disk. After the disk dispersal, we examine the long term orbital stability of the configurations by integrating the systems up to \SI{10}{Gyr}.

\citetalias{Wang2020} applied the above strategy for the HL Tau, and concluded that the resulting three-planet systems are mostly stable up to \SI{10}{Gyr} from its current epoch. This paper generalized and improved the work by \citetalias{Wang2020} in four major aspects. First, we take into account different mass assignment possibilities for embedded planets following \citetalias{Wang2021}. Second, we consider the pebble accretion, in addition to the gas accretion, during the planet-disk interaction phase.  Third, we examine the stability of planetary systems by including the effect of perturbative gravity due to surrounding planetesimals after the disk dispersal, in addition to those from the central star and other planets. Finally, we apply the methodology to twelve ALMA disks with clear multi-gap structures.

Our main finding of the present paper is summarized as follows.
\begin{enumerate}
\item After the disk stage evolution, the planet population exhibits a wide distribution in both semi-major axis and mass (Figure \ref{fig:comp_config_fidu}) and starts to overlap with the observed planets. Based on their evolution tracks, the planets can be roughly categorised into four groups (Figure \ref{fig:comp_config_zoom}). (i) A large fraction of planets with initial mass greater than $ M_{\rm iso} $ experience efficient mass growth however inefficient migration. These planets become distant super Jupiters or larger brown dwarfs with semi-major axis greater than \SI{30}{au}, which resemble planets in HR 8799, $ \beta $ Pictoris and PDS 70 systems discovered via direct imaging. (ii) Planets with mass slightly below $ M_{\rm iso} $ undergo faster inward migration first followed by rapid run-away gas accretion. They eventually become Jupiter or Saturn size planets around \SI{20}{au}. (iii) Some sub-$ M_{\rm iso} $ planets failed to trigger run-away gas accretion and thus migrates inward without significant mass growth. A few planets in these group are close to the Neptune in terms of orbits and mass. (iv) Planets that are initially small ($ \sim \text{a few } \si{M_{\oplus}} $) and distant ($ \sim \SI{100}{au} $) suffer from both inefficient migration and accretion, and they remain largely unchanged.

\item We found both initial planetary mass assignments and $ \alpha $ viscosity are important in shaping the architecture of the planetary systems. Planetary systems initialised with the same orbits but different planetary masses can result in distinct final configurations with qualitative difference in semi-major axis and mass (section~\ref{sec:mass_assign_evo}, Figure \ref{fig:evo_fidu}). Increasing the $ \alpha $ parameter also results in more massive, compact and inner planetary systems due to enhanced migration and accretion (section \ref{sec:alpha_dependence}, Figure \ref{fig:comp_config_LSalp}). The planetary systems also tend to become less stable at the disk stage when $ \alpha $ increases.

\item At the end of the disk stage, the majority of the adjacent planet pairs have period ratios smaller than four, and a small fraction of the pairs have period ratio smaller than two (section \ref{sec:rslt_period_ratio}). Pairs with large period ratios has a small inner planet that undergoes fast migration first and then cause the pair to separate apart (Figure \ref{fig:period_hist}). The evolution of the resonant angles show that planets in compact planetary systems (e.g., DL Tau) are in a chain of MMR due to the convergent migration and higher order MMR (Figure \ref{fig:res_angle}).

\item Our stability analysis shows most of the systems can stay stable for at least \SI{10}{Gyr} when the perturbative strength $ \eta_{\rm per} < \num{e-6}$ (Figure \ref{fig:instfratio}). Majority of our results are consistent with the predicted instability time given by \cite{Morrison2016} (section~\ref{sec:discussion_consistency}). The stabilities of a few planetary systems, particularly AS~209 and DL Tau, have been underestimated largely because of the far-away third planet from the closest planet pair as well as the chain of MMR (Figure \ref{fig:comp_instime}).
\end{enumerate}

We also discuss the implication of our results on future observations. Our predictions suggest the potential existence of a large number of distant gas giants that resemble the observed giant planetary systems (e.g., HR 8799), which may be observed by the next generation telescope such as the upcoming \textit{JWST}. Our results also suggest that additional protoplanets may exist in the inner disk region in order to explain the formation of close-in gas giants via migration and accretion, whose signatures may be resolved by future ALMA observation or \textit{ngVLA} on the nearby PPDs.

\section*{Acknowledgement}

We thank an anonymous referee for numerous important comments.
The numerical simulations were carried out on \texttt{nekoya} cluster
offered by the Institute for Physics of Intelligence, University of
Tokyo and the general calculation node from Center for Computational
Astrophysics (CfCA), National Astronomical Observatory of Japan. We
compiled the \texttt{C} code using the standard \texttt{gcc} compiler,
and processed the results in \texttt{Python}
environment. S.W. acknowledges support from the Chinese Scholarship
Council (CSC). This work is supported partly by the Japan Society for
the Promotion of Science (JSPS) Core-to-Core Program “International
Network of Planetary Sciences”, and also by JSPS KAKENHI grant
Nos. JP18H01247 and JP19H01947 (Y.S.), JP17H01103 and JP19K14779
(K.D.K).

\appendix
\label{sec:appendix}

\section{Evolution of unstable systems at the disk stage}
\label{app:unst_exp}

\begin{figure}
	\begin{minipage}{.4\linewidth}\vspace{0pt}
		\includegraphics[width=\linewidth]{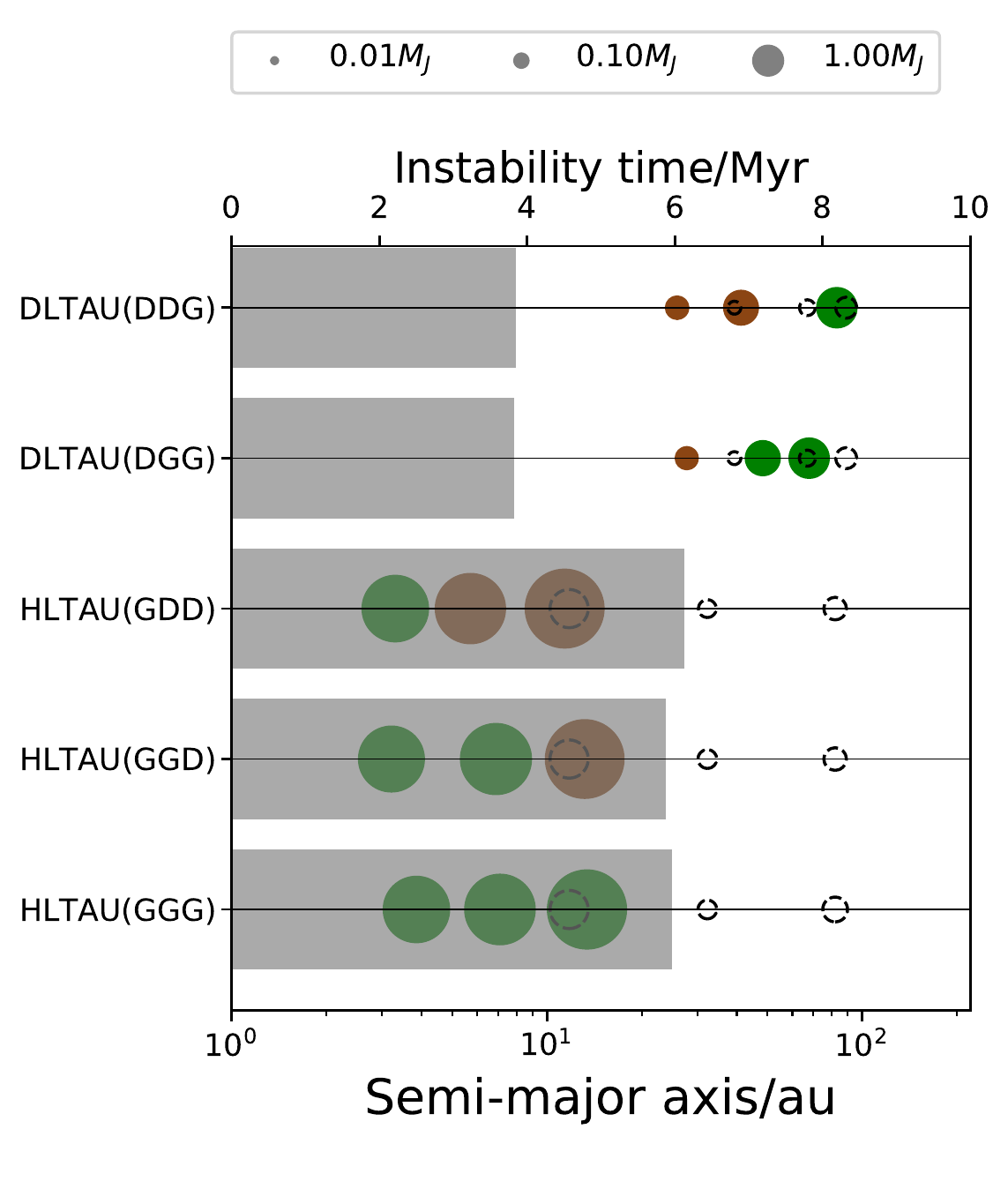}
	\end{minipage}
	\hfill
	\begin{minipage}{.6\linewidth}\vspace{0pt}\raggedright
		\begin{minipage}[t]{\linewidth}\vspace{0pt}\raggedright
			\includegraphics[width=\linewidth]{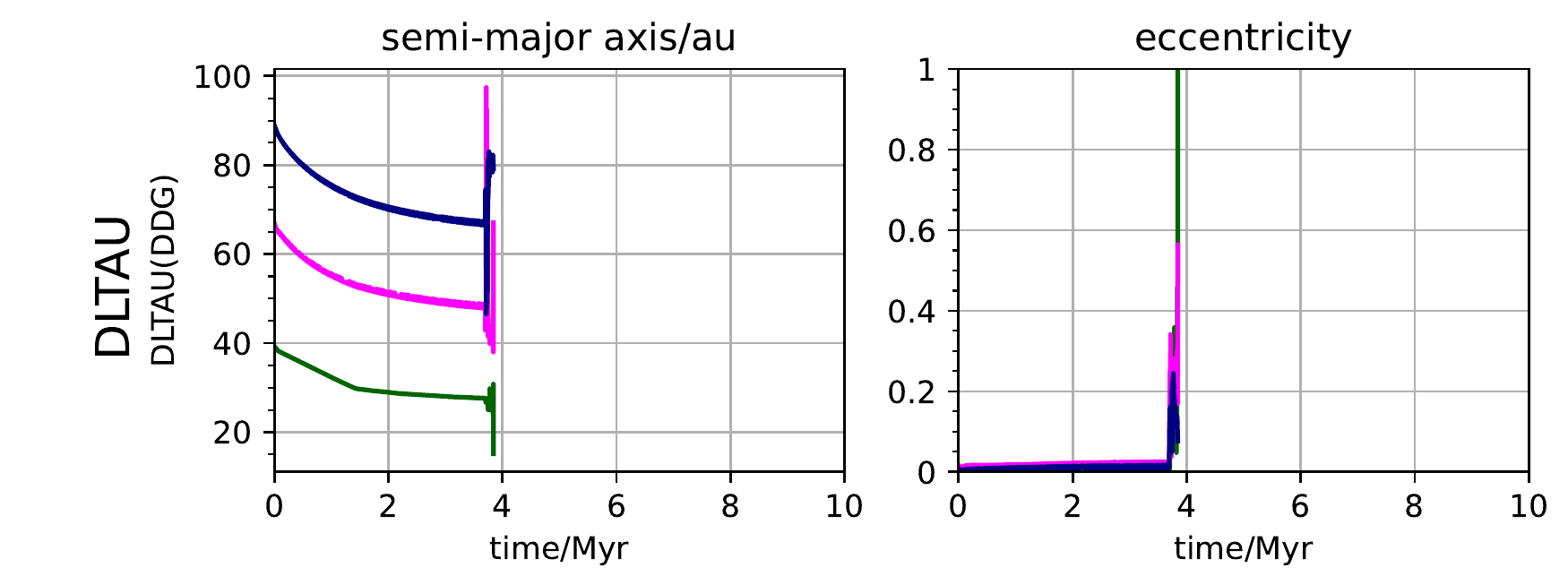}
		\end{minipage}
		\begin{minipage}[t]{\linewidth}\vspace{0pt}\raggedright
			\includegraphics[width=\linewidth]{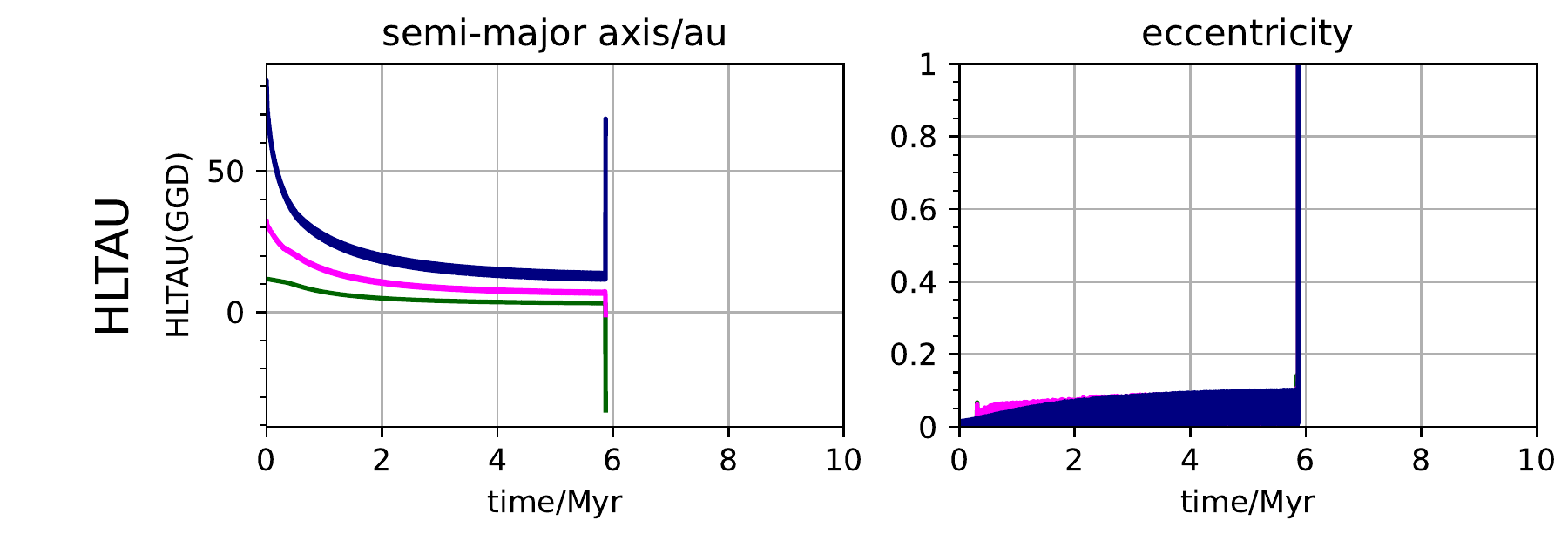}
		\end{minipage}
	\end{minipage}
	\caption{Left: configurations (just before instability) as well as the instability time of the systems that become unstable at the disk stage, when $ \alpha = \num{2e-3} $. The dashed circles represent the initial location and mass of the planetary systems. Right:Evolution of DL Tau(DDG) and HL Tau(GGD), which are both unstable at the disk stage when $ \alpha =\num{2e-3} $.}
	\label{fig:unstable_examples}
\end{figure}

At the disk stage, we evolve the systems with three different $ \alpha $ values, $ \num{5e-4} $, $ \num{e-3} $ and $ \num{2e-3} $. No system becomes unstable when $ \alpha $ is set to be $ \num{5e-4} $ and $ \num{e-3} $, while five runs from DL Tau and HL Tau become unstable during the disk stage when the large $ \alpha = \num{2e-3}$ is adopted. In Figure \ref{fig:unstable_examples}, we plot the final configurations of the planetary systems just before they get unstable. On the same figure, we also plot the time when instability happens, as shown by the grey bars. Both sets of DL Tau become unstable at around \SI{4}{Myr}, while the three HL Tau sets become unstable at around \SI{6}{Myr}. All systems are much compact and massive than their initial states, as represented by the dashed circles.

Figure \ref{fig:unstable_examples} also plots the evolution of two example planetary systems, DL Tau(DDG) and HL Tau(GGD), in terms of semi-major axis, mass, eccentricity and period ratio. Both of the systems become unstable due to the planet-planet scattering that is triggered by the close approach of the two planets. In the case of DL Tau, the instability happens at around \SI{4}{Myr}, with the innermost planet collide with the star. Similarly, for the HL Tau system, the outermost planet is scattered and ejected at around \SI{6}{Myr}. The scattering events in both systems are associated with strong excitation of eccentricity.

Since our numerical framework cannot accommodate these unstable cases, the simulation stops once the system becomes unstable. Such gravitationally excited systems may be important to form the Hot Jupiters and/or misaligned planets, as discussed in section \ref{sec:caveats}. 

\section{Dependence of time averaged $ \eta_{\rm per} $}
\label{app:eta_mass_number_dep}
\begin{figure}
	\gridline{\fig{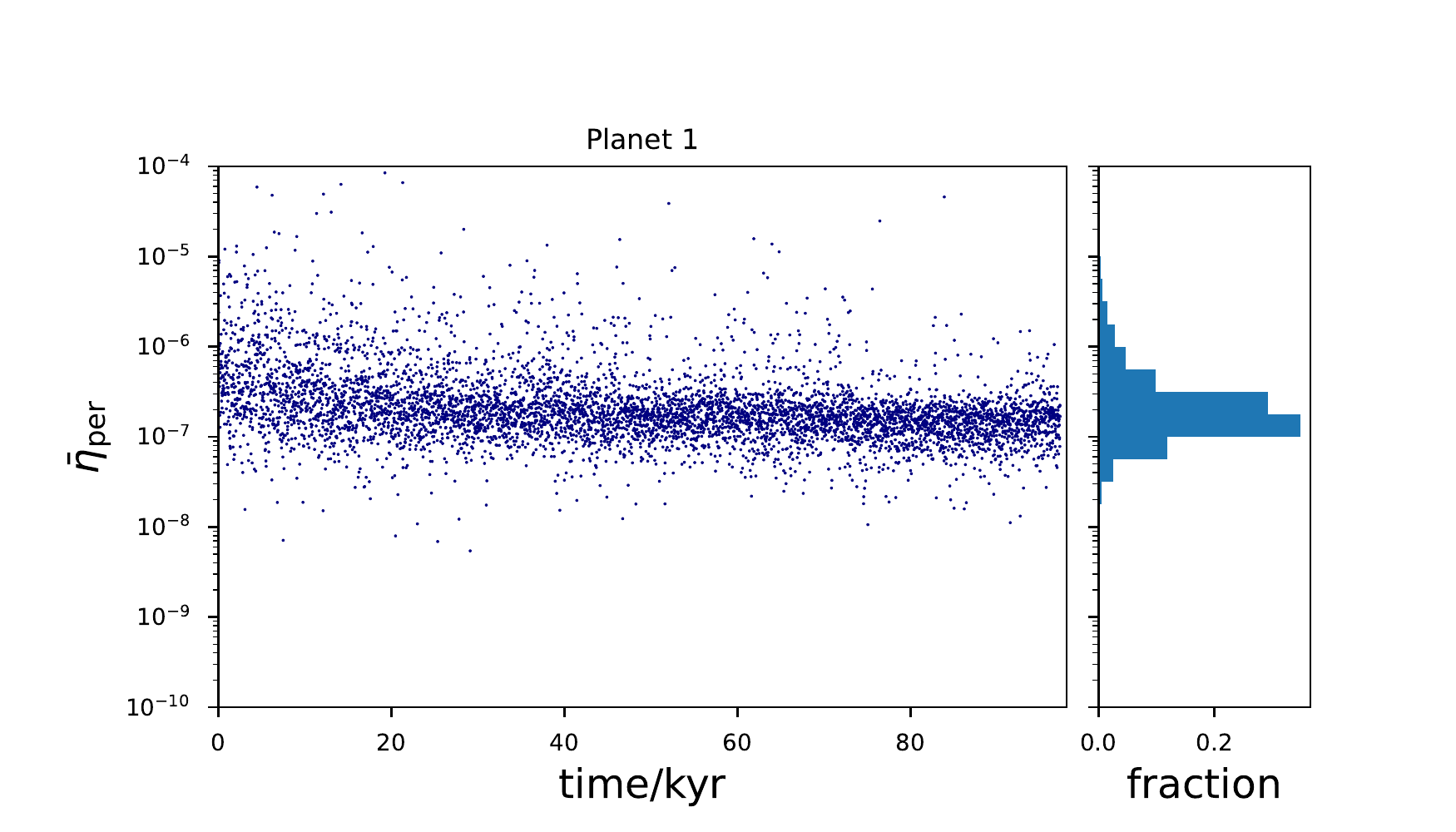}{.5\linewidth}{(a)}
	\fig{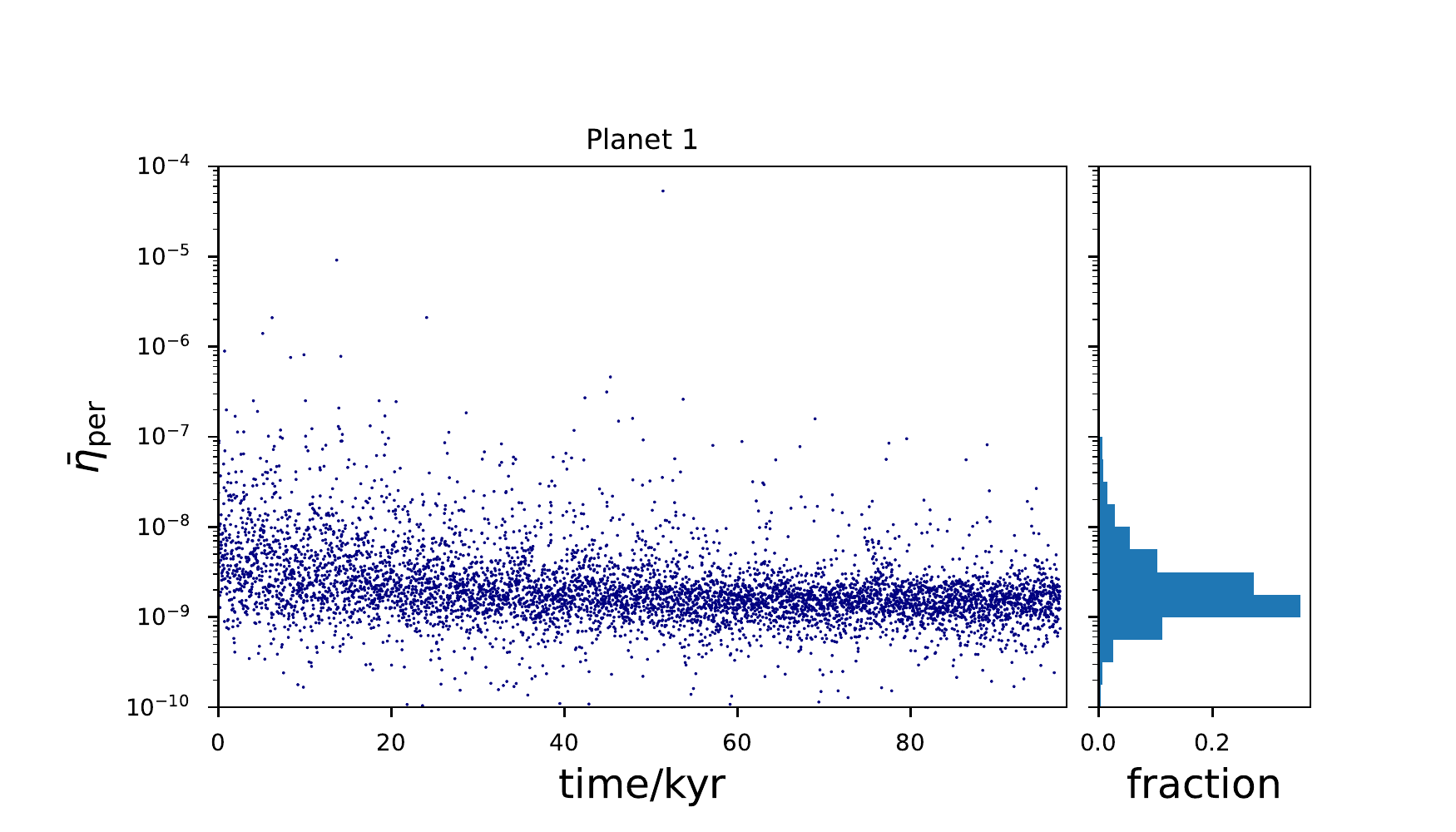}{.5\linewidth}{(b)}}
	\gridline{\fig{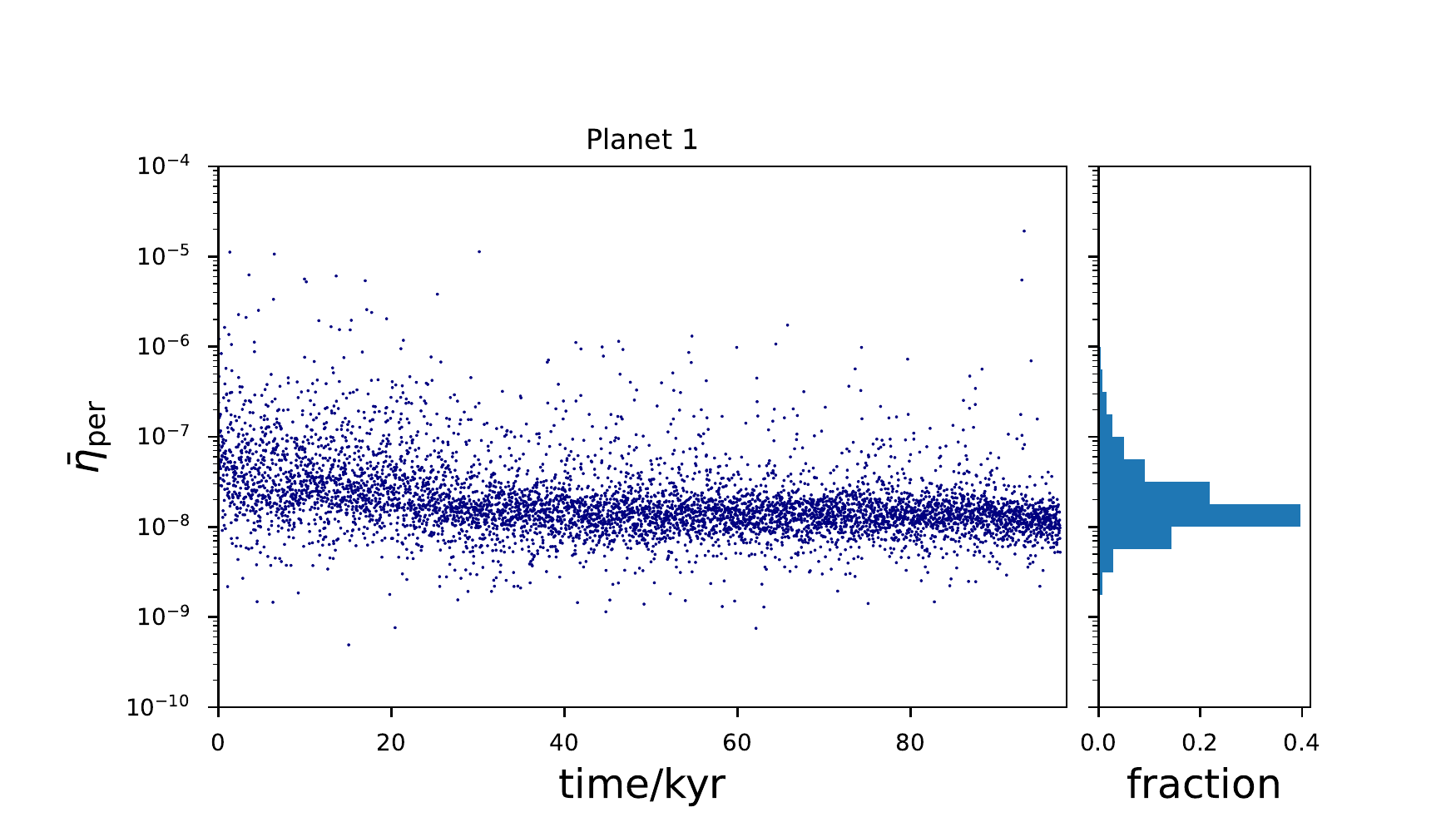}{.5\linewidth}{(c)}}
	\caption{Same as Figure \ref{fig:avg_perturb_force} (right panel) but with: (a) 10 times disk mass; (b) $ 10 \% $ disk mass; (c) $ 2000 $ planetesimal particles.}
	\label{fig:dep_pltmal_eta}
\end{figure}

We carry out simulations to explore how the time averaged $ \eta_{\rm per} $ depends on the planetesimal disk mass and number of particles, and the result for Planet 1 is plotted in Figure \ref{fig:dep_pltmal_eta}. By setting the planetesimal disk mass to be $ 10 $ times and $ 10 \% $ of the fiducial value while keeping the other parameters as the same, we found the peak of the histogram shift linearly with the change of the disk mass: when the disk is $ 10 $ times as massive as the fiducial case, the peak of $ \tilde{\eta}_{\rm per}  $ is close to \num{e-7}, which is about $ 10 $ times of that in the fiducial case $ \tilde{\eta}_{\rm per} (= \num{e-8}) $. Similarly,  when the disk mass is set to be $ 10 \% $ of the fiducial disk mass, the peak of $ \tilde{\eta}_{\rm per}  $ shifts to \num{e-9}. To investigate whether the number of particles can affect $ \tilde{\eta}_{\rm per}  $, we adopt the same disk mass as the fiducial case but introduce $ 2000 $ particles, twice as many as that in the fiducial case. However, as shown by Figure \ref{fig:dep_pltmal_eta} (c), the peak and distribution of $ \bar{\eta}_{\rm per} (= \num{e-8}) $ do not change with the number of particles.

\clearpage

\end{document}